\documentstyle[psfig,epsf,epsfig,psfrag]{article}
\def\photonatomrightt{\begin{picture}(3,1.5)(0,0)
                               \put(0,-0.75){\tencirc \symbol{2}}
                               \put(1.5,-0.75){\tencirc \symbol{1}}
                               \put(1.5,0.75){\tencirc \symbol{3}}
                               \put(3,0.75){\tencirc \symbol{0}}
                     \end{picture}
                    }

\def\photonrightthalf{\begin{picture}(15,1.5)(0,0)
                    \multiput(0,0)(3,0){5}{\photonatomrightt}
                 \end{picture}
                }

\def\fermionrighthalf{\begin{picture}(15,1)(0,0)  
                            \put(0,0){\line(1,0){7.5}}
                          \put(7.5,0){\line(1,0){7.5}}
                      \end{picture}
                     }
\def\fermionul{\begin{picture}(15,15)(0,0)
                        \put(0,0){\vector(-1,1){7.5}}
                        \put(-7.5,7.5){\line(-1,1){7.5}}
                  \end{picture}
                 }

\def\fermionur{\begin{picture}(15,15)(0,0)
                        \put(-15,-15){\vector(1,1){7.5}}
                        \put(-7.5,-7.5){\line(1,1){7.5}}
                  \end{picture}
                 }

\def\gaugebosonurhalf{\begin{picture}(15,15)(0,0)
                            \put(0,0){\line(1,1){15.0}}
                  \end{picture}
                 }

\def\gaugebosondrhalf{\begin{picture}(15,15)(0,0)
                            \put(0,0){\line(1,-1){15}}
                  \end{picture}
                 }
\def\gaugebosondrhalff{\begin{picture}(7.5,7.5)(0,0)
                            \put(0,0){\line(1,-1){7.5}}
                  \end{picture}
                 }

\newenvironment{Feynman}[3]{\begin{center}
                            \setlength{\unitlength}{#3 mm}
                            \begin{picture}(#1)(#2)
                            \thicklines
                           }{\end{picture} \end{center}}

\newcommand{\ee}{\mbox{$\mathrm{e}^{+}\mathrm{e}^{-}$}}
\newcommand{\ra}{\mbox{$\rightarrow$}}

\newcommand{\Zo} {{\mathrm {Z}}}
\newcommand{\db}    {{d_{\rm B}}}%
\newcommand{\dgz}  {{\Delta g_1^{\Zo}}}%
\newcommand{\dkg}   {{\Delta \kappa_\gamma}}%

\begin{document}

\thispagestyle{empty}
\def\thefootnote{\fnsymbol{footnote}}       

\begin{center}
\mbox{ }

\end{center}
\begin{flushright}
\Large
\mbox{\hspace{10.2cm} hep-ph/0502002} \\
\end{flushright}
\begin{center}
\vskip 1.0cm
{\Huge\bf
Higgs Physics: from LEP to a
}
\vspace{2mm}

{\Huge\bf
Future Linear Collider
}
\vskip 1cm
{\LARGE\bf Andr\'e Sopczak}\\
\smallskip
\Large Lancaster University

\vskip 2.5cm
\centerline{\Large \bf Abstract}
\end{center}

\vskip 3.0cm
\hspace*{-0.5cm}
\begin{picture}(0.001,0.001)(0,0)
\put(,0){
\begin{minipage}{\textwidth}
\Large
\renewcommand{\baselinestretch} {1.2}
New preliminary combined results from the LEP experiments
on searches for the Higgs boson beyond the Standard Model 
are presented. The new determination of the top quark mass
at the Tevatron in 2004 influences the interpretations of the 
LEP results in both, the Standard Model, and the Minimal 
Supersymmetric extension of the Standard Model. Higgs boson
physics will also be a major research area at the future 
Linear Collider. A review including new preliminary results on 
the potential for precision measurements is given.
\renewcommand{\baselinestretch} {1.}

\normalsize
\vspace{5.5cm}
\begin{center}
{\sl \large
Presented at the XVIIIth International Workshop on High Energy Physics and 
Quantum Field Theory (QFTHEP), 2004, St. Petersburg, Russia; and at the\\ 
Xth Workshop on Nuclear Physics, WONP'2005, Havana, Cuba.
\vspace{-6cm}
}
\end{center}
\end{minipage}
}
\end{picture}
\vfill


\newpage
\thispagestyle{empty}
\mbox{ }
\newpage
\setcounter{page}{1}

\begin{center}
{\Large \bf Higgs Physics: from LEP to a Future Linear Collider} \\

\vspace{4mm}

Andr\'e Sopczak\\
Lancaster University, UK. E-mail: andre.sopczak@cern.ch
\end{center}

\begin{abstract}
New preliminary combined results from the LEP experiments
on searches for the Higgs boson beyond the Standard Model 
are presented. The new determination of the top quark mass
at the Tevatron in 2004 influences the interpretations of the 
LEP results in both, the Standard Model, and the Minimal 
Supersymmetric extension of the Standard Model. Higgs boson
physics will also be a major research area at the future 
Linear Collider. A review including new preliminary results on 
the potential for precision measurements is given.
\end{abstract}

The LEP experiments took data between August 1989 and November 2000
at center-of-mass energies first around the Z resonance (LEP-1) and from 1996
up to 209 GeV (LEP-2). In 2000 most data was taken around 206 GeV.
The LEP accelerator operated very successfully and 
a total luminosity of ${\cal L} = 2461$ pb$^{-1}$ was accumulated
at LEP-2 energies.
Data-taking ended on 3 November 2000, although some data excess was 
observed in searches for the Standard Model (SM) Higgs boson with 115~GeV mass.
In the following sections, these aspects are addressed:
1) SM candidates and mass limits;
2) coupling limits;
3) the Minimal Supersymmetric extension of the SM (MSSM):
   dedicated searches, three-neutral-Higgs-boson hypothesis,
   benchmark and general scan mass limits;
4) CP-violating models;
5) invisible Higgs boson decays;
6) neutral Higgs bosons in the general two-doublet Higgs model;
7) Yukawa Higgs boson processes $\rm b\bar b h$ and $\rm b\bar b A$;
8) singly-charged Higgs bosons;
9) doubly-charged Higgs bosons;
10) fermiophobic Higgs boson decays $\rm h\ra WW, ZZ, \gamma\gamma$;
11) uniform and stealthy Higgs boson scenarios;
12) summary of LEP limits;
13) the International Linear Collider (ILC);
14) SM Higgs physics;
15) beyond the SM.

While the results from Standard Model Higgs boson searches are final~\cite{sm},
the results of searches in extended models are mostly preliminary~\cite{summer2004}.
Limits are given at 95\% CL. The Linear Collider aspects are based on
Refs.~\cite{ichep02,paris04}.

\vspace*{-0.2cm}
\section{SM Candidates and Mass Limits}
\vspace*{-0.2cm}

\begin{figure}[htb]
\vspace*{-0.5cm}
\begin{center}
\includegraphics[scale=0.26]{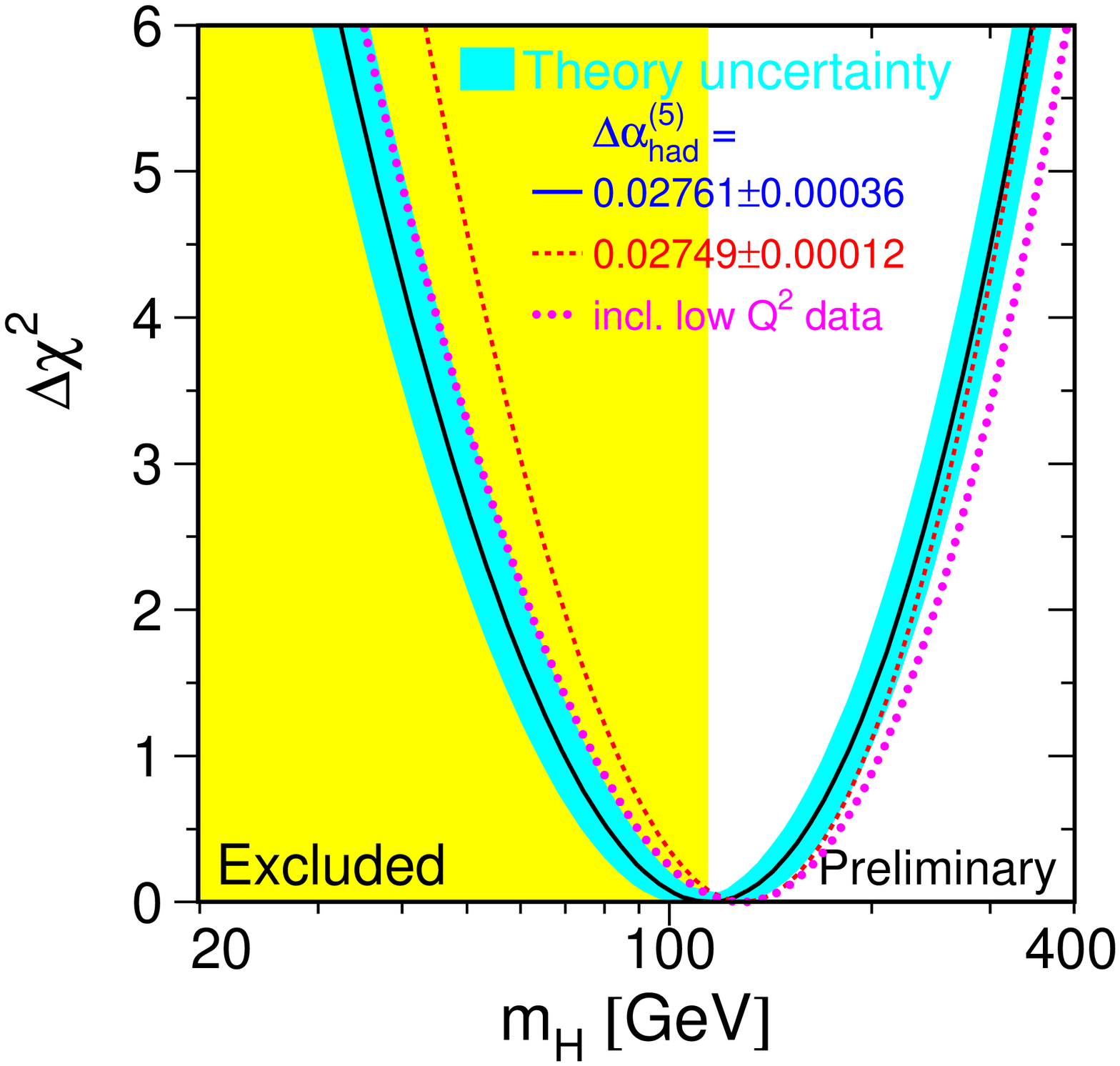}\hfill
\includegraphics[scale=0.26]{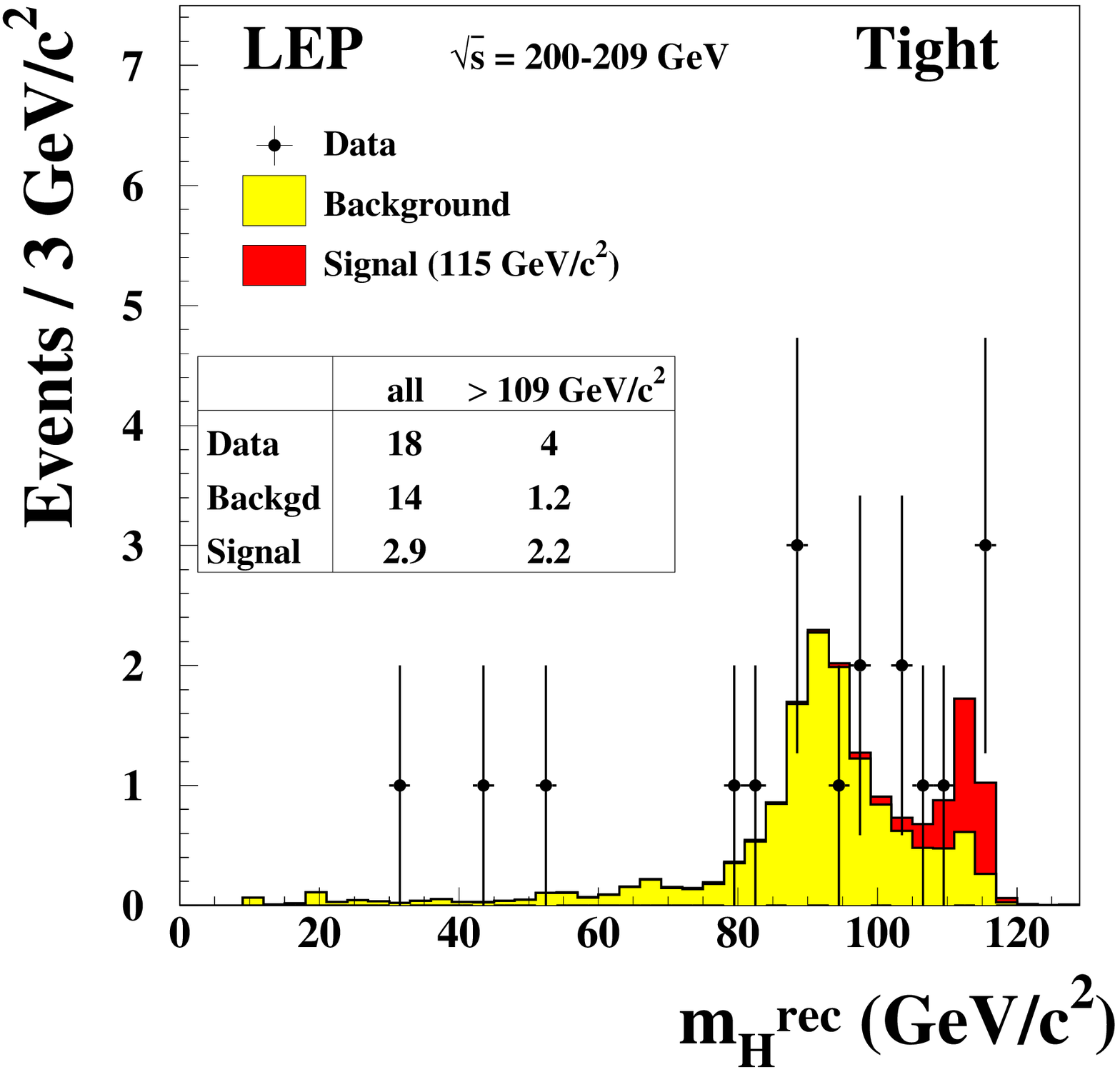}\hfill
\includegraphics[scale=0.26]{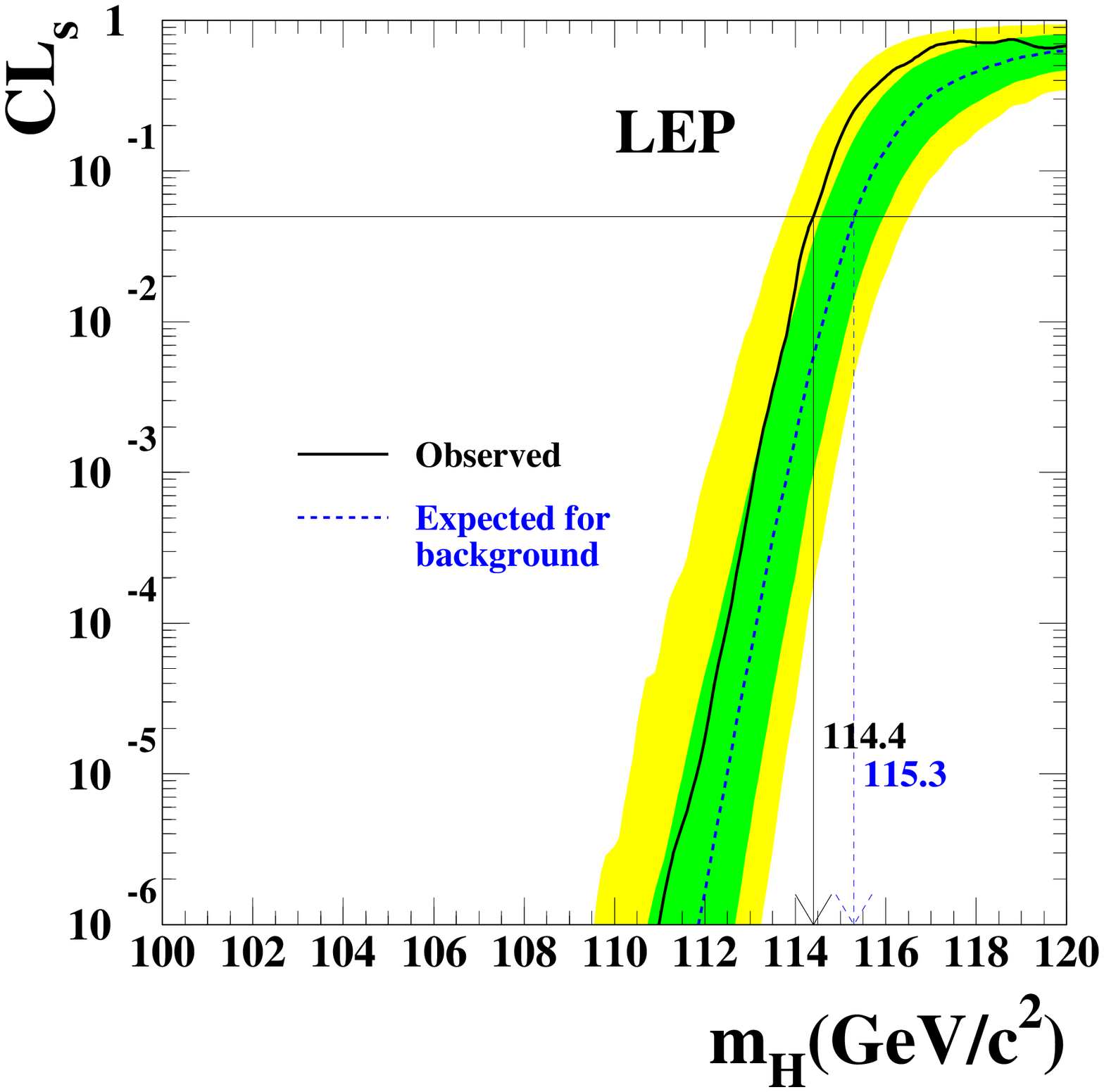}
\end{center}
\vspace*{-0.6cm}
\caption{
Left: $\Delta \chi^2 = \chi^2 - \chi^2_{\rm min}$ vs. Higgs boson mass.
The band represents an estimate of the theoretical errors due to
higher order corrections. The dashed curves reflect the evaluation for
different $\Delta(\alpha_{\rm had})$.
Center: reconstructed mass of SM Higgs boson candidates
           for tight selection cuts.
Right: SM Higgs boson mass limit.
The 1$\sigma$ and 2$\sigma$ error bands on the expected 
limit for background are indicated (shaded areas).
\label{fig:sm_limit}}
\vspace*{-0.5cm}
\end{figure}

From precision electro-weak measurements, the Higgs boson of the SM has a stringent upper 
mass limit as shown in Fig.~\ref{fig:sm_limit} (left plot).
The preferred value for the SM Higgs boson mass is 114 GeV. This value increased
due to the new top mass measurements~\cite{newtop04}
 and it is now very close to mass limit from direct
searches.
Including both the experimental and the theoretical uncertainties, the mass of 
the SM Higgs boson is lower than about 260 GeV (one-sided 95\% upper CL).
The figure shows also the reconstructed mass of the Higgs boson 
candidates for a 115~GeV Higgs boson hypothesis. In addition, the observed
SM Higgs boson mass limit and the expected limit for a background-only
hypothesis are shown.

\clearpage

\vspace*{-0.2cm}
\section{Coupling Limits}
\vspace*{-0.2cm}

\begin{figure}[htb]
\vspace*{-0.85cm}
\begin{center}
\includegraphics[scale=0.26]{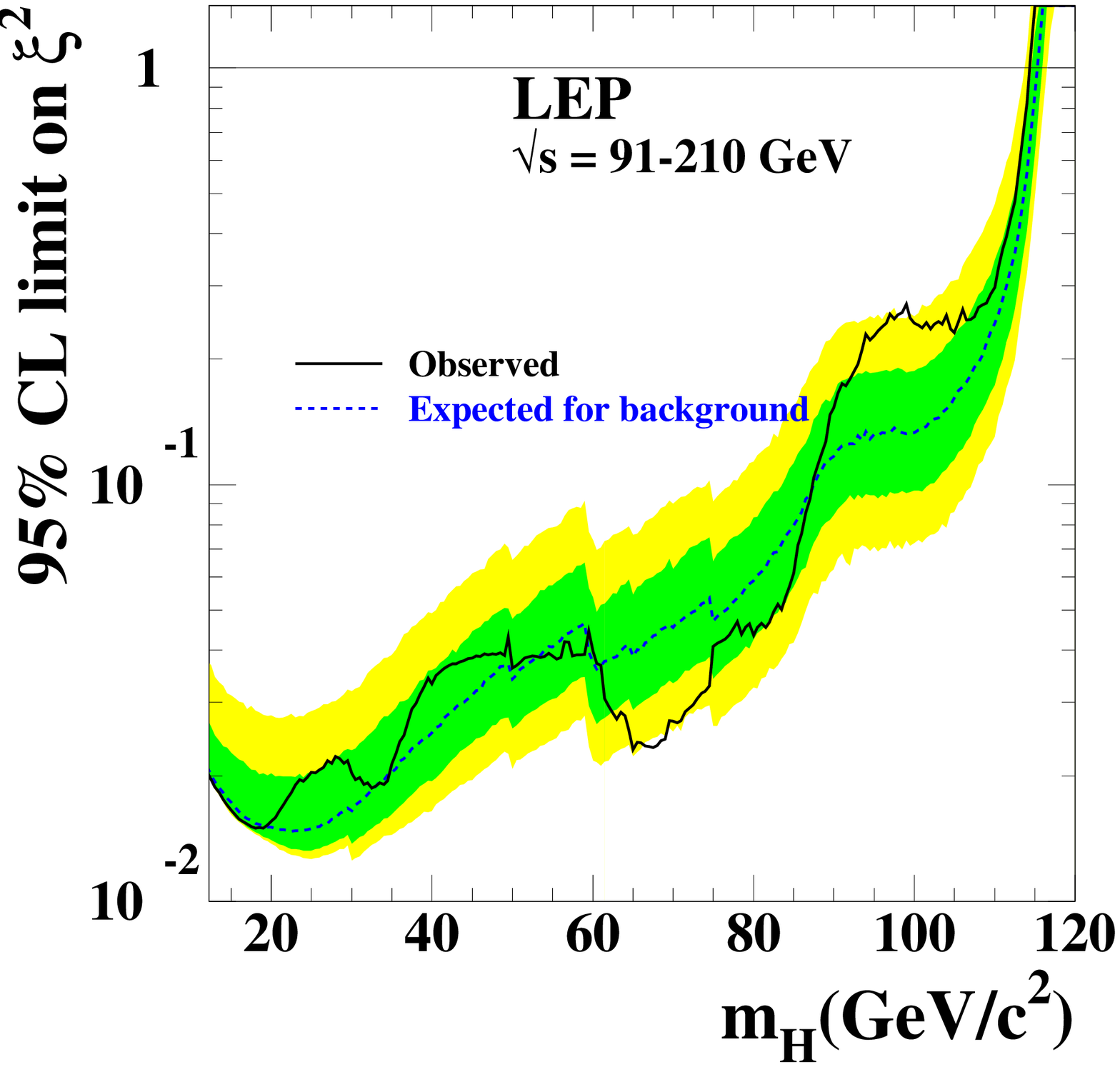}\hfill
\includegraphics[scale=0.26]{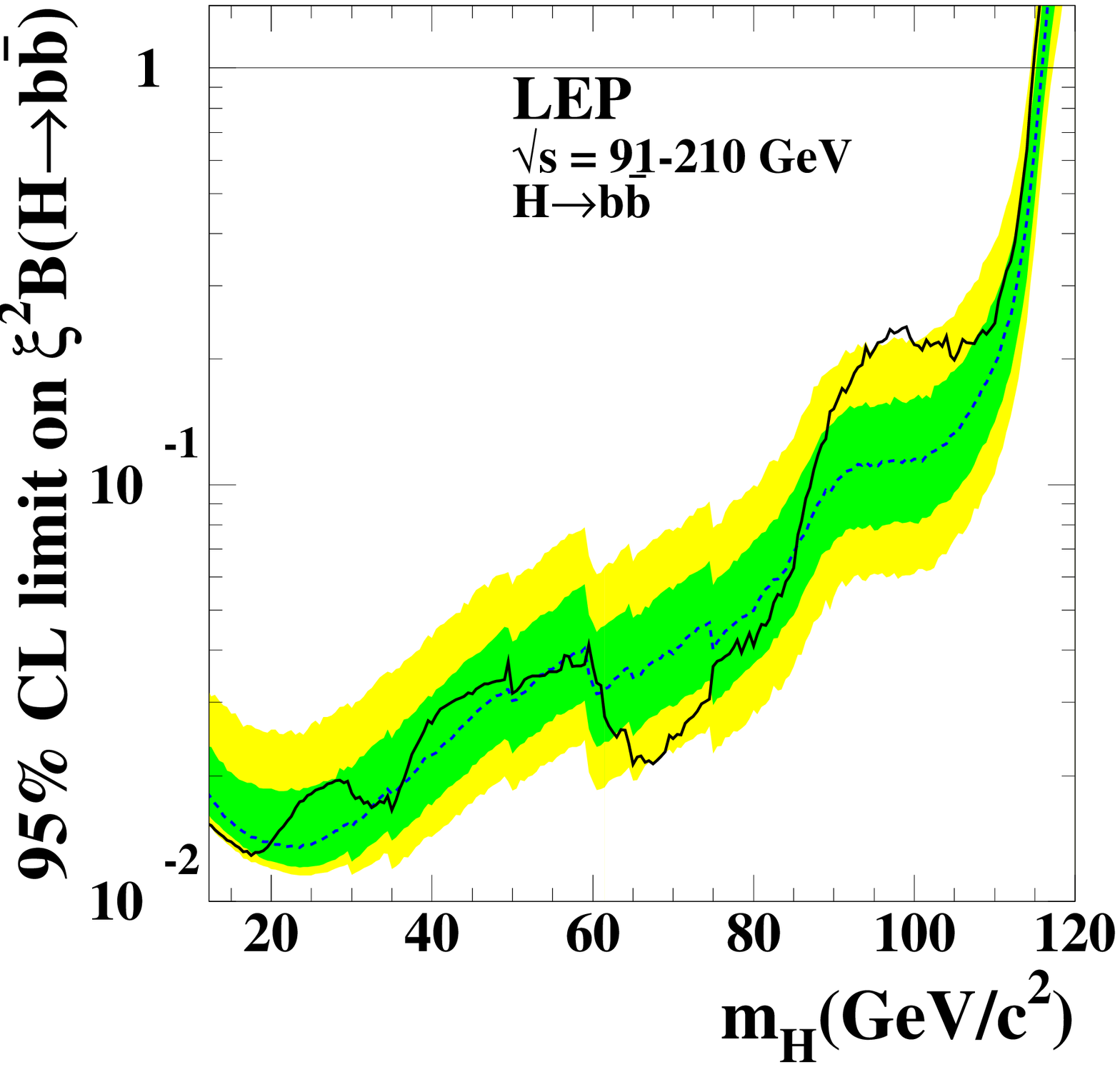}\hfill
\includegraphics[scale=0.26]{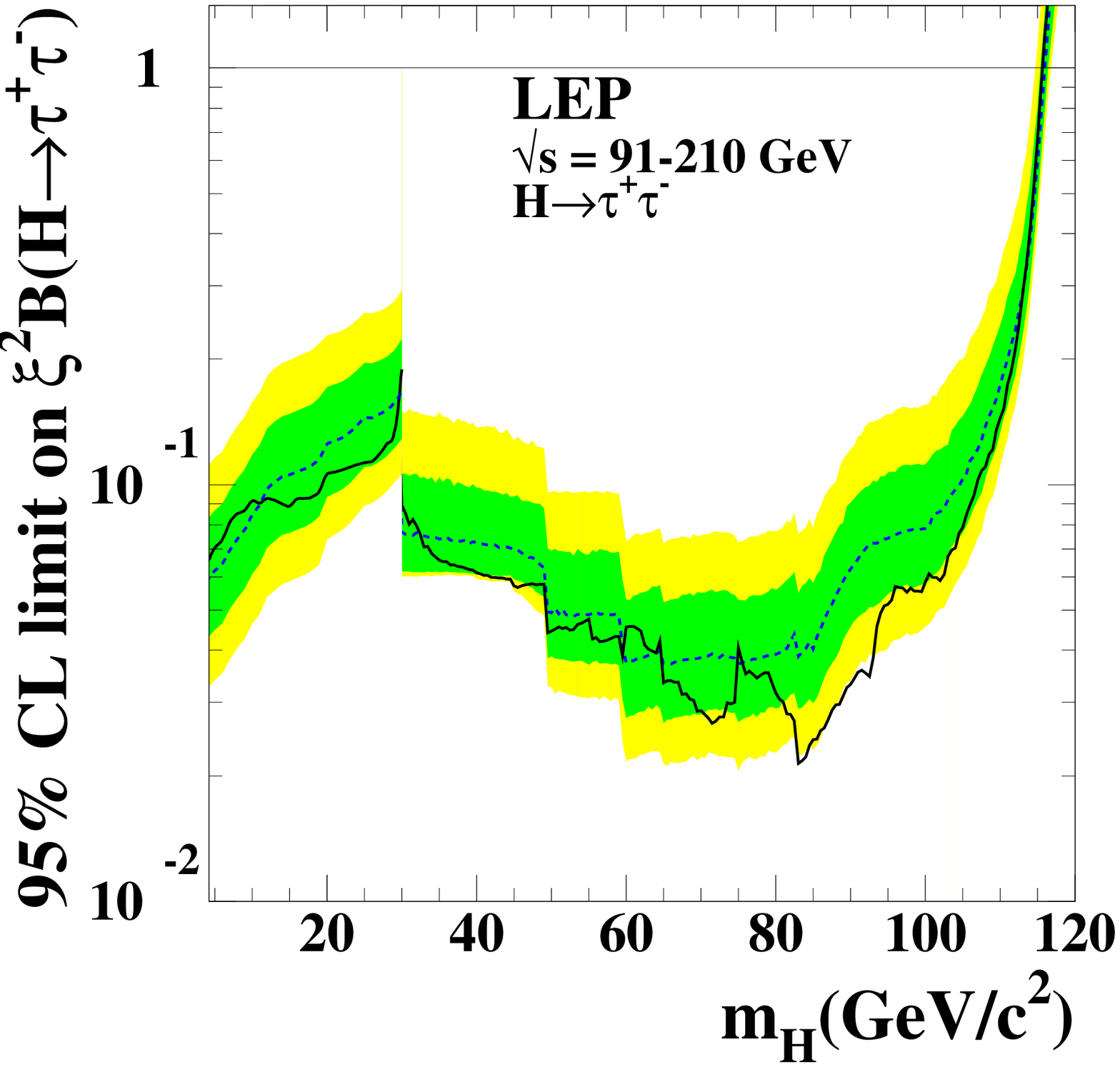}
\end{center}
\vspace*{-0.8cm}
\caption{Coupling limits.
Left: SM decay mode.
Center: b-quark decay mode. Right: $\tau$-lepton\,decay mode. The 1$\sigma$ and 2$\sigma$ 
error bands on the expected limit for background are indicated (shaded areas).
\label{fig:limits}}
\end{figure}

Figure~\ref{fig:limits} shows coupling limits assuming the Higgs boson decays with 
SM branching fractions and a SM production rate reduced by 
$\xi^2 = (g_{\rm HZZ}/g_{\rm HZZ}^{\rm SM})^2$. Coupling limits are also 
presented for b-quark and $\tau$-lepton decay modes.

In searches for flavor-independent hadronic Higgs boson decays,
Fig.~\ref{fig:flind} (left and center plot) shows no indication of a signal
for the process
$\rm hZ\ra q\bar q \ell^+\ell^-$ above the 
$\rm ZZ\ra q\bar q \ell^+\ell^-$ background.
The figure shows also combined LEP limits.
Remarkably, these mass limits for flavor-blind hadronic decays 
are close to the SM decay mode limits.
Figure~\ref{fig:aleph_had} (left and center plot) shows results from
a related analysis for flavor-independent hadronic Higgs boson decays.
The data agrees well with the expected background from W and Z production
and resulting limits are shown.
Anomalous couplings to the Higgs boson can be parametrized 
with $\xi$, $d,~\db,~\dgz,~\dkg$.
In addition to the $\xi$ limits, the parameters $d,~\db,~\dgz,~\dkg$
are constrained as shown in Fig.~\ref{fig:aleph_had} (right plot).

\newcommand{\mm}      {\mbox{$\rm \mu^+ \mu^- $}}

\begin{figure}[htb]
\vspace*{-0.6cm}
\begin{center}
\includegraphics[width=0.3\textwidth,clip]{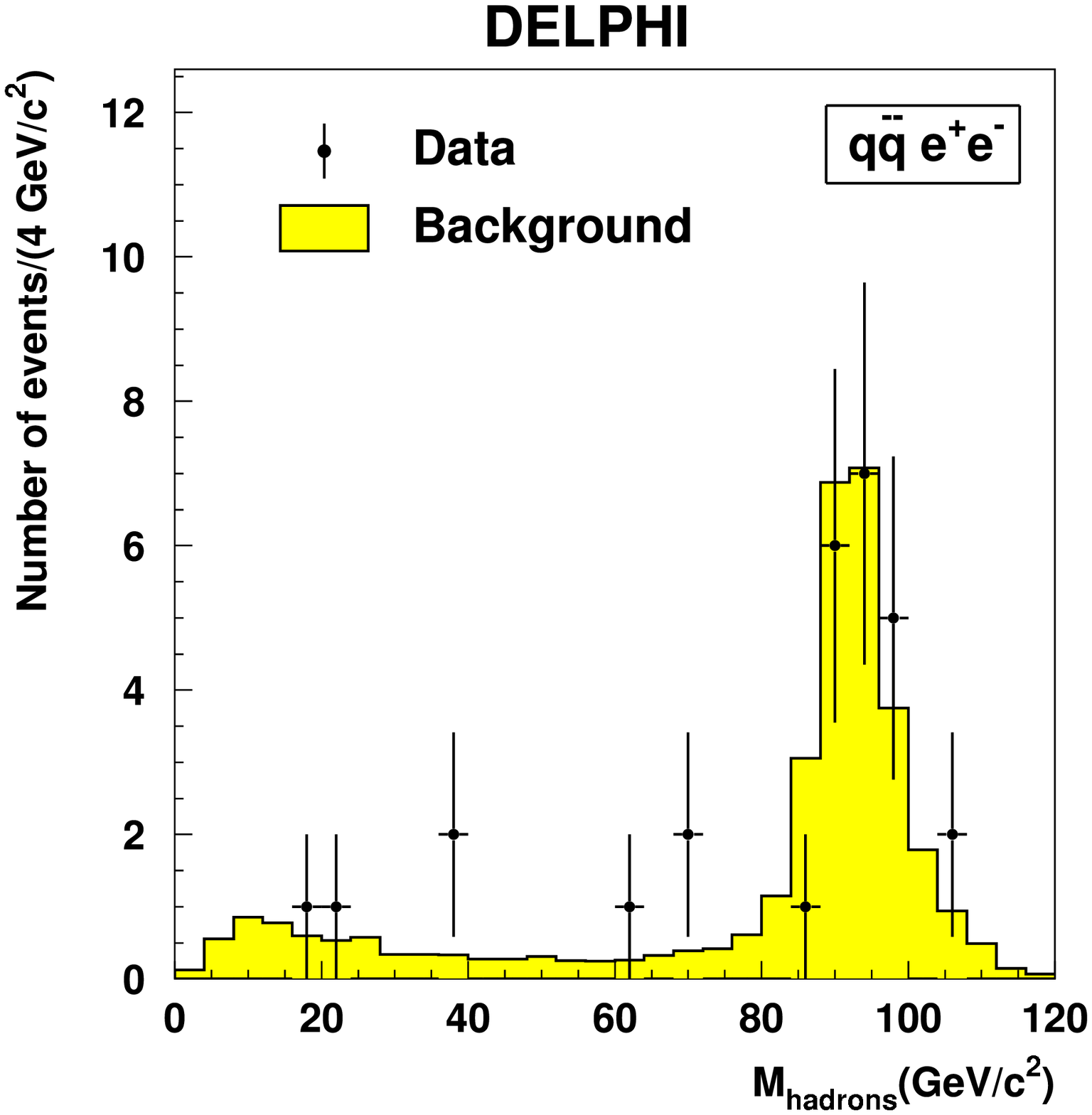}\hfill
\includegraphics[width=0.3\textwidth,clip]{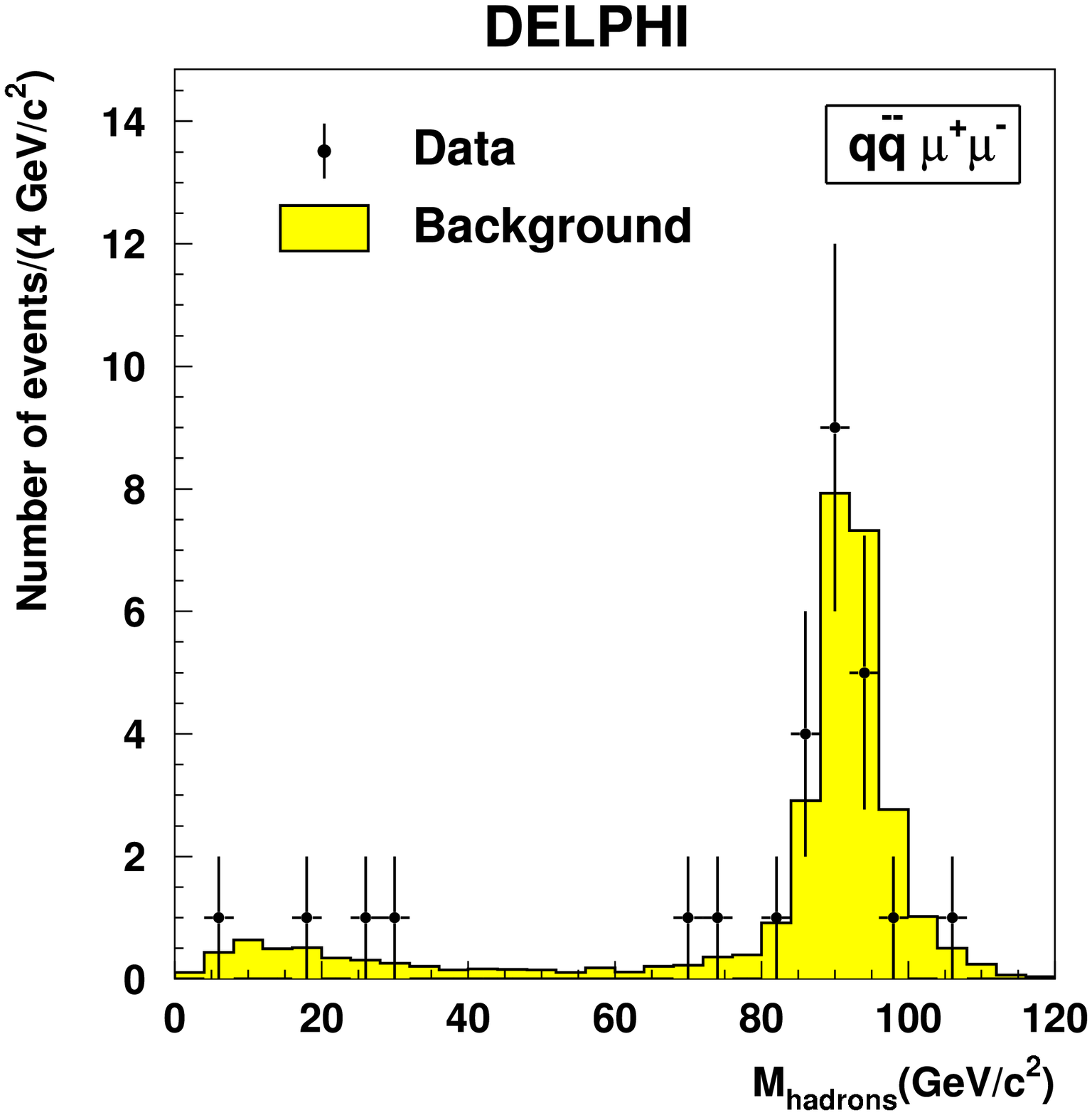}\hfill
\includegraphics[scale=0.25]{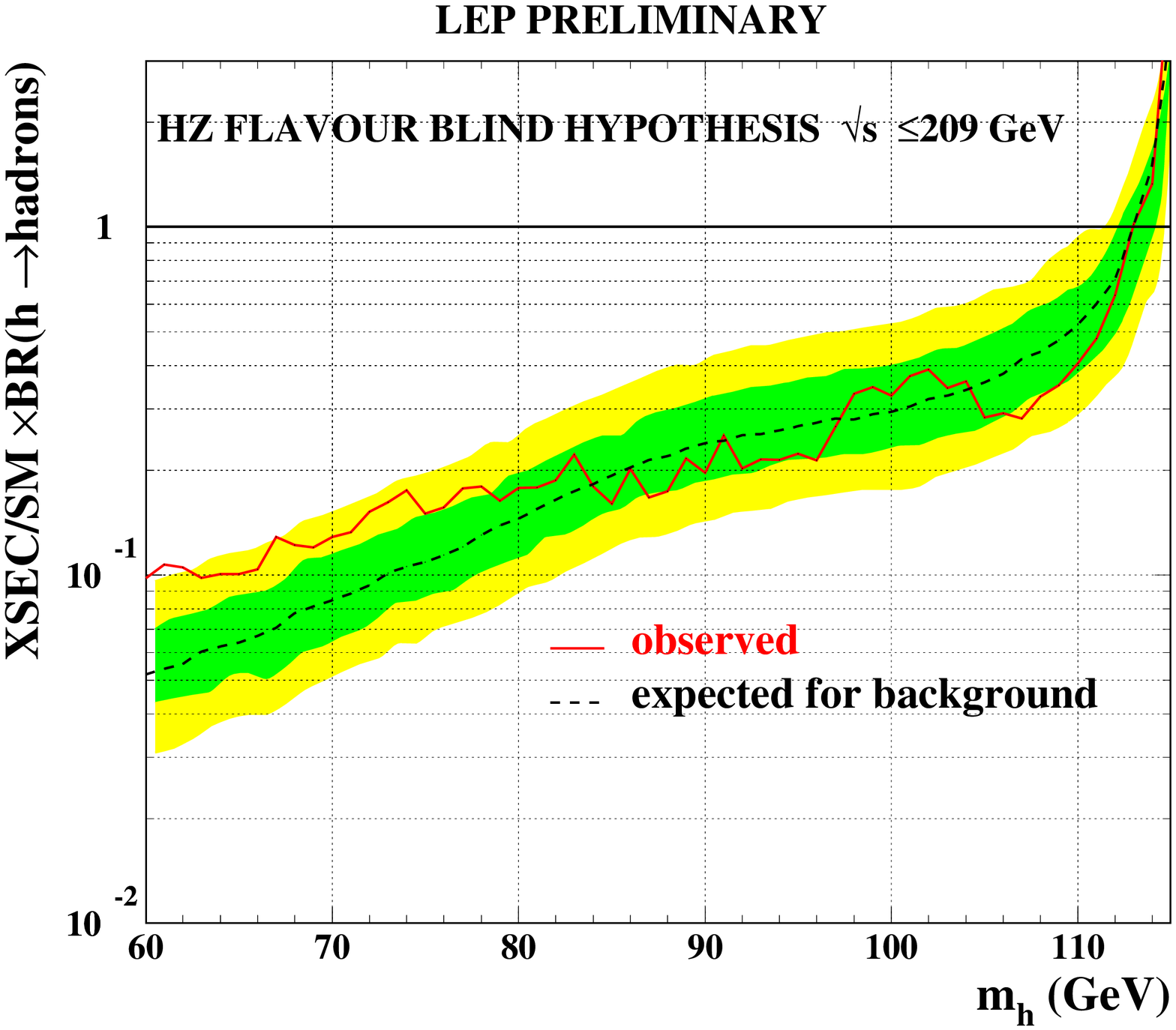}
\end{center}
\vspace*{-0.7cm}
\caption{Left and center: 
         no indication of a signal for the process 
         $\rm hZ\ra q\bar q \mm, q\bar q \ee$ above the 
         $\rm ZZ\ra q\bar q \mm, q\bar q \ee$ background.
Right: combined LEP flavor-independent limits from searches for 
        hadronic Higgs boson decays.
\label{fig:flind}}
\vspace*{-0.5cm}
\end{figure}

\newcommand{\mctwo}   {\,GeV}
\newcommand{\xitau} {$\xi^{2}_\tau$}
\newcommand{\xihad} {$\xi_{\rm {had}}^2$}

\begin{figure}[htb]
\vspace*{-3cm}
\begin{center}
\begin{picture}(140,140)
\psfrag{a1}[t][b][.9]{~\hspace{-23mm}$m_{\rm H}$ (\mctwo)} 
\psfrag{a2}[b][t][1]{~\hspace{-0mm} \xitau}
\psfrag{a3}[t][b][0.9]{~\hspace{-23mm}$m_{\rm H}$ (\mctwo)} 
\psfrag{a4}[b][t][1]{~\hspace{-3mm} \xihad}
\psfrag{b1}[t][b][.9]{~\hspace{-23mm}$m_{\rm H}$ (\mctwo)} 
\psfrag{b2}[b][t][1]{~\hspace{-30mm} number of events}

\put(-10,-70){\epsfxsize 0.35\textwidth\epsfbox{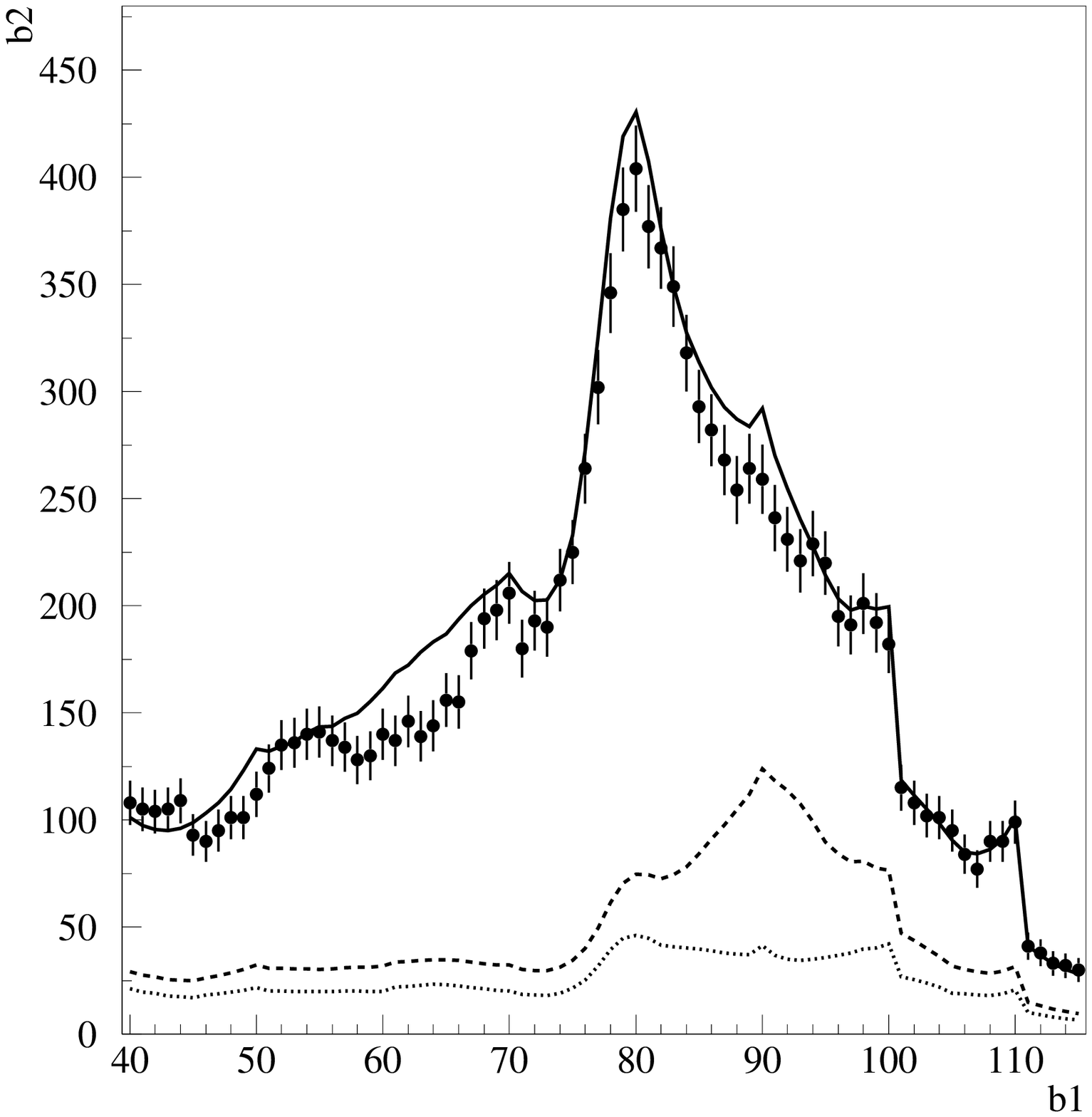}}
\put(25,30){ALEPH } 
\end{picture}\hfill
\begin{picture}(140,140)
\psfrag{a1}[t][b][.9]{~\hspace{-23mm}$m_{\rm H}$ (\mctwo)} 
\psfrag{a2}[b][t][1]{~\hspace{-0mm} \xitau}
\psfrag{a3}[t][b][0.9]{~\hspace{-23mm}$m_{\rm H}$ (\mctwo)} 
\psfrag{a4}[b][t][1]{~\hspace{-3mm} \xihad}
\psfrag{b1}[t][b][.9]{~\hspace{-23mm}$m_{\rm H}$ (\mctwo)} 
\psfrag{b2}[b][t][1]{~\hspace{-30mm} number of events}

\put(0,-70){\epsfxsize 0.35\textwidth\epsfbox{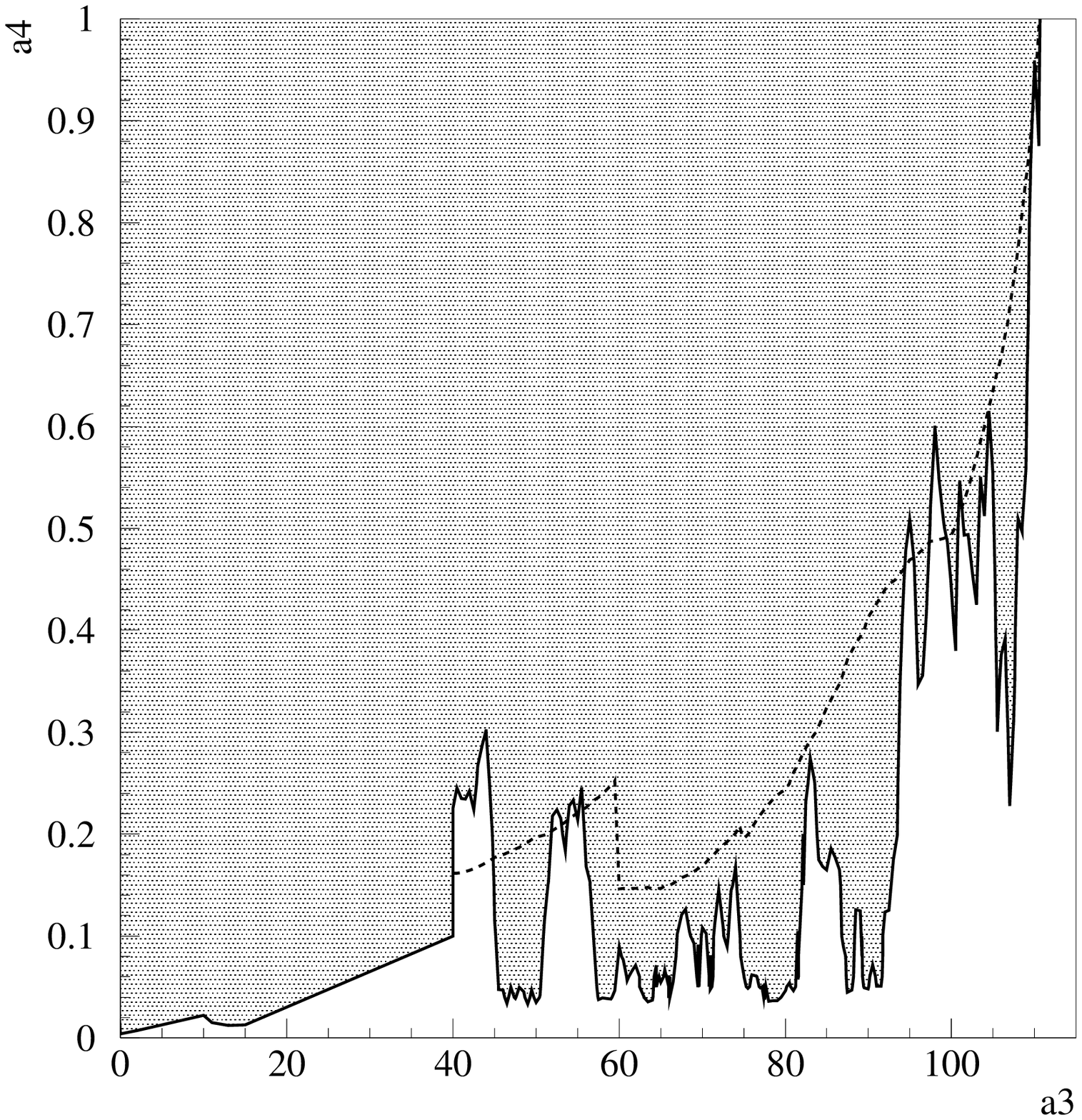}}
\put(35,30){ALEPH } 
\end{picture}\hfill
\begin{tabular}{cc}
   \includegraphics[width=0.15\textwidth]{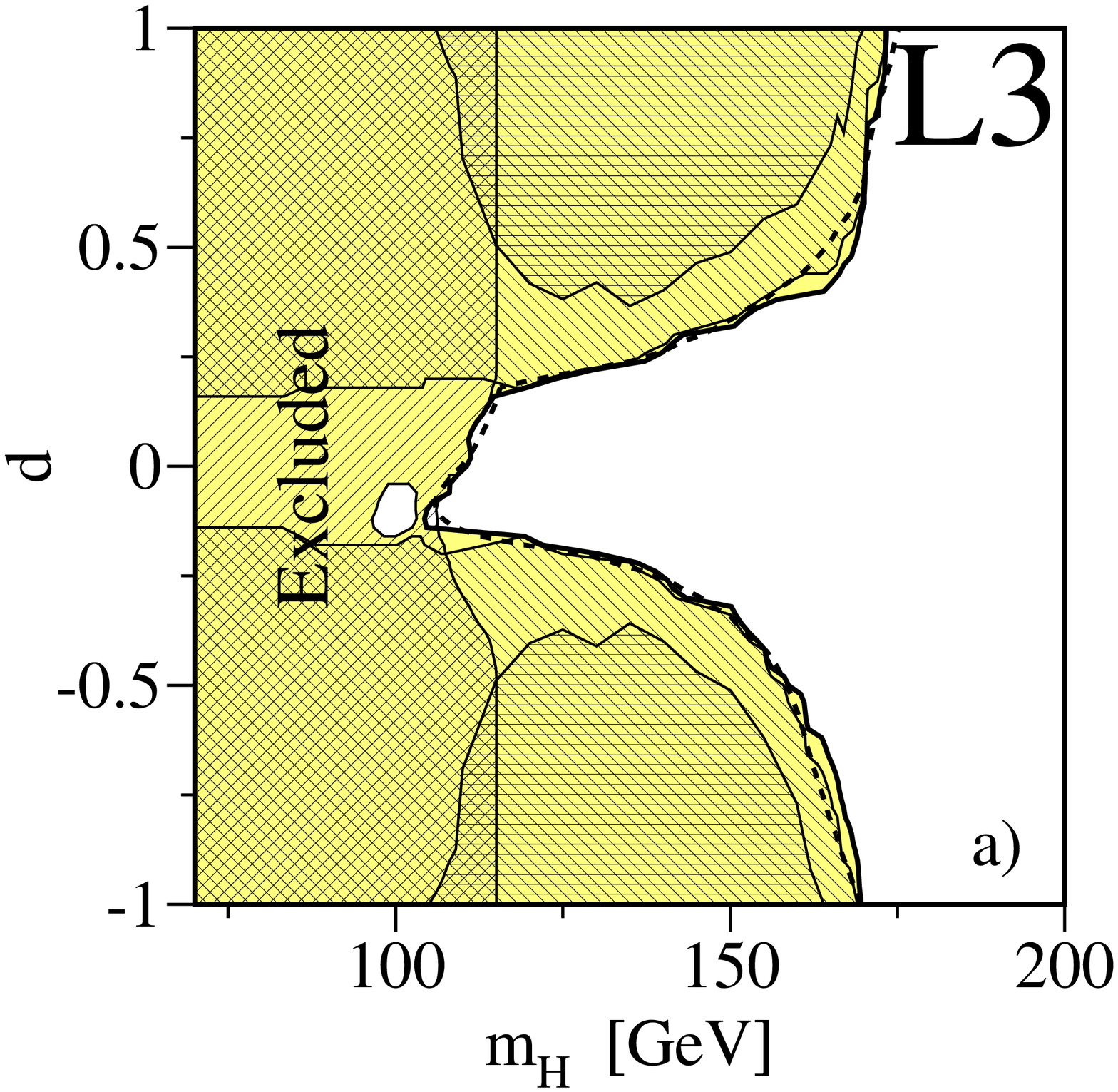} &
   \hspace{-4mm}\includegraphics[width=0.15\textwidth]{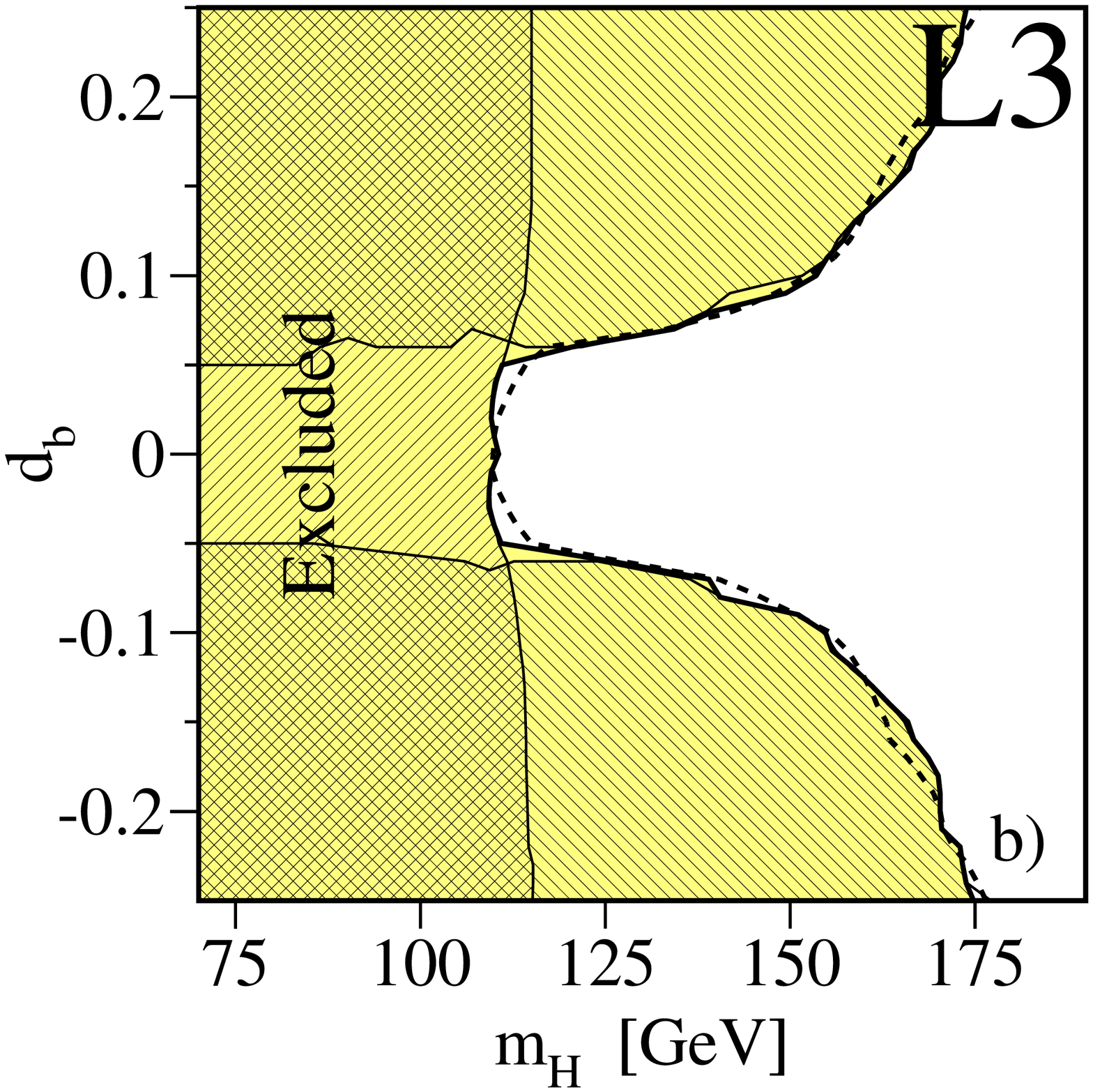}\hspace{-4mm} \\
   \includegraphics[width=0.15\textwidth]{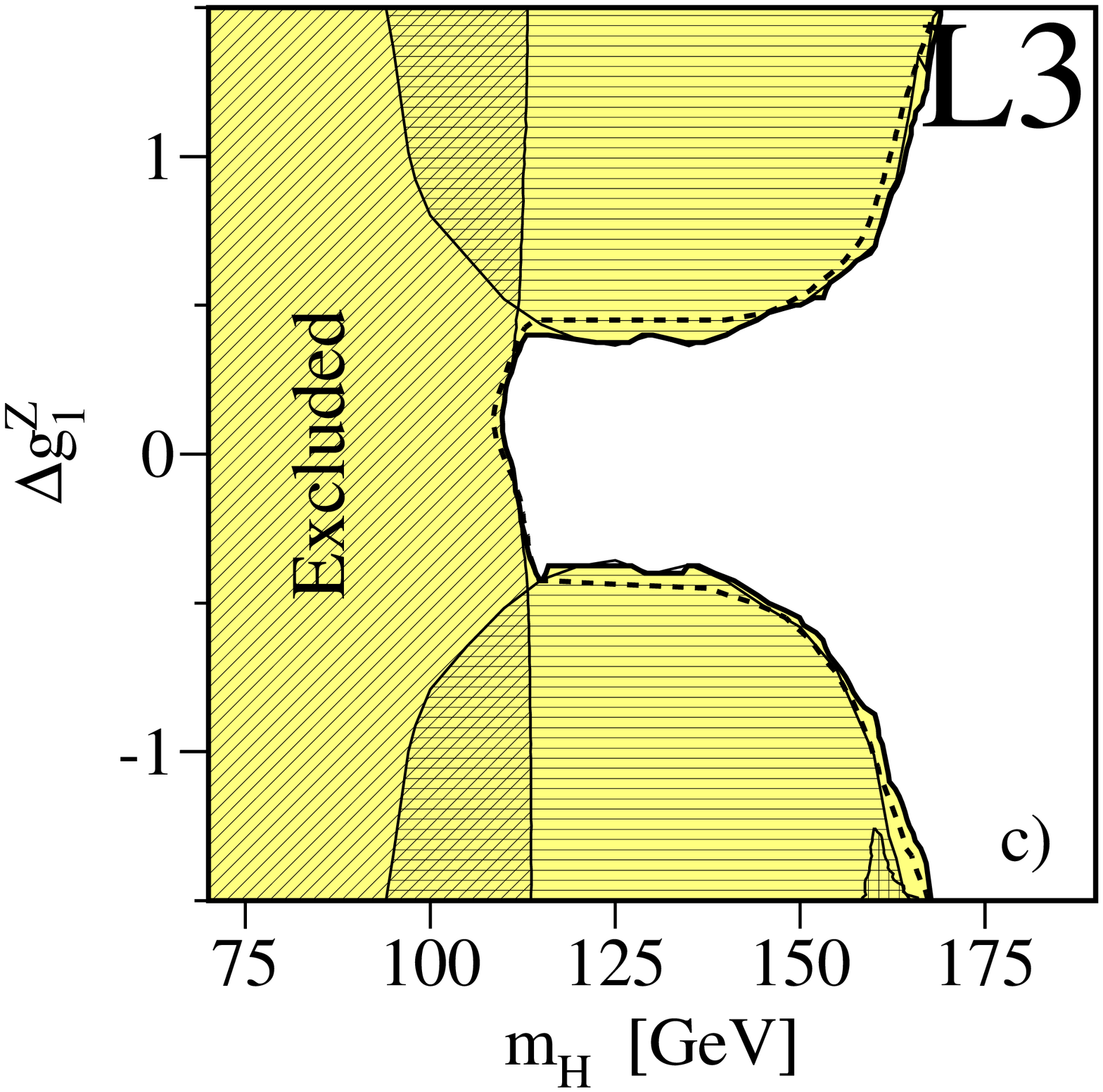} &
   \hspace{-4mm}\includegraphics[width=0.15\textwidth]{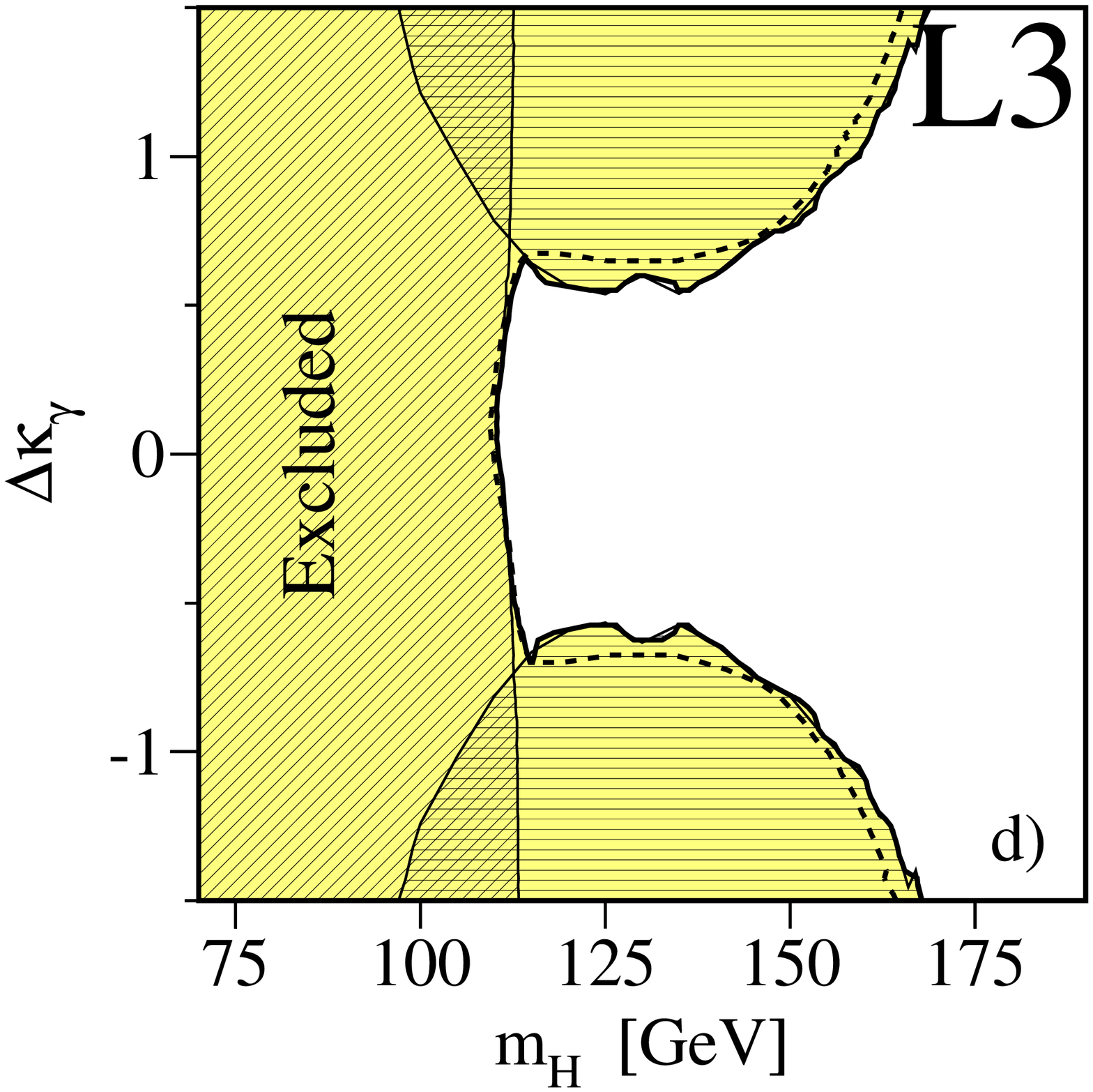}\hspace{-4mm} \\
\end{tabular}
\end{center}
\vspace*{-0.5cm}
\caption
{SM Higgs boson. Left: hadronic Higgs boson decay mode: 
 reconstructed mass for data, W and Z background.
 Center: hadronic Higgs boson decay mode: coupling limits. 
 Right: limits from searches for anomalous couplings.
\label{fig:aleph_had}}
\end{figure}

\newcommand{\MWnew} {{m_{\mathrm{W}}}}%
\newcommand{\W}  {{\mathrm{W}}}%

\section{Minimal Supersymmetric Extension of the SM (MSSM)}

\subsection{Benchmark Limits and Dedicated Low \boldmath$m_{\rm A}$ and $\rm h\ra AA$\unboldmath\ Searches}

\begin{figure}[htb]
\vspace*{-0.7cm}
\begin{center}
\includegraphics[scale=0.22]{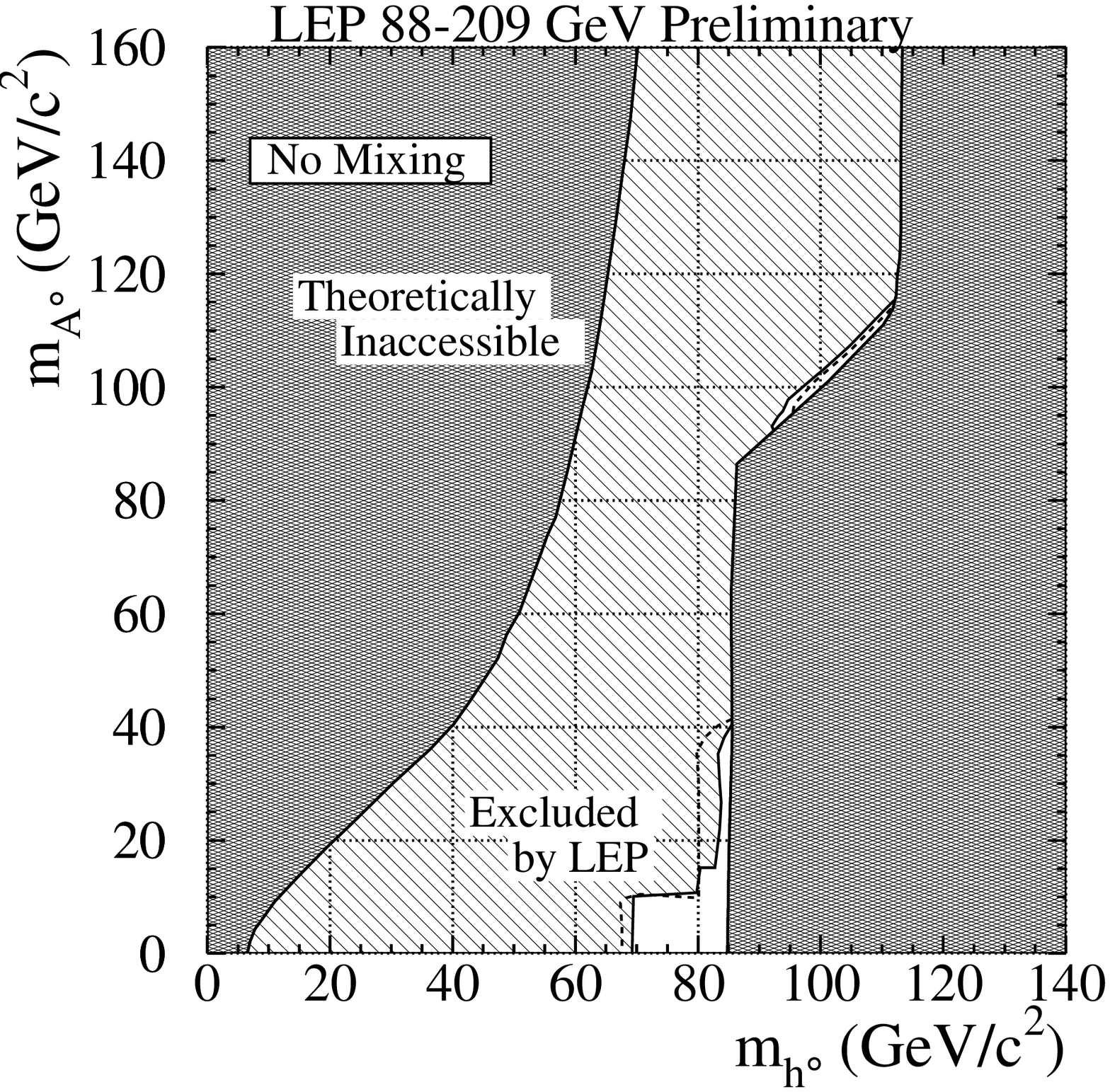}\hfill
\includegraphics[scale=0.22]{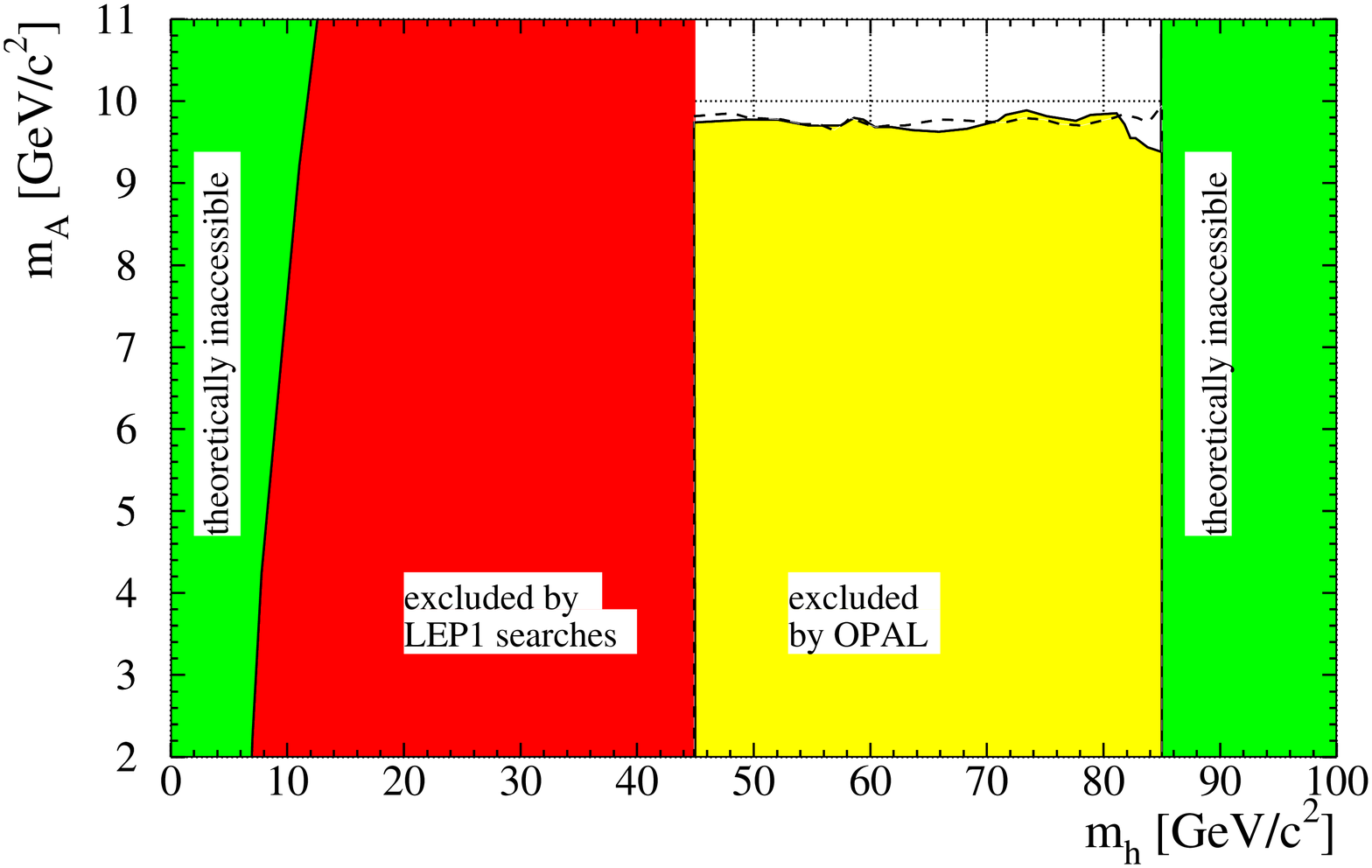}\hfill
\includegraphics[scale=0.29]{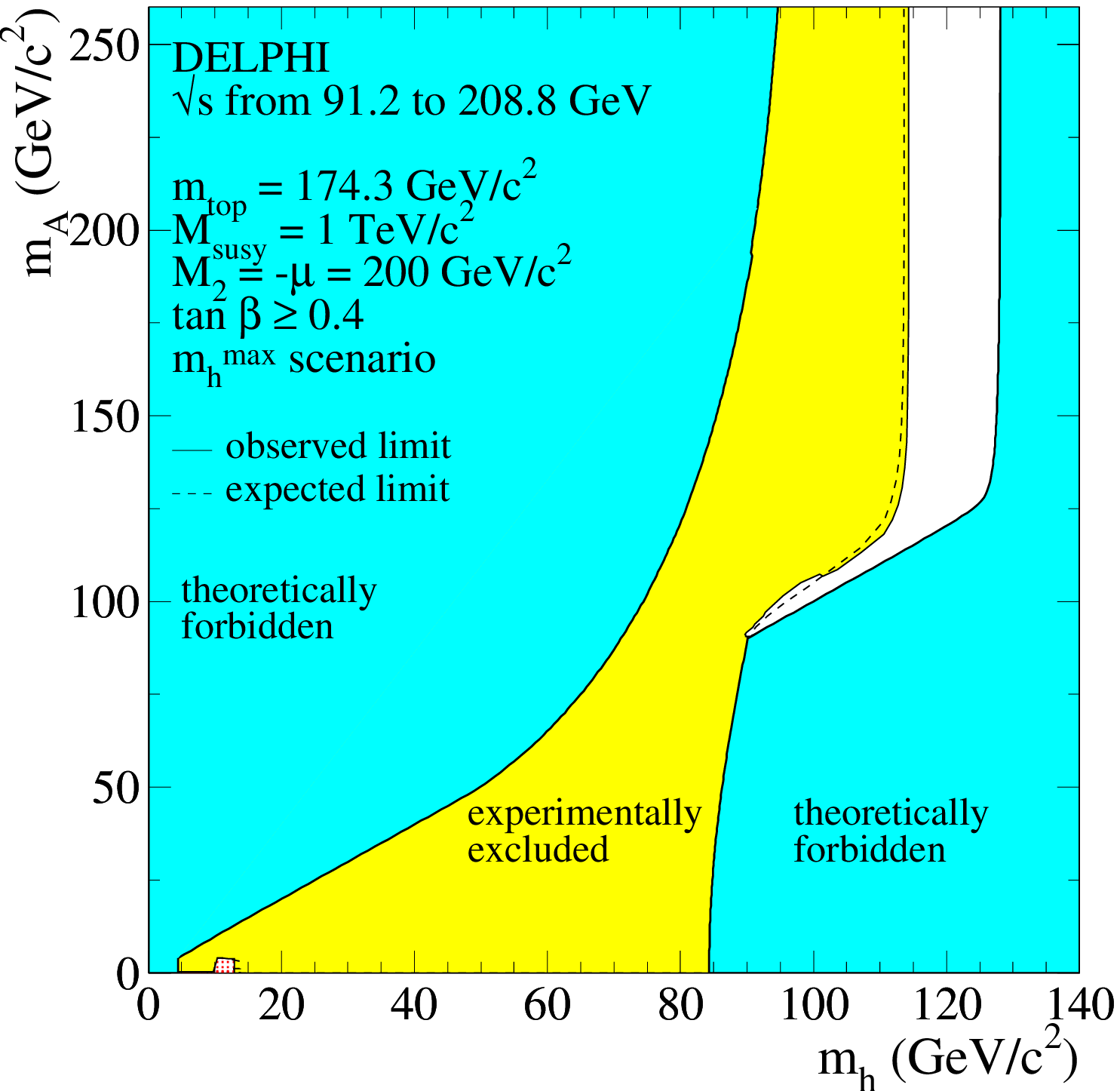}
\end{center}
\vspace*{-0.7cm}
\caption{MSSM. Left: unexcluded mass region for a light A boson
           in the no-mixing scalar top benchmark scenario.
           Center: excluded mass region by dedicated searches
           for a light A boson.
           Right: mass limits in the maximum h-mass benchmark scenario.                
\label{fig:light} }
\end{figure}

\begin{figure}[tp]
\vspace*{-0.5cm}
\epsfig{file=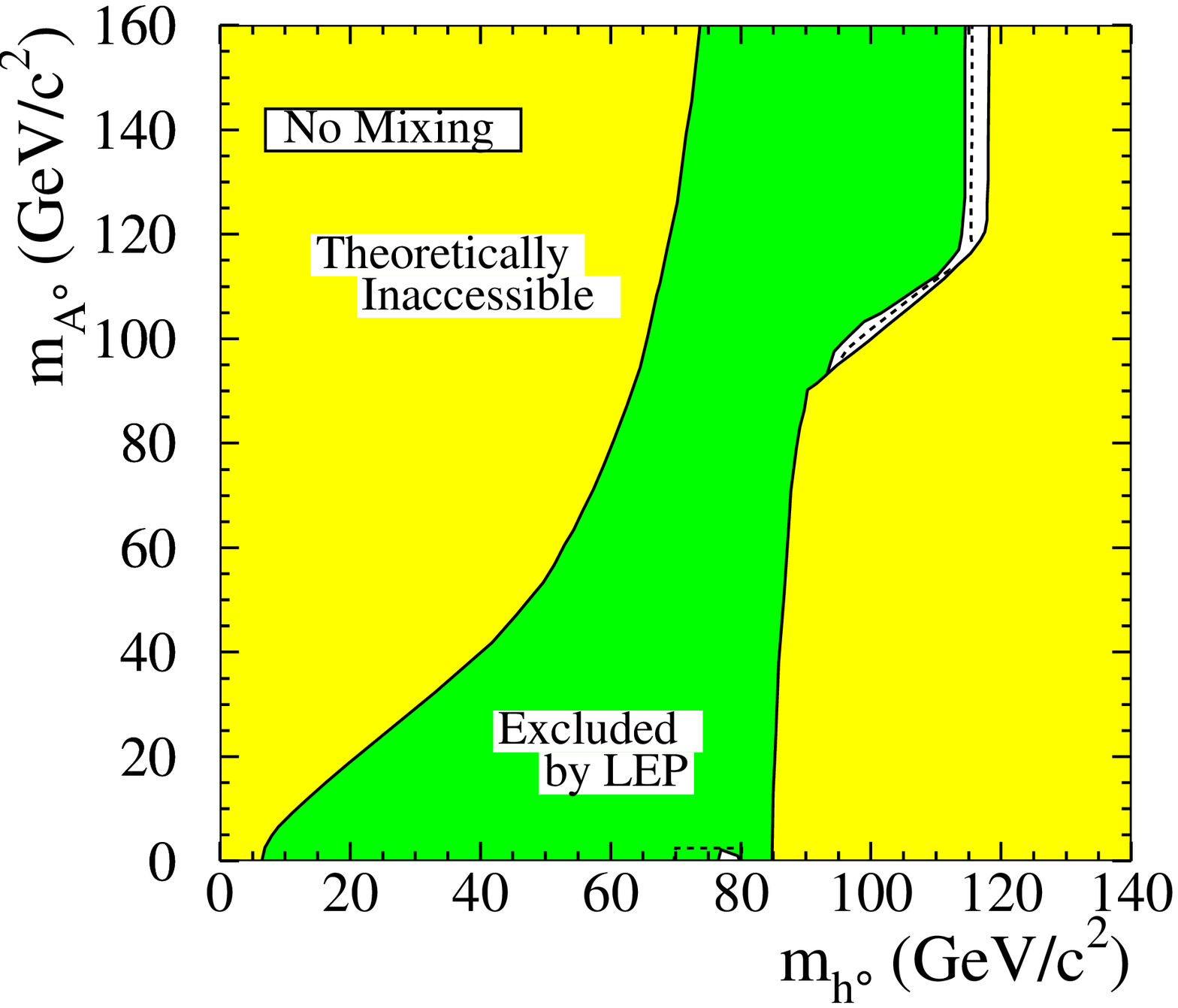,width=0.33\textwidth}
\epsfig{file=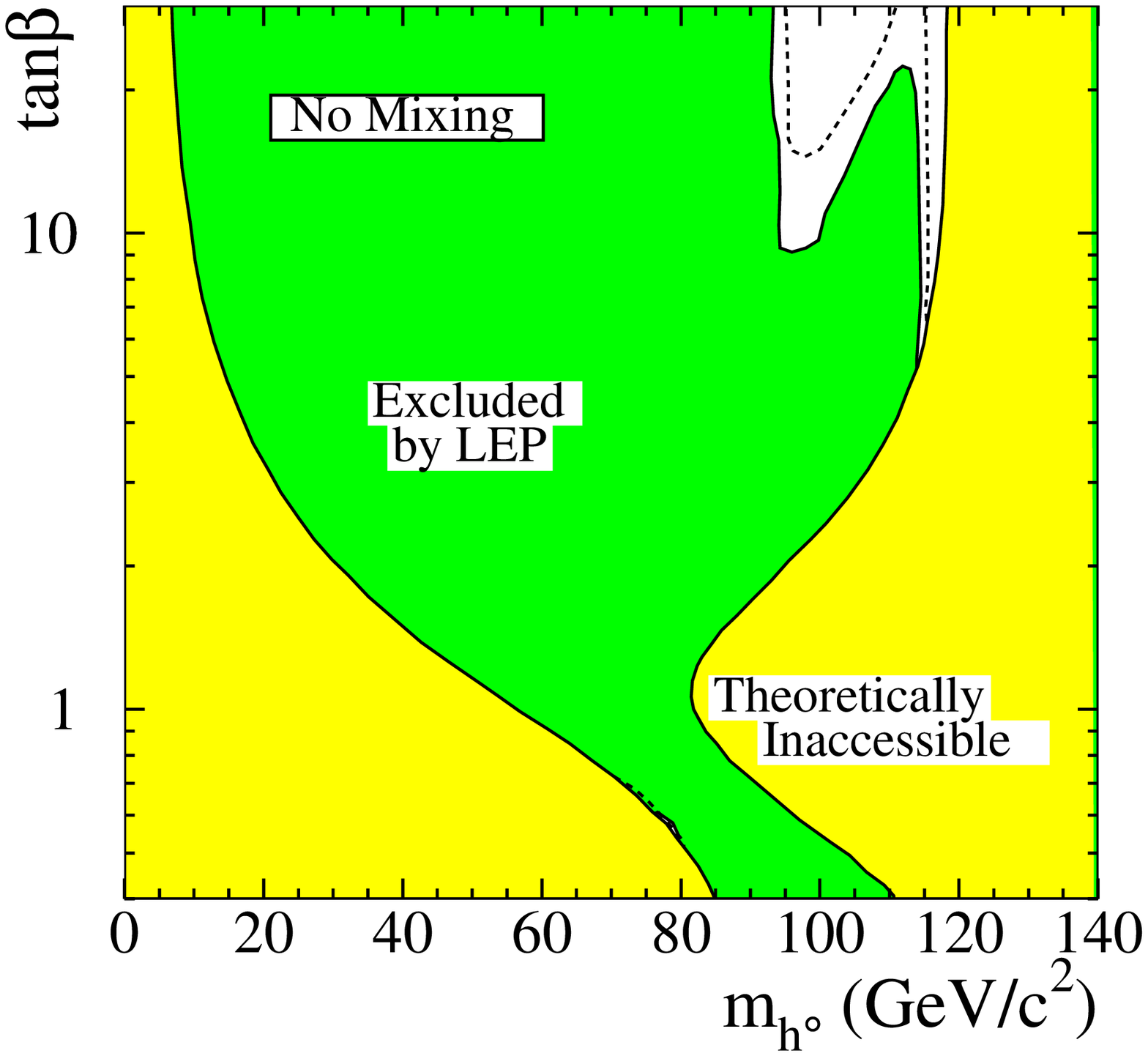,width=0.33\textwidth}
\epsfig{file=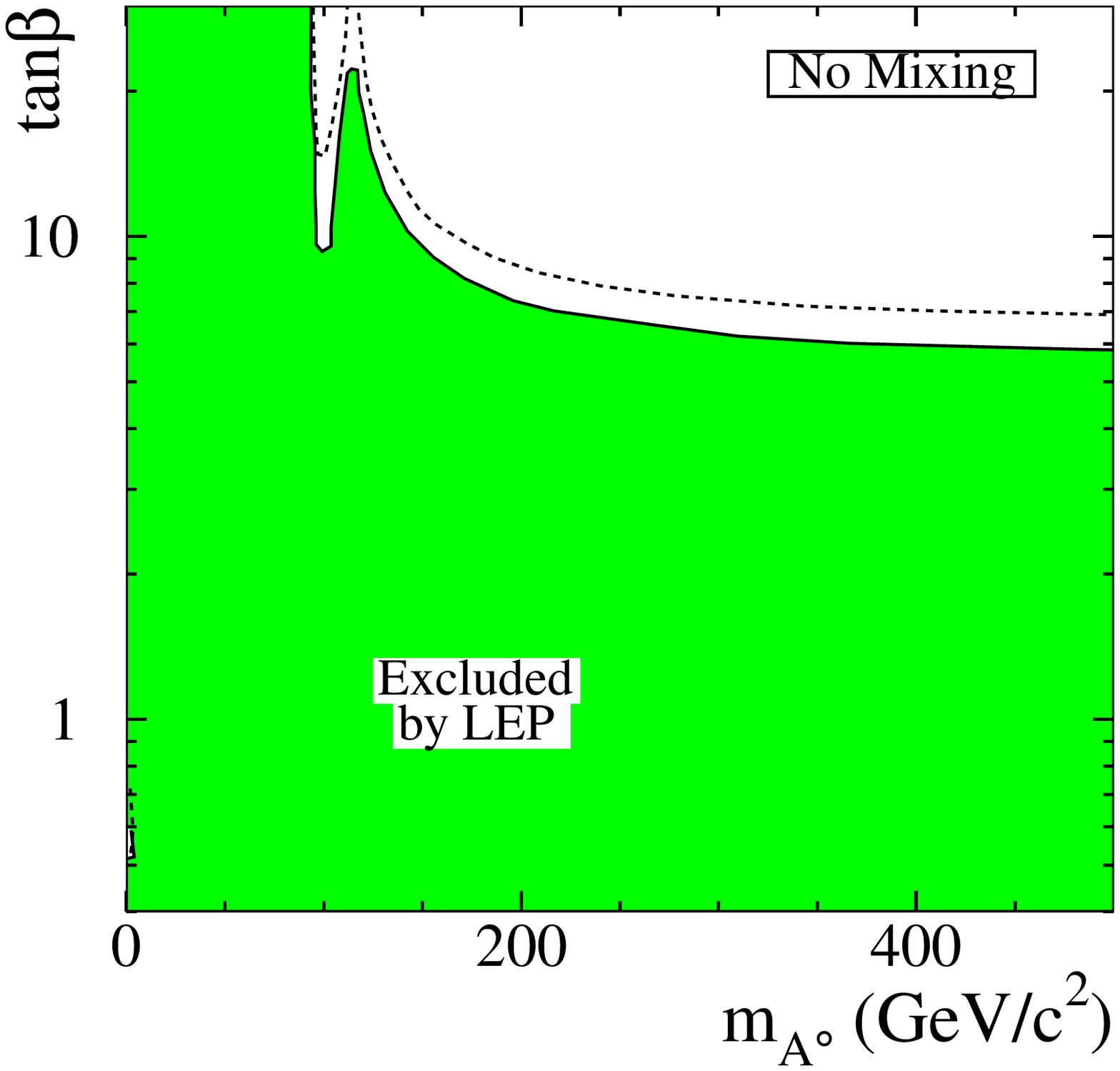,width=0.33\textwidth}
\epsfig{file=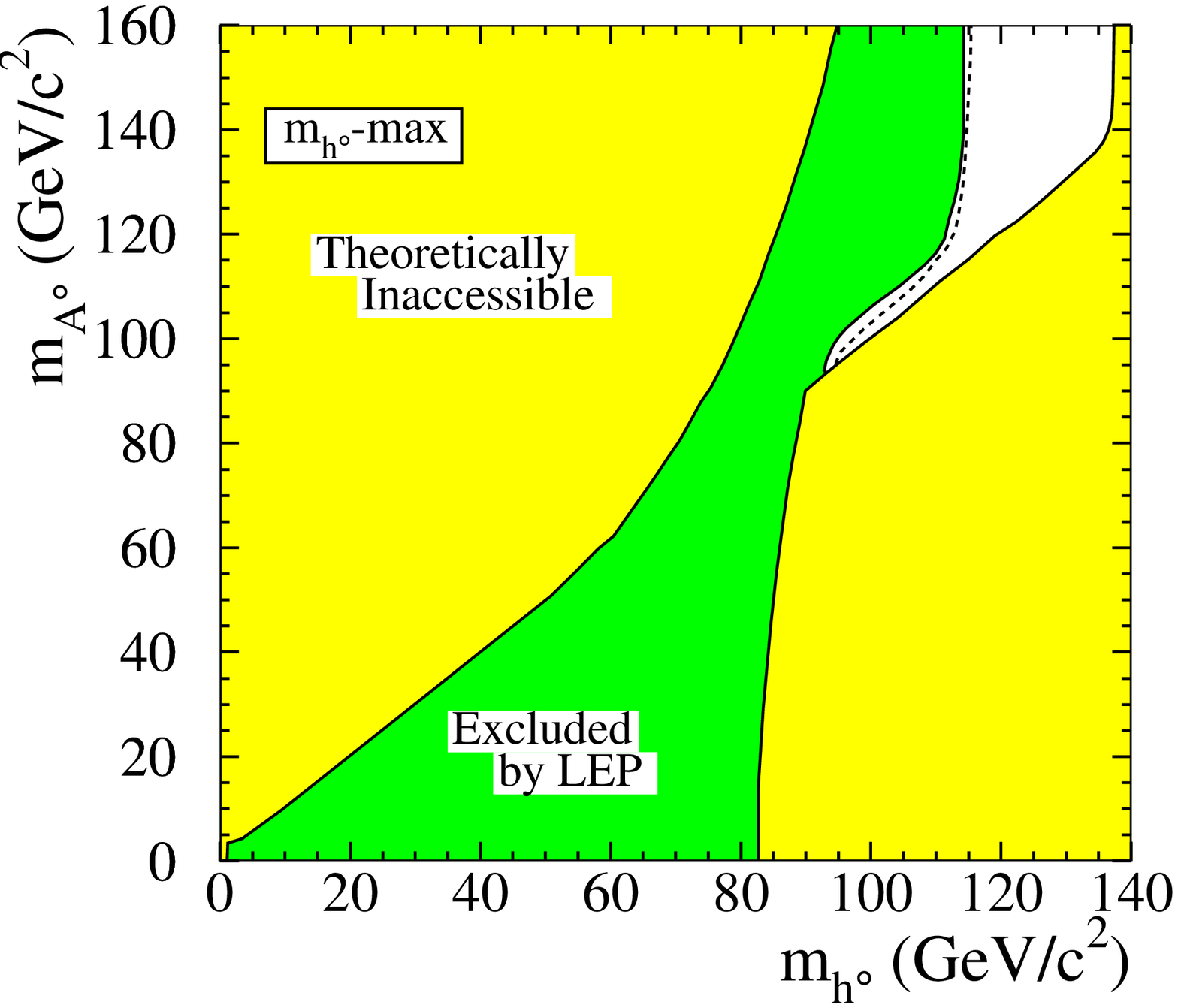,width=0.33\textwidth}
\epsfig{file=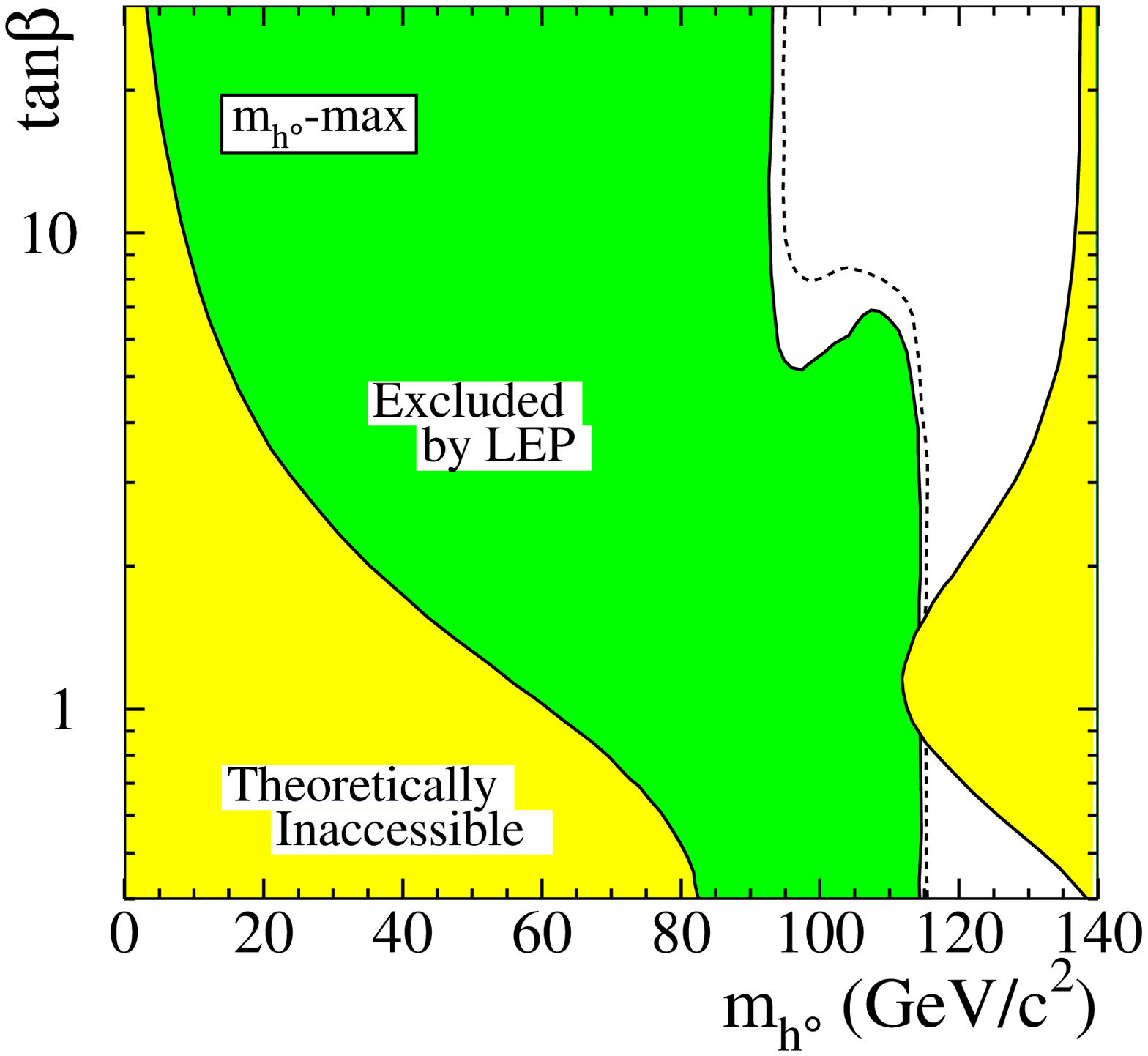,width=0.33\textwidth}
\epsfig{file=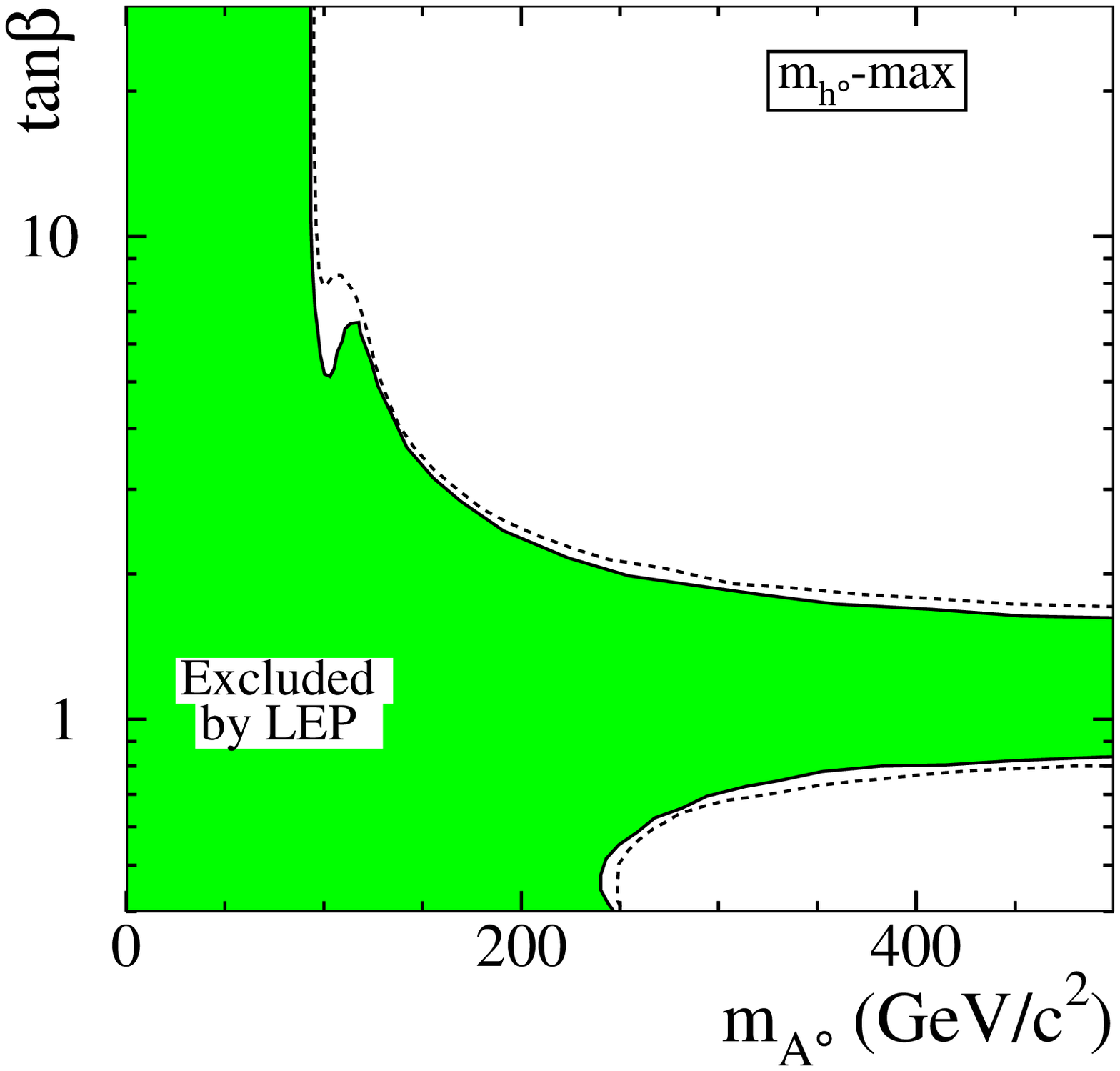,width=0.33\textwidth}
\vspace*{-0.8cm}
\caption{
\label{fig:nomix}
  MSSM: limits in the no-mixing and maximum h-mass benchmark scenarios
  with $m_{\rm t}=179.3$~GeV.
}
\end{figure}

Figure~\ref{fig:light} (left plot) shows a small previously unexcluded mass region
for light A masses in the no-mixing scalar top benchmark scenario.
This region is mostly excluded by new dedicated searches for 
a light A boson (center plot). Limits for the maximum h-mass benchmark 
scenario, including results from dedicated searches for the 
reaction $\rm h\ra AA$ are also shown (right plot).
Benchmark limits with $m_{\rm t}=179.3$~GeV are shown in Fig.~\ref{fig:nomix}.

\clearpage

\subsection{Three-Neutral-Higgs-Boson Hypothesis and a MSSM Parameter Scan}

\begin{figure}[htb]
\vspace*{-1.0cm}
\begin{center}
\includegraphics[scale=0.24]{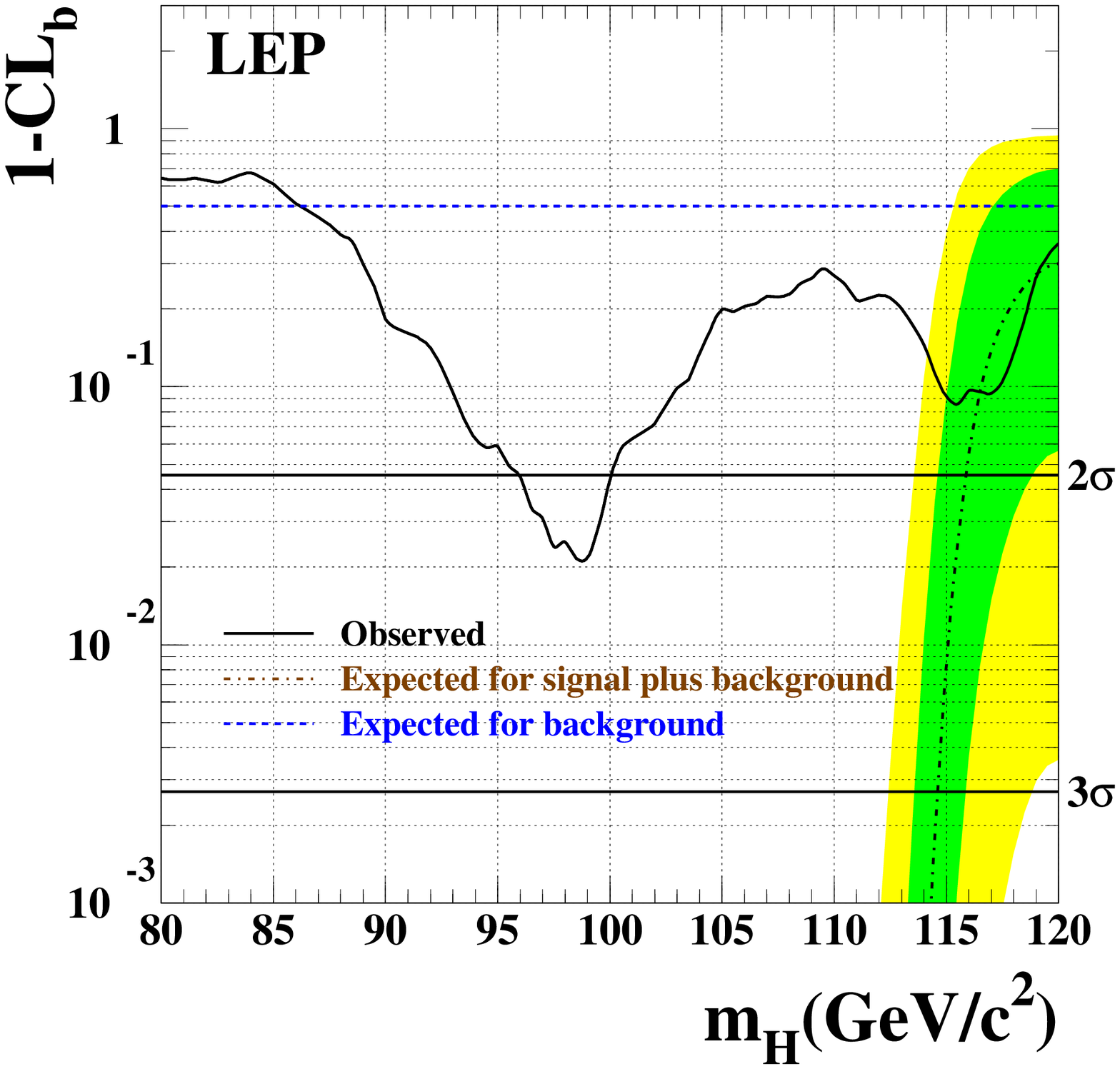}       \hfill
\includegraphics[scale=0.35]{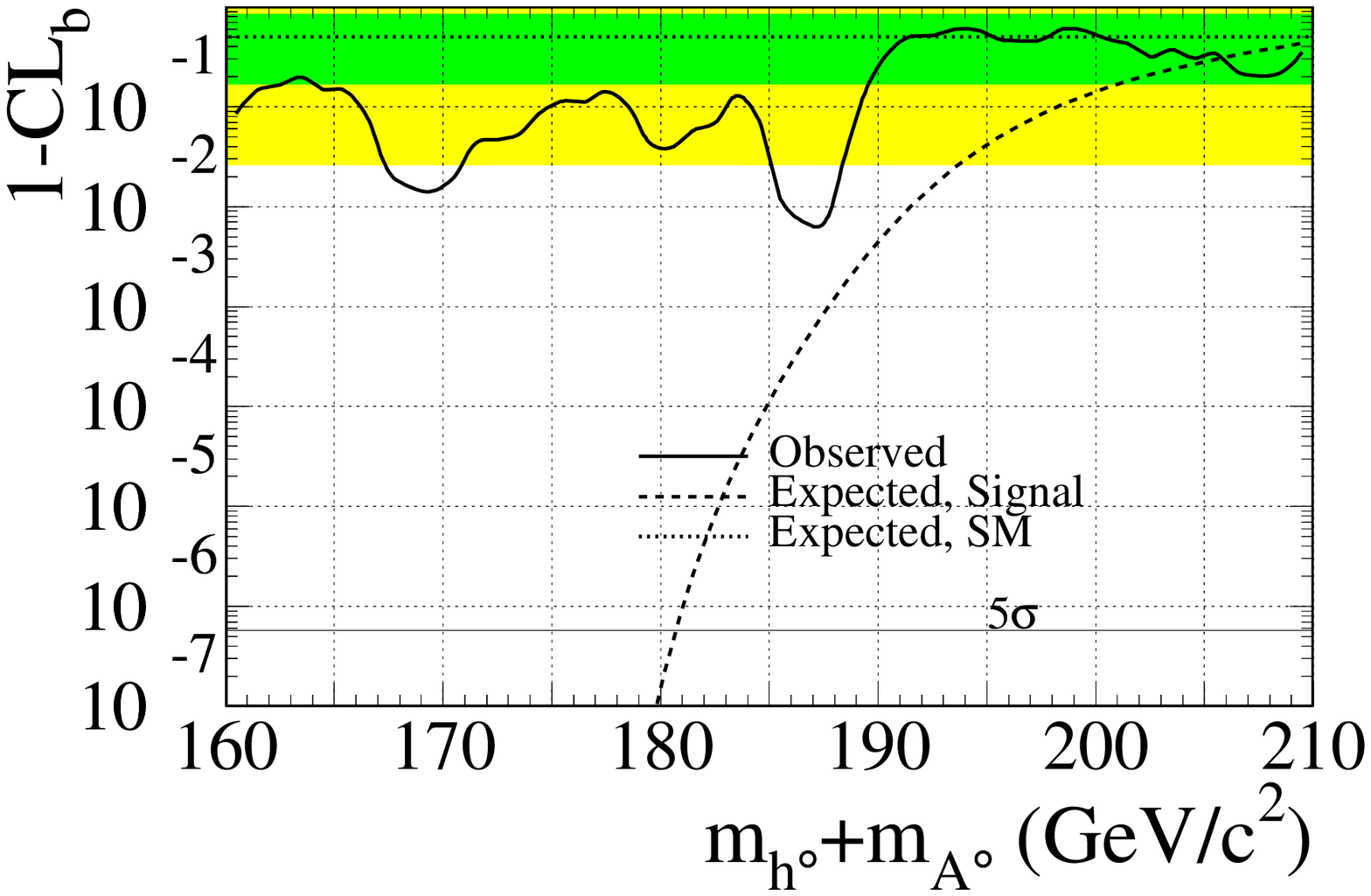} \hfill
\epsfig{file=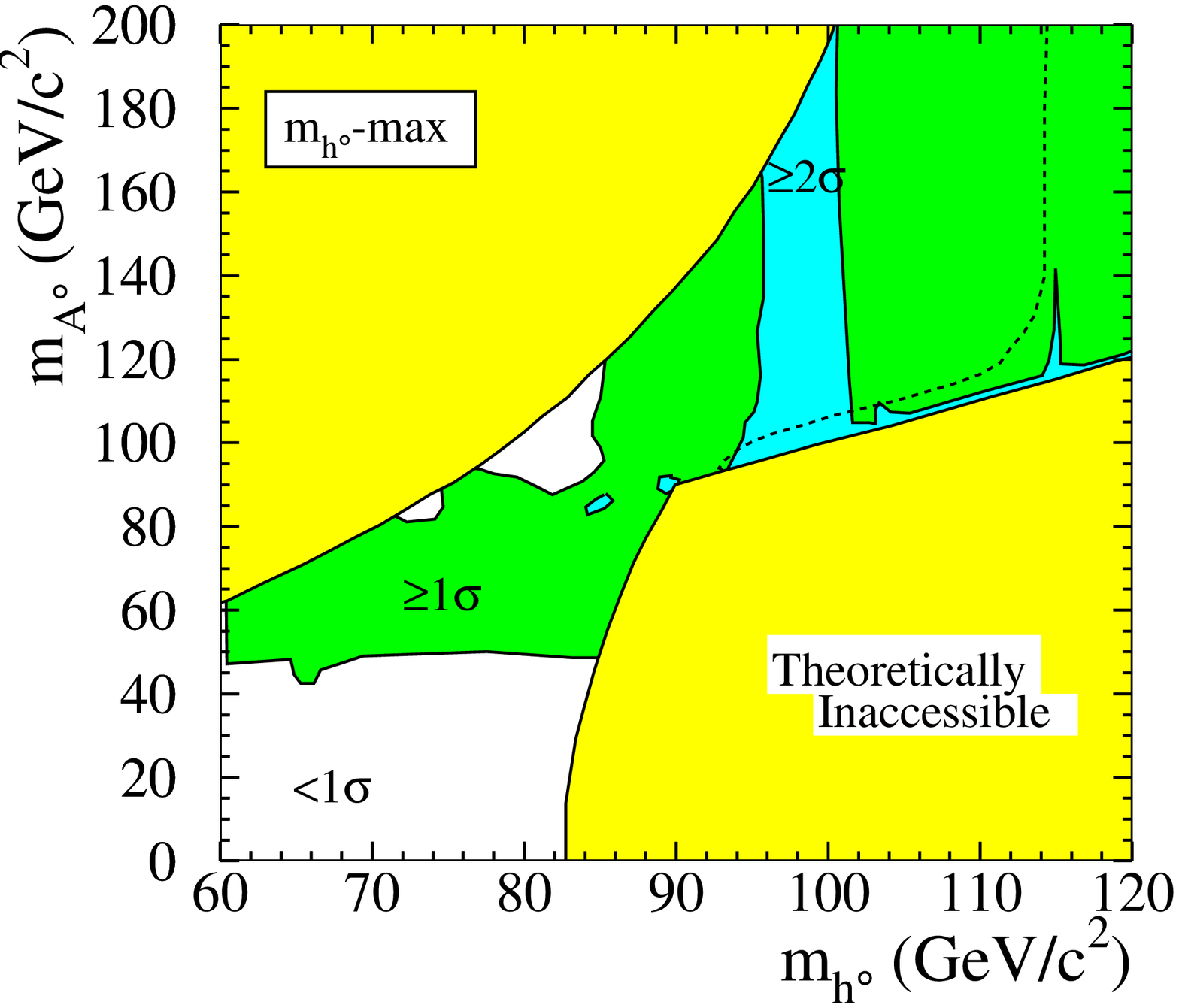,width=0.31\textwidth}
\end{center}
\vspace*{-0.7cm}
\caption{
  MSSM. 
  Left: small data excess at 99 GeV and 116 GeV in hZ/HZ searches.
  Center: small data excess at 
          $m_{\rm h}+m_{\rm A}=187$~GeV in hA searches.
  Right: contours of $1-CL_{\rm b}$ for the maximum h-mass benchmark scenario.
$1-CL_{\rm b}$ expresses the incompatibility of the observation
with the background-only hypothesis.
\label{fig:three}}
\vspace*{-0.2cm}
\end{figure}

The hypothesis of three-neutral-Higgs-boson production, via hZ, HZ and hA
is compatible with the data excess seen in Fig.~\ref{fig:three}.
For the reported MSSM parameters~\cite{as2000} reduced hZ production 
near 100 GeV and HZ production near 115 GeV is compatible with the 
data (left plot). 
For $m_{\rm h} \approx m_{\rm A}$, hA production is also compatible 
with the data (center plot).
The parameters have been obtained from a general MSSM parameter scan.
The overall agreement of data and background is shown in the right plot.

\subsection{A General MSSM Parameter Scan}

Mass limits from a general MSSM parameter scan are shown in 
Fig.~\ref{fig:largetgb} (left plot). 
After the inclusion of searches for invisible Higgs boson decays, 
mass limits are only slightly reduced compared to the benchmark results.

\subsection{Large \boldmath$\tan\beta$\unboldmath\ Scenario}

\begin{figure}[bp]
\vspace*{-1cm}
\begin{center}
\includegraphics[scale=0.26]{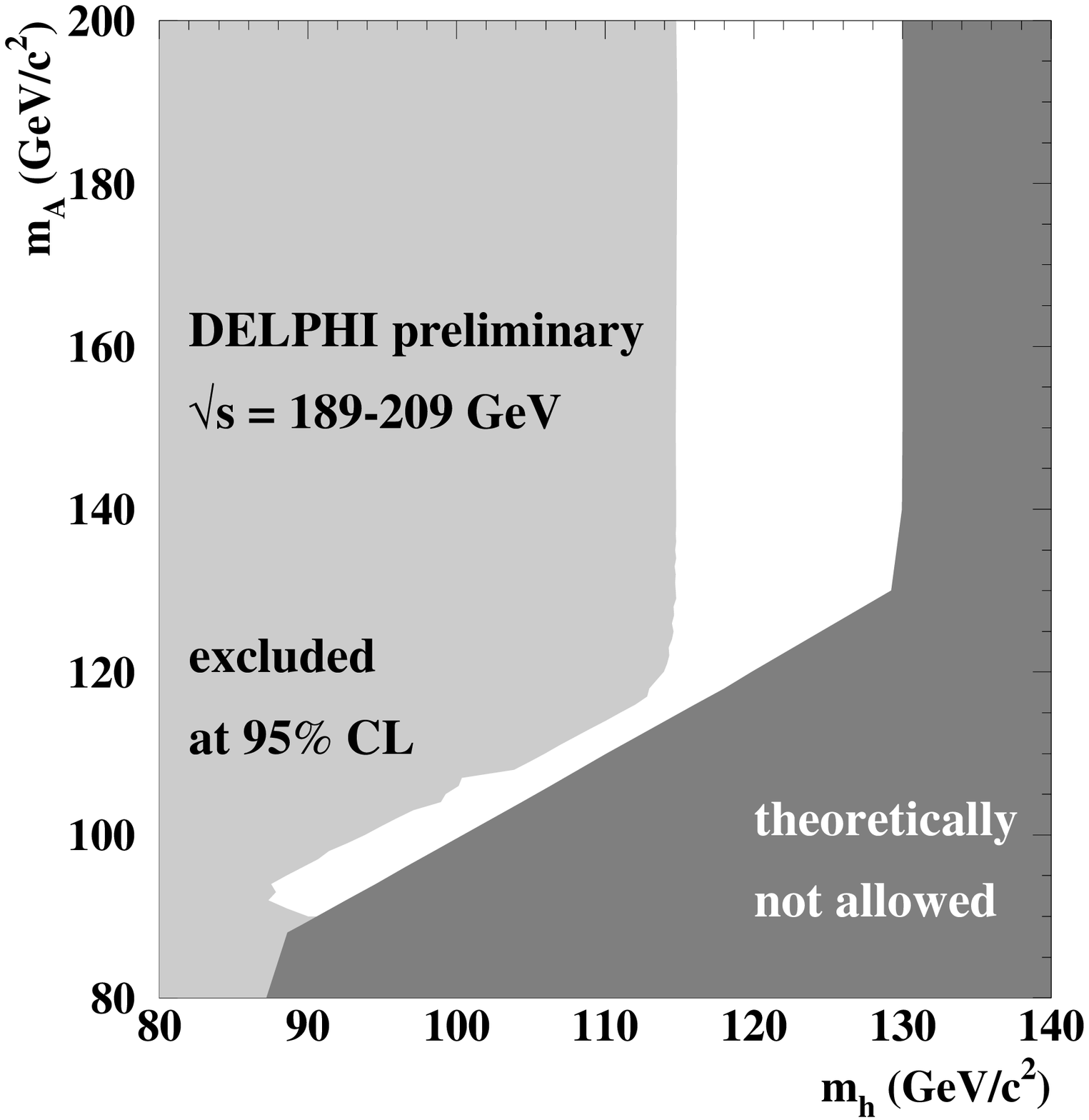}
\includegraphics[width=0.32\textwidth,clip]{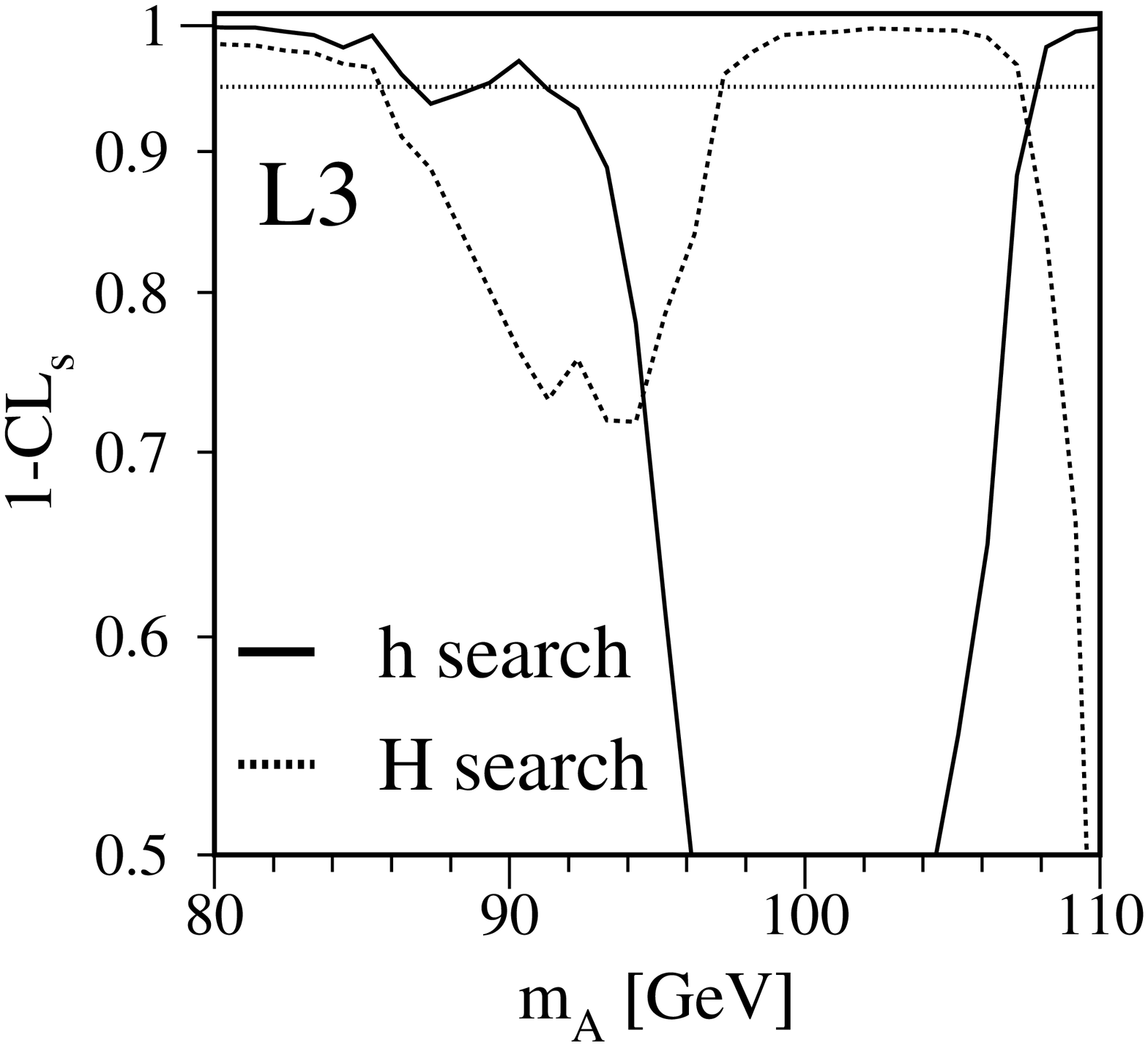}
\includegraphics[width=0.3\textwidth,bbllx=27pt,bblly=2pt,bburx=258pt,bbury=238pt,clip=]{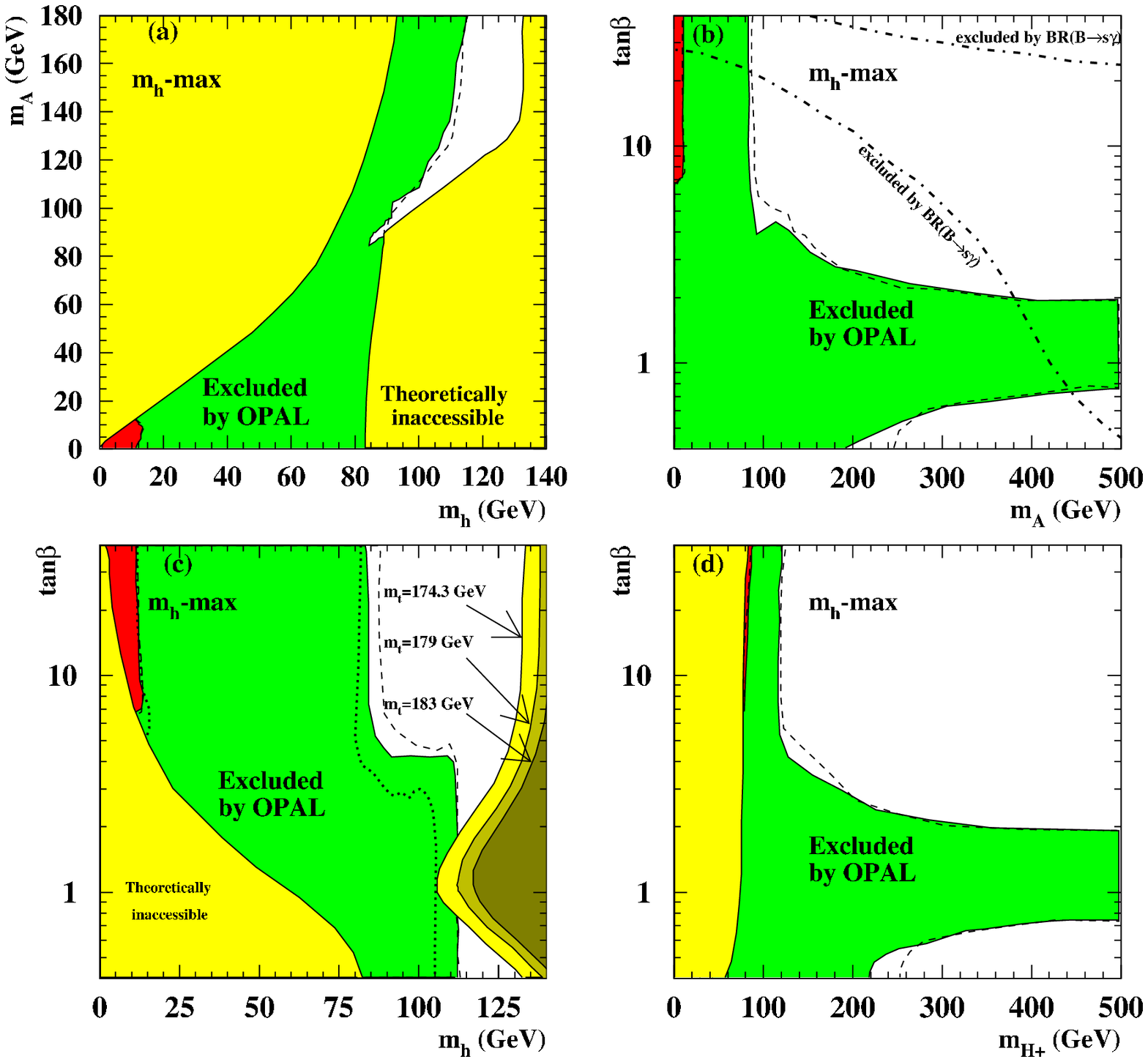}
\end{center}
\vspace*{-0.6cm}
\caption{MSSM. 
Left: mass limits from a general MSSM parameter scan.
Center: excluded mass region for large $\tan\beta$ from h and H searches. 
Right: much reduced $\tan\beta$ limits due to the larger top-quark mass.
       The overlap of the shaded regions is reduced for larger top-quark masses. 
\label{fig:largetgb}}
\vspace*{-0.2cm}
\end{figure}

For large values of $\tan\beta$, MSSM parameter regions exist, where hA production is
inaccessible kinematically and hZ production is suppressed by small $\sin(\beta-\alpha)$ values,
thus HZ production is dominant.
The importance of the HZ contribution is shown in Fig.~\ref{fig:largetgb} (center plot)
where the mass region
between 96 and 107~GeV is excluded from the non-observation of HZ production.

\subsection{Large Effect from Increased Top-Quark Mass}

An increased top-quark mass of $178.0\pm4.3$~GeV has been reported~\cite{newtop04}.
This mass is about 4~GeV larger compared to the previous value. This larger top-quark mass has an 
important effect of the Higgs boson search, as a larger top-quark mass results in
a larger maximum for the Higgs boson mass. Therefore, in particular the previous 
limits on $\tan\beta$ are reduced as illustrated in 
Fig.~\ref{fig:largetgb} (right plot).

\section{CP-Violating Models}

\begin{figure}[htb]
\vspace*{-0.3cm}
\epsfig{figure=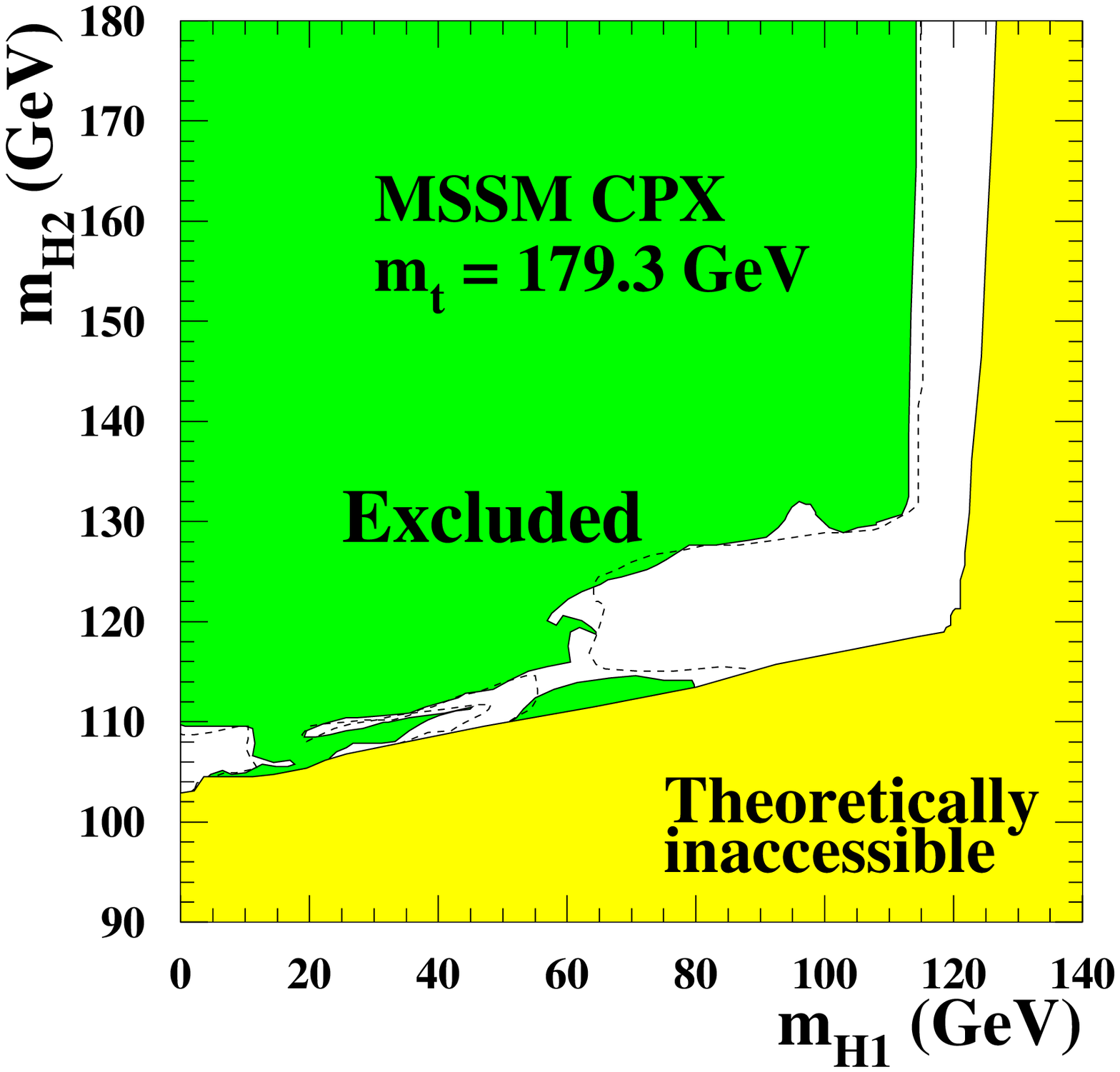,width=0.32\textwidth} \hfill
\epsfig{figure=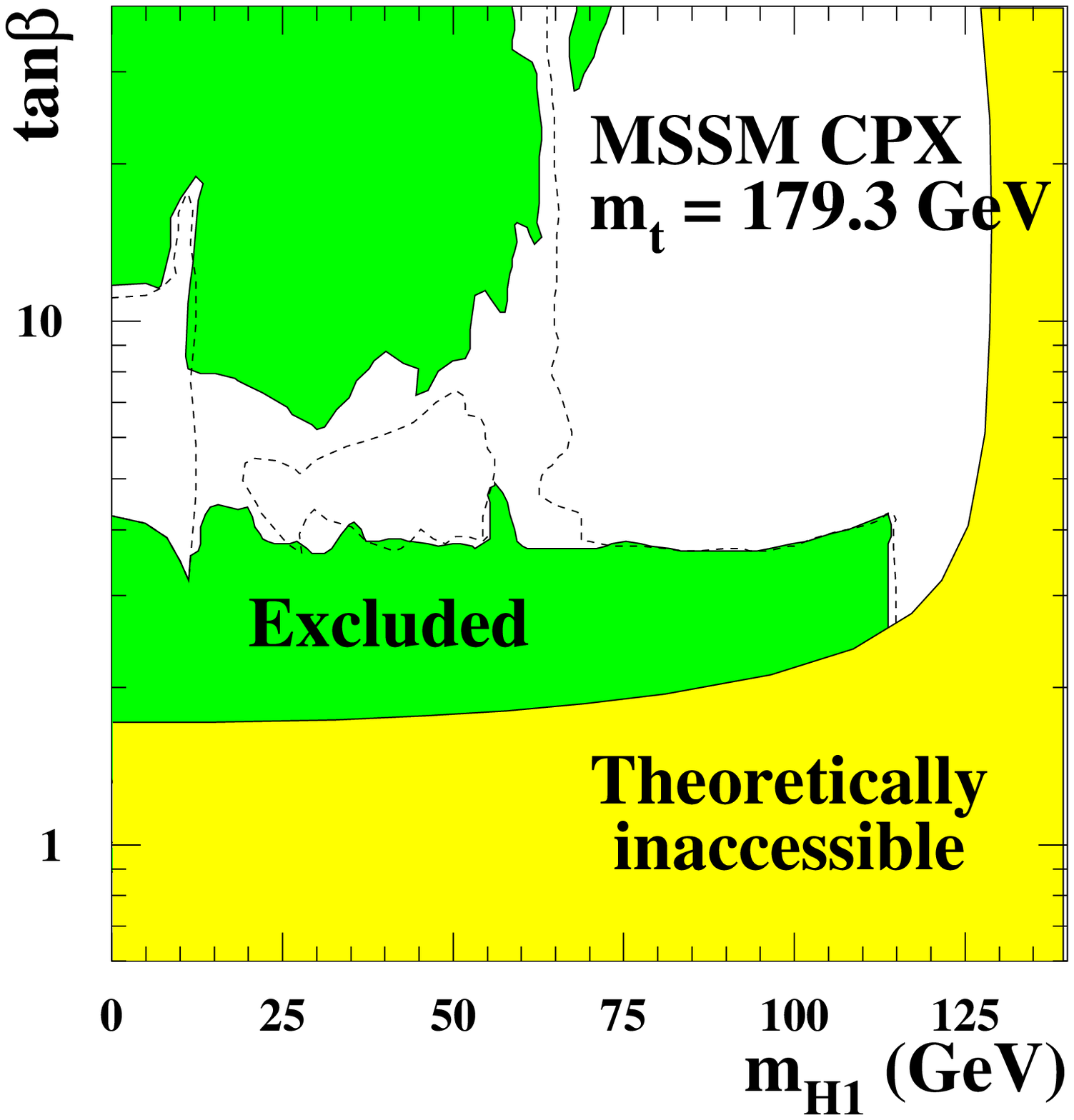,width=0.32\textwidth} \hfill
\epsfig{figure=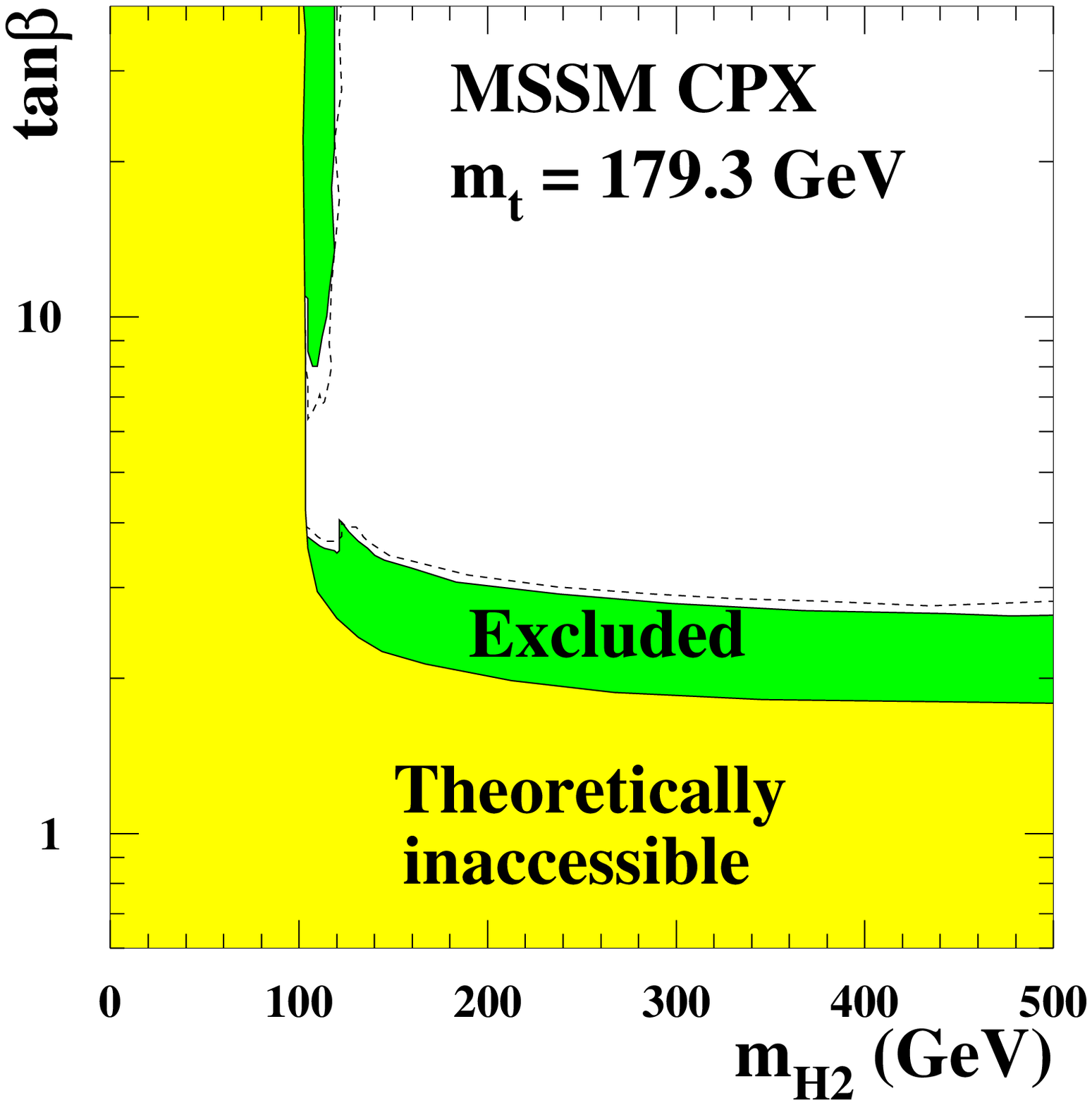,width=0.32\textwidth}
\vspace*{-0.2cm}
\caption{CP-violation models. Limits for full CP-mixing with $m_{\rm t}=179.3$~GeV.
\label{fig:cpx-179}}
\end{figure}

Instead of h, H and A, the Higgs bosons are named $\rm H_1, H_2$ and $\rm H_3$.
In general, CP-mixing reduces the MSSM mass limits significantly~\cite{dis04}.
Combined LEP limits for $m_{\rm t}=179.3$~GeV are shown in Fig.~\ref{fig:cpx-179}.

\section{Invisible Higgs Boson Decays}

\begin{figure}[htb]
\vspace*{-0.5cm}
\begin{center}
\includegraphics[scale=0.34]{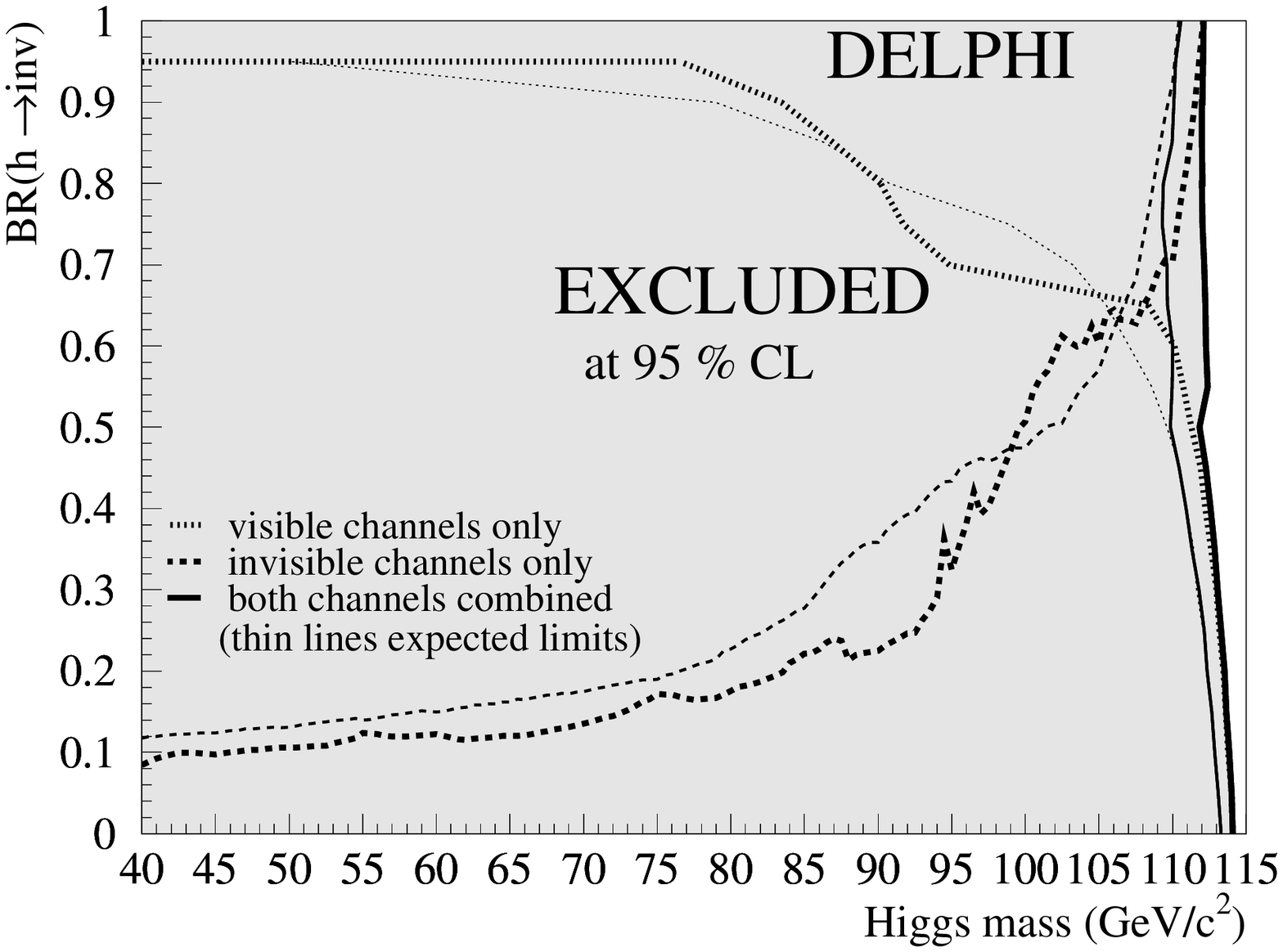}\hfill
\includegraphics[scale=0.26]{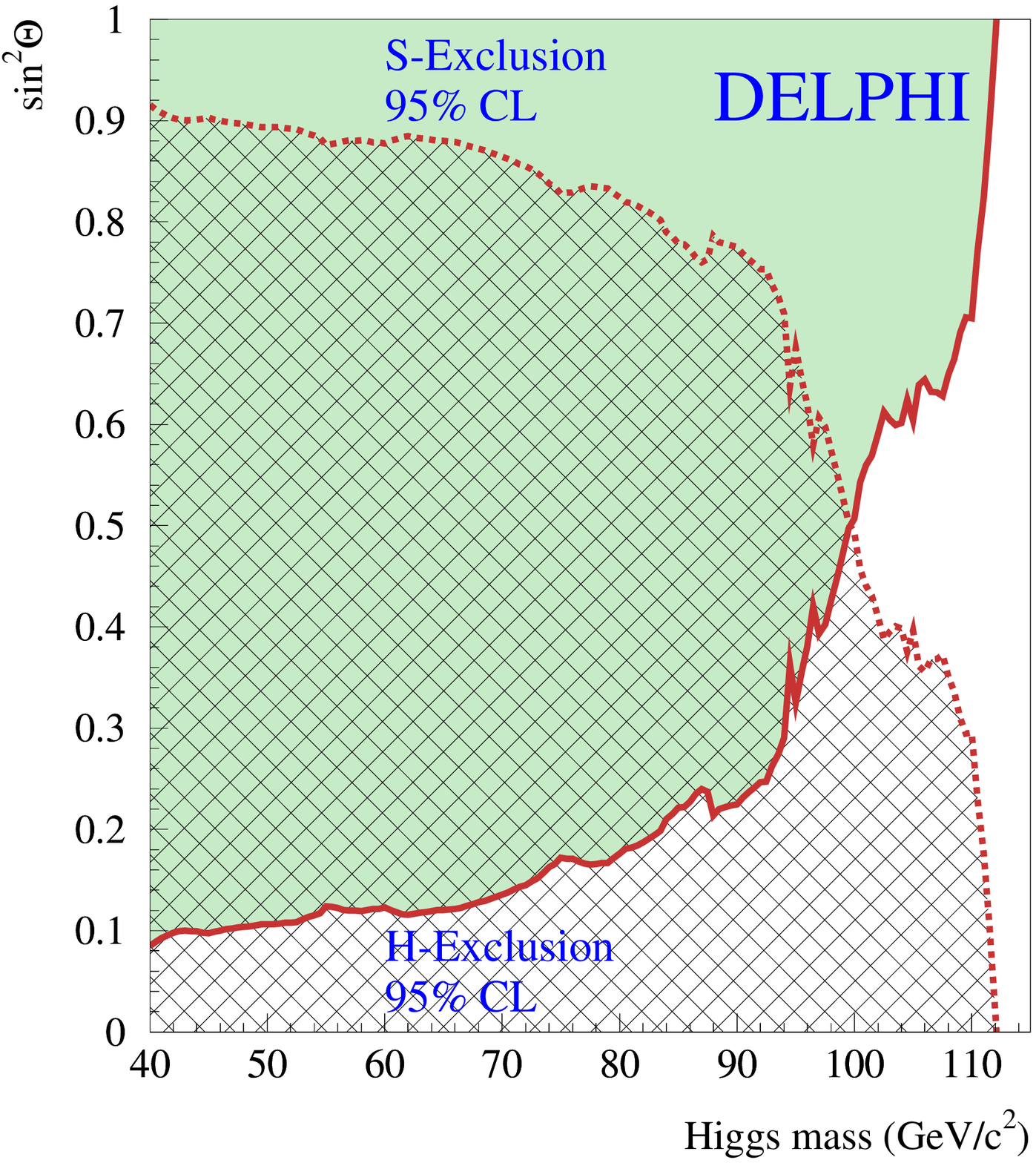} \hfill
\includegraphics[scale=0.26]{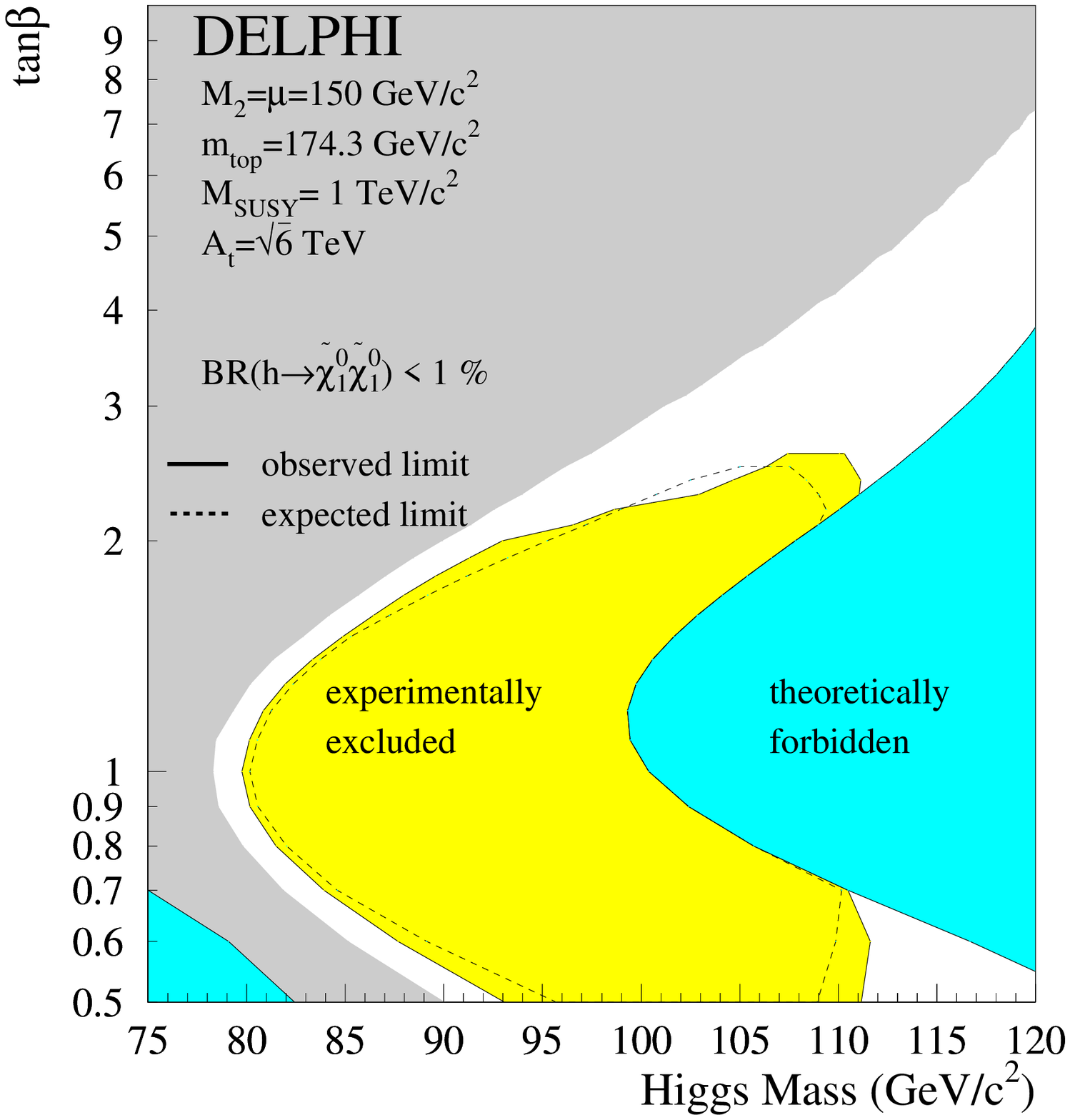}
\vspace*{-0.8cm}
\end{center}
\caption{Left: mass limits for SM and invisible Higgs boson decays combined.  
           Center: mass limits in Majoron models with an extra complex singlet,
           $\rm H/S\ra JJ$, where J escapes undetected. $\sin\theta$ is the H/S
           mixing angle.
           Right: mass limits in the MSSM for $\rm h\to \tilde\chi^0_1\tilde\chi^0_1$.
\label{fig:invlimit}}
\end{figure}

No indication of invisibly-decaying Higgs bosons is observed.
Figure~\ref{fig:invlimit} shows mass limits for SM and invisible Higgs boson decays combined.
The results are also interpreted in a 
Majoron model with an extra complex singlet, $\rm H/S\ra JJ$,
where J escapes undetected. In addition, mass limits are shown in the MSSM
for $\rm h\to \tilde\chi^0_1\tilde\chi^0_1$.

\clearpage

\section{Neutral Higgs Bosons in the General Two-Doublet Higgs Model}

\begin{figure}[htb]
\vspace*{-0.4cm}
\begin{center}
\includegraphics[scale=0.36]{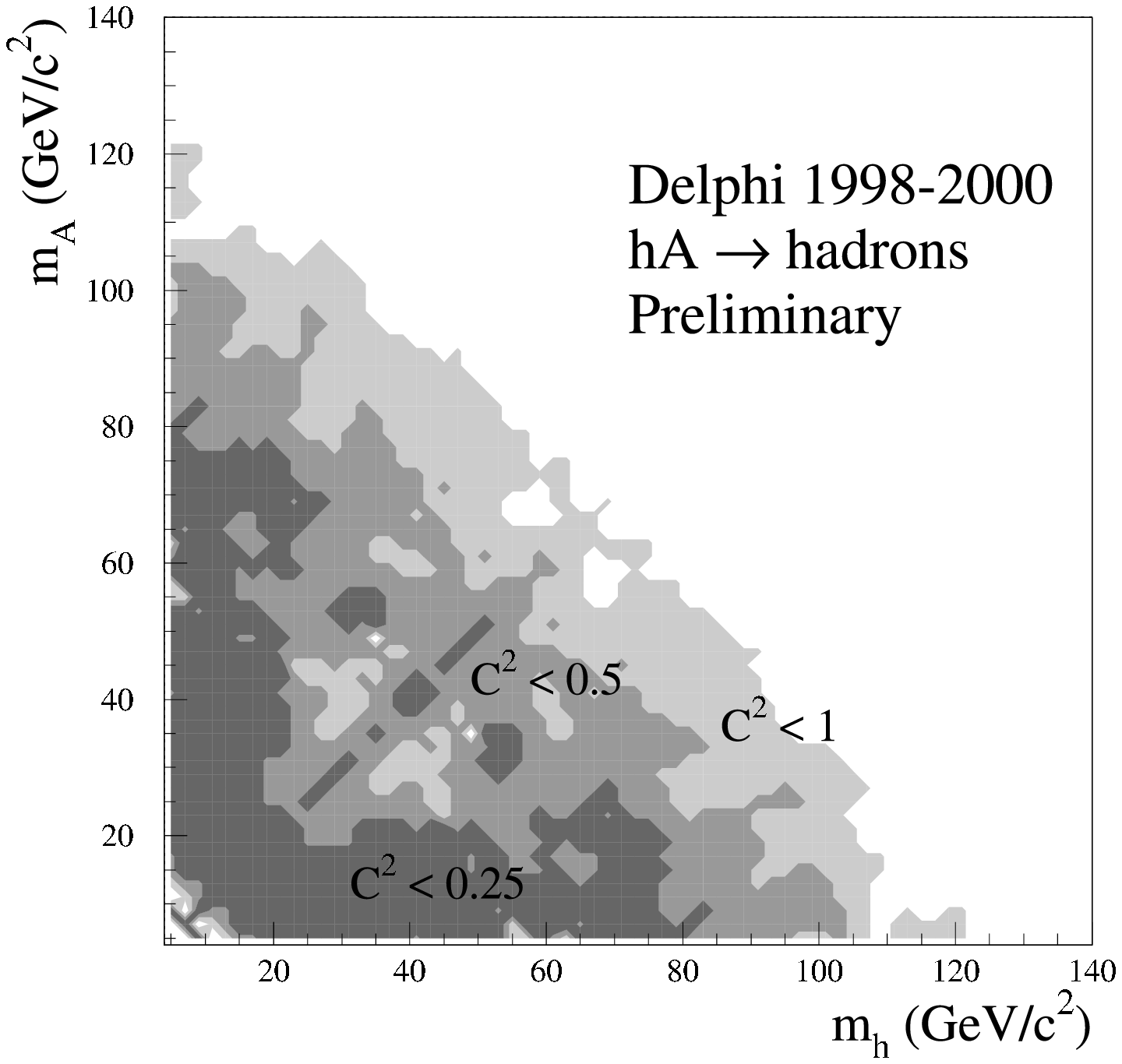}
\includegraphics[scale=0.24]{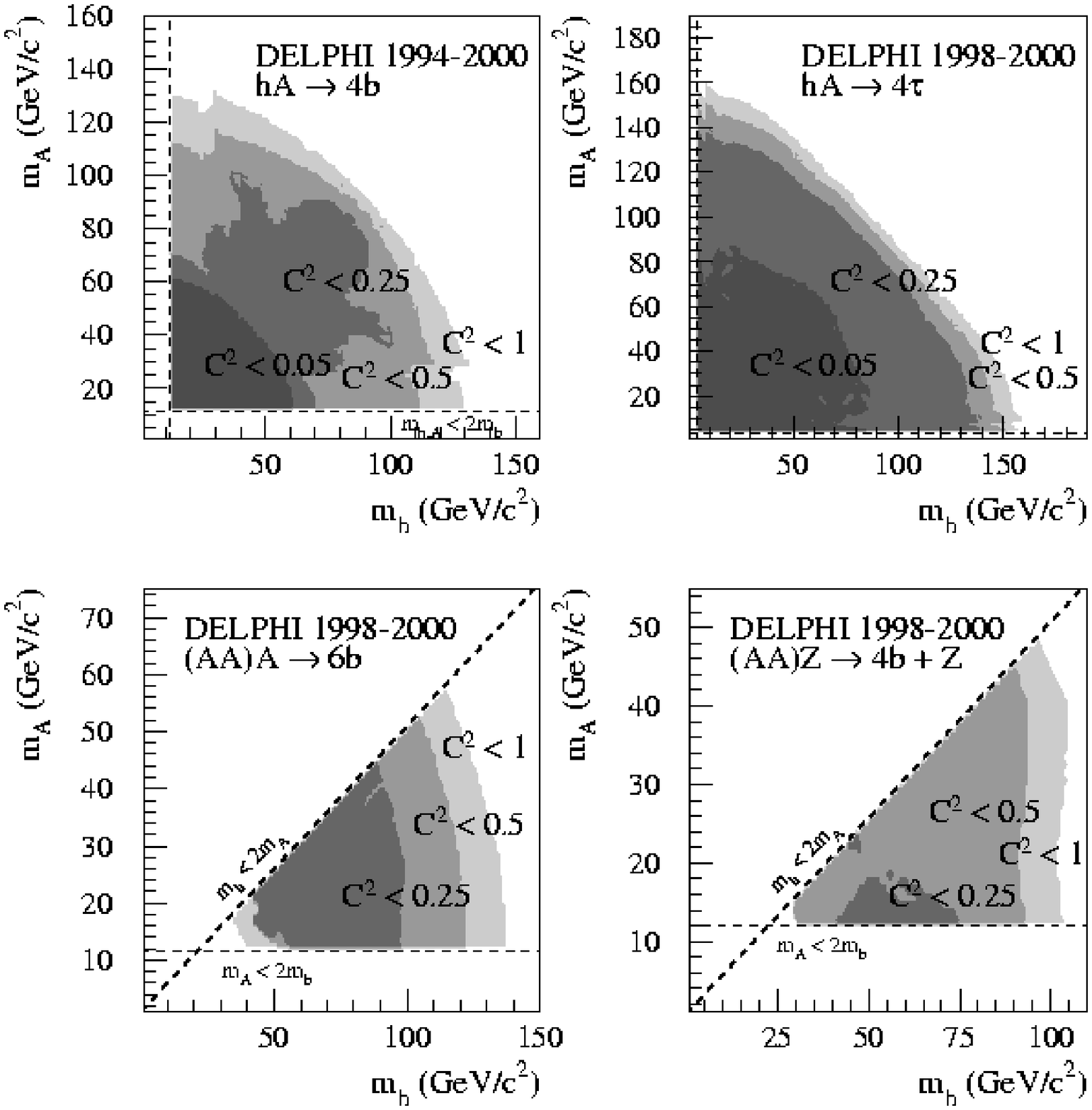}\hfill
\includegraphics[scale=0.27]{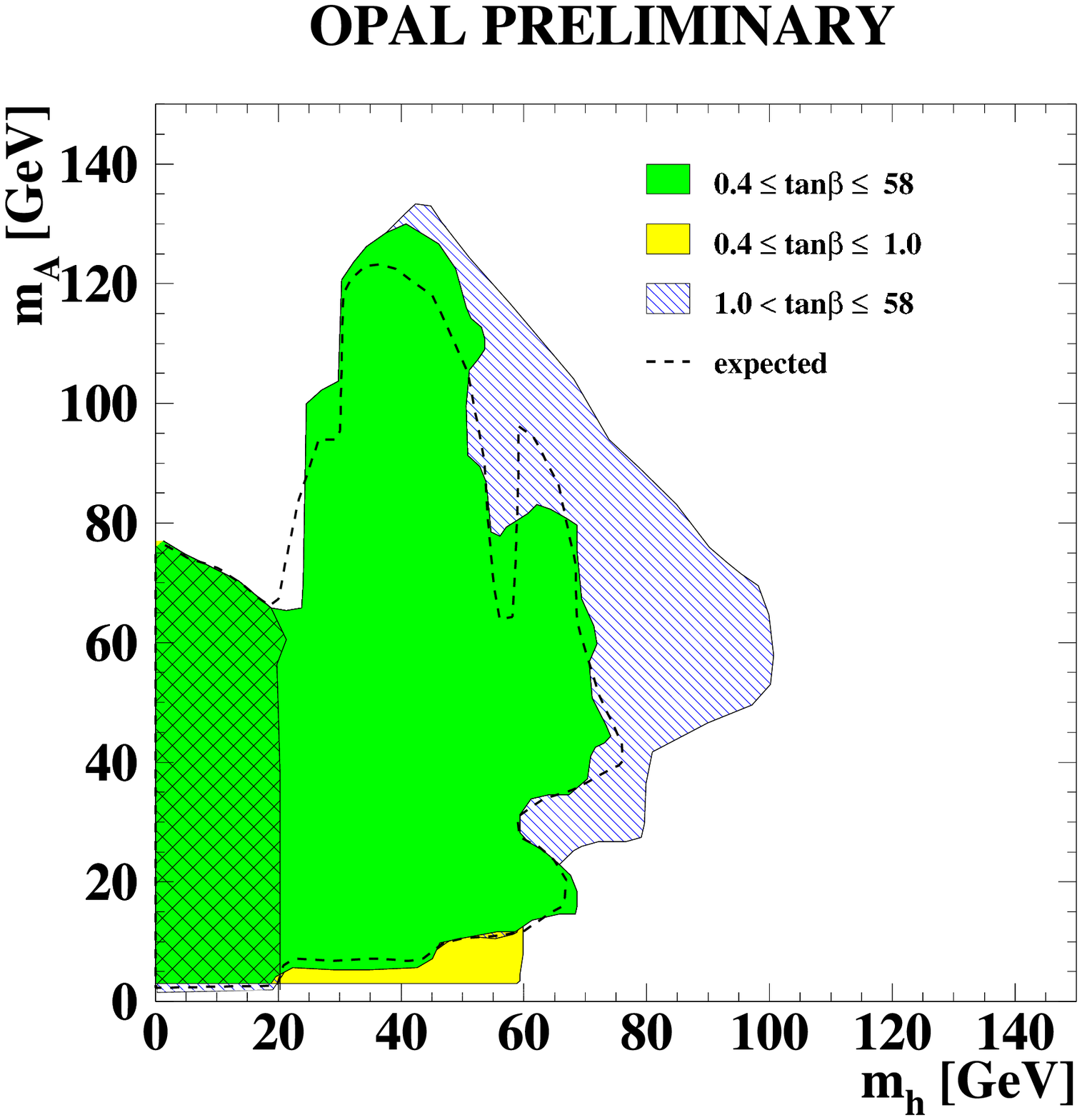}
\vspace*{-0.6cm}
\end{center}
\caption{Two-doublet Higgs model. 
           Left: Flavor-independent limits.            
           $C^2$ is the reduction factor on the maximum production cross section.
           Center:  Limits from dedicated searches for hA production.
           Right: mass limits from a general parameter scan in the two-doublet Higgs model.
\label{fig:2dhm}}
\vspace*{-0.3cm}
\end{figure}

\begin{figure}[htb]
\vspace*{-0.3cm}
\epsfig{figure=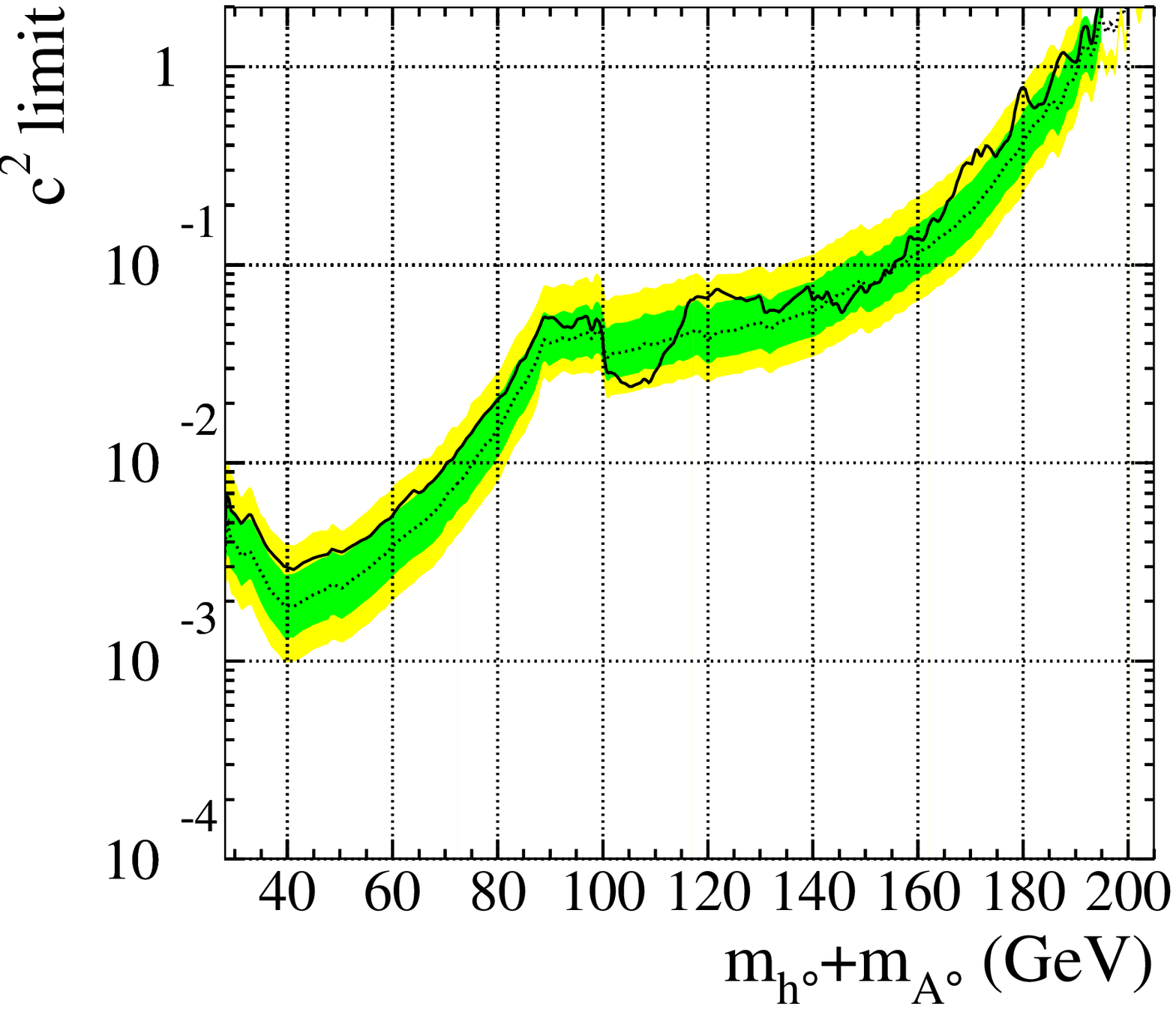,width=0.32\textwidth}  \hfill
\epsfig{figure=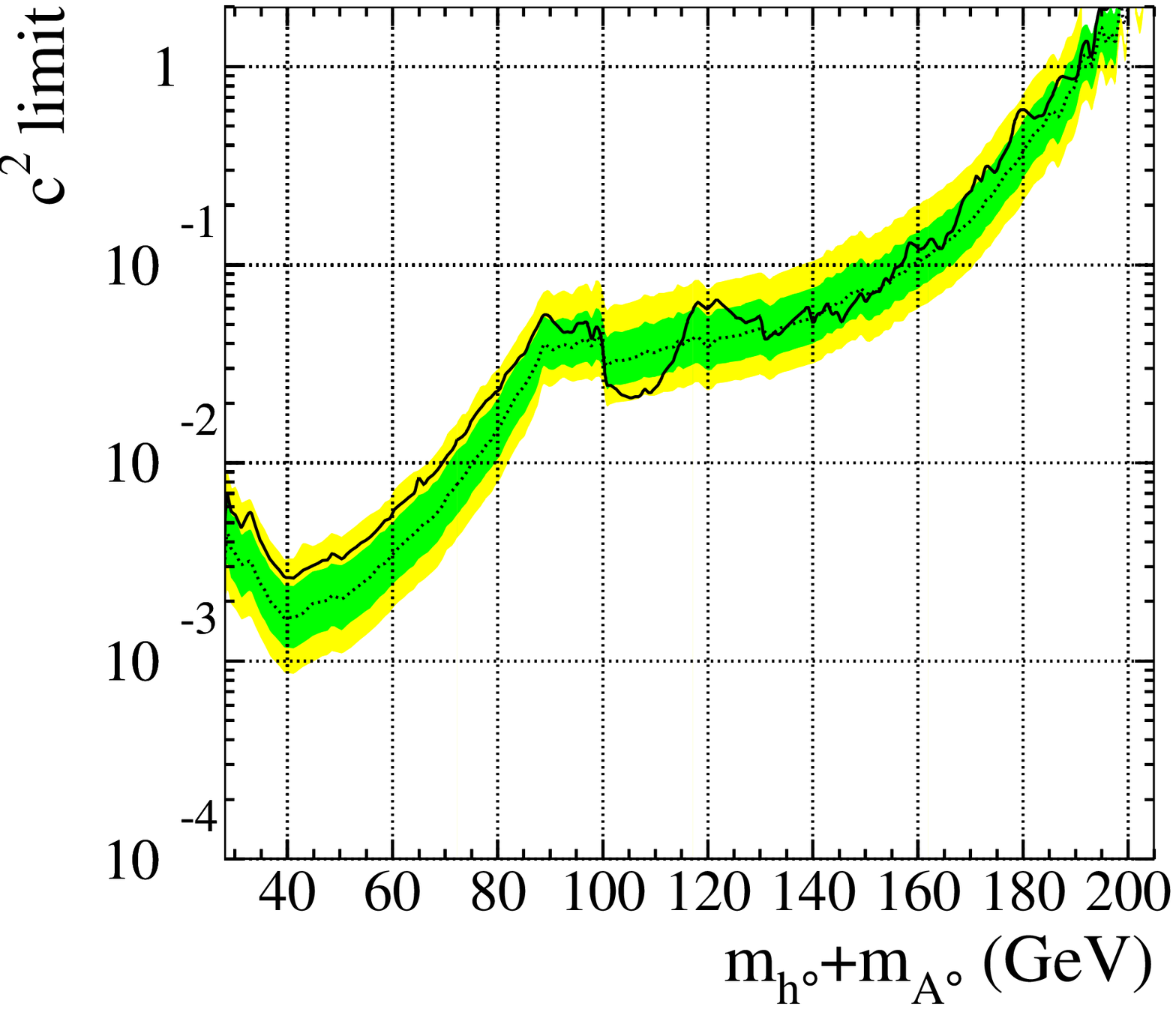,width=0.32\textwidth}  \hfill   
\epsfig{figure=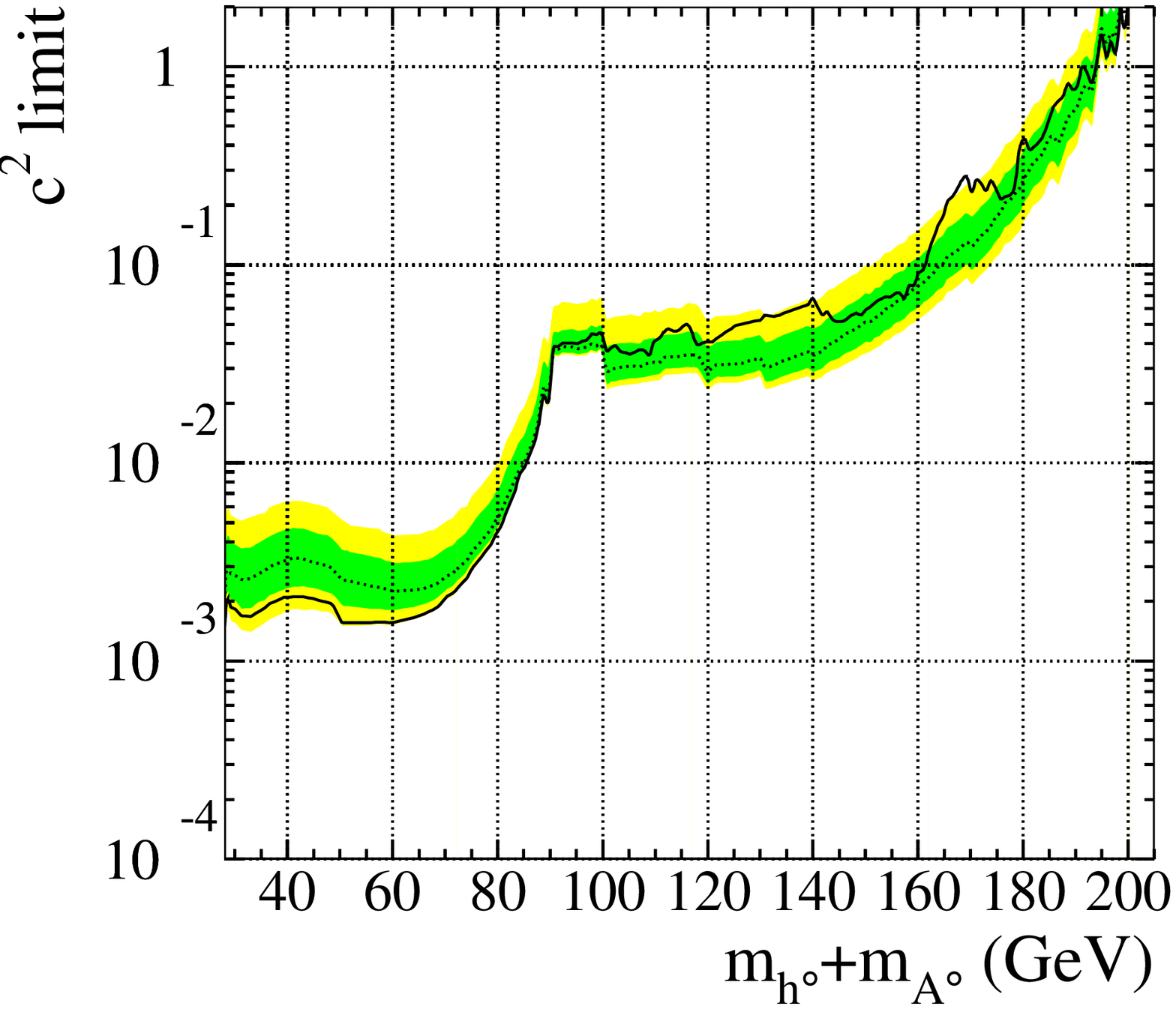,width=0.32\textwidth} 
\vspace*{-0.3cm}
\caption{Upper bound on the scale factor $C^2$, assuming $m_{\rm h} \approx m_{\rm A}$
Left: typical branching ratios.
Center: decay exclusively into b-quark pairs. 
Right: one decay into a b-quark pair and the other into a $\tau$-lepton pair.
\label{mssm-c2}}
\end{figure}

Figures~\ref{fig:2dhm} and~\ref{mssm-c2} show mass limits from flavor-independent and
dedicated searches for hA production, and from a parameter scan.
The scan combines searches with b-tagging and flavor-independent searches.

\section{Yukawa Higgs Boson Processes 
\boldmath$\rm b\bar b h$ and $\rm b\bar b A$\unboldmath}

\begin{figure}[htb]
\vspace*{-0.4cm}
\begin{center}
\includegraphics[scale=0.53]{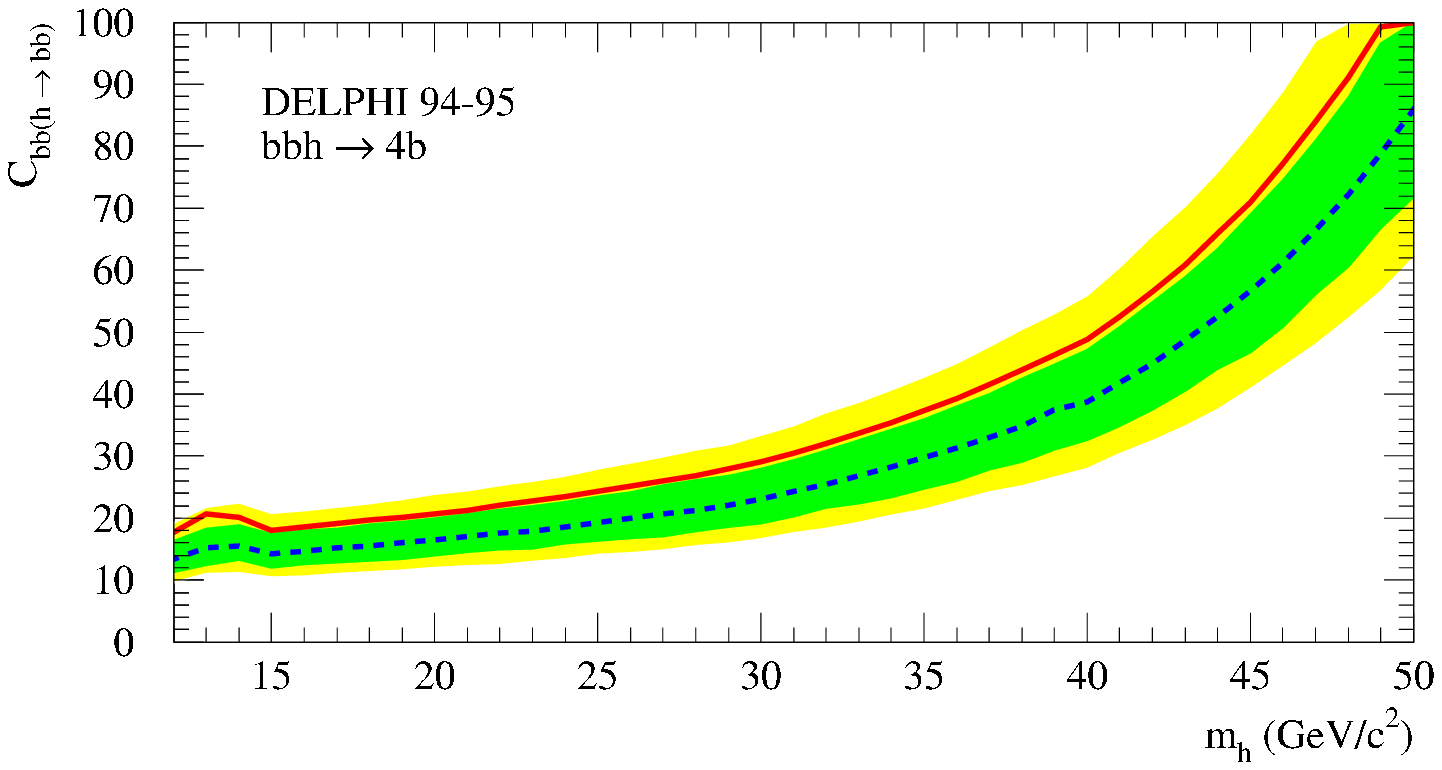}\hfill
\includegraphics[scale=0.53]{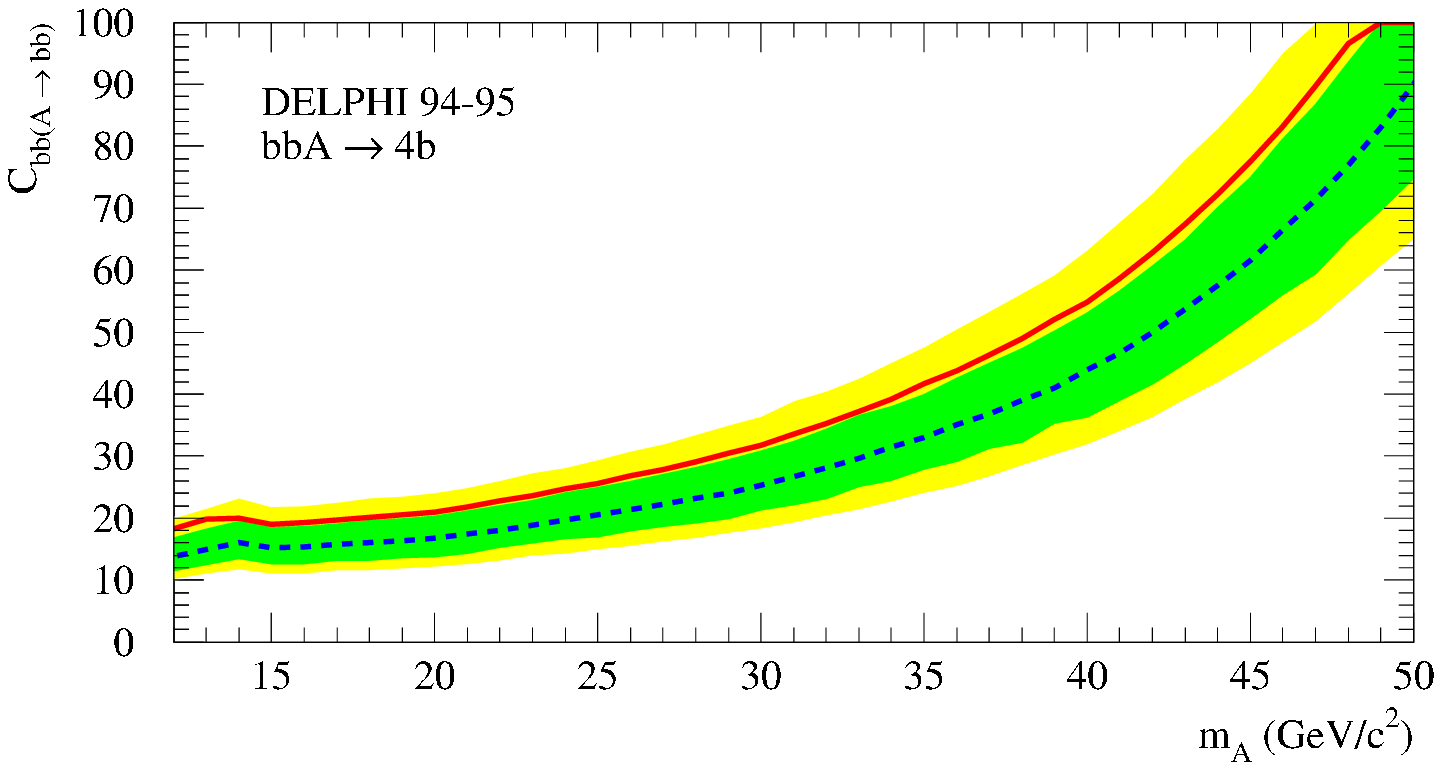}
\includegraphics[scale=0.53]{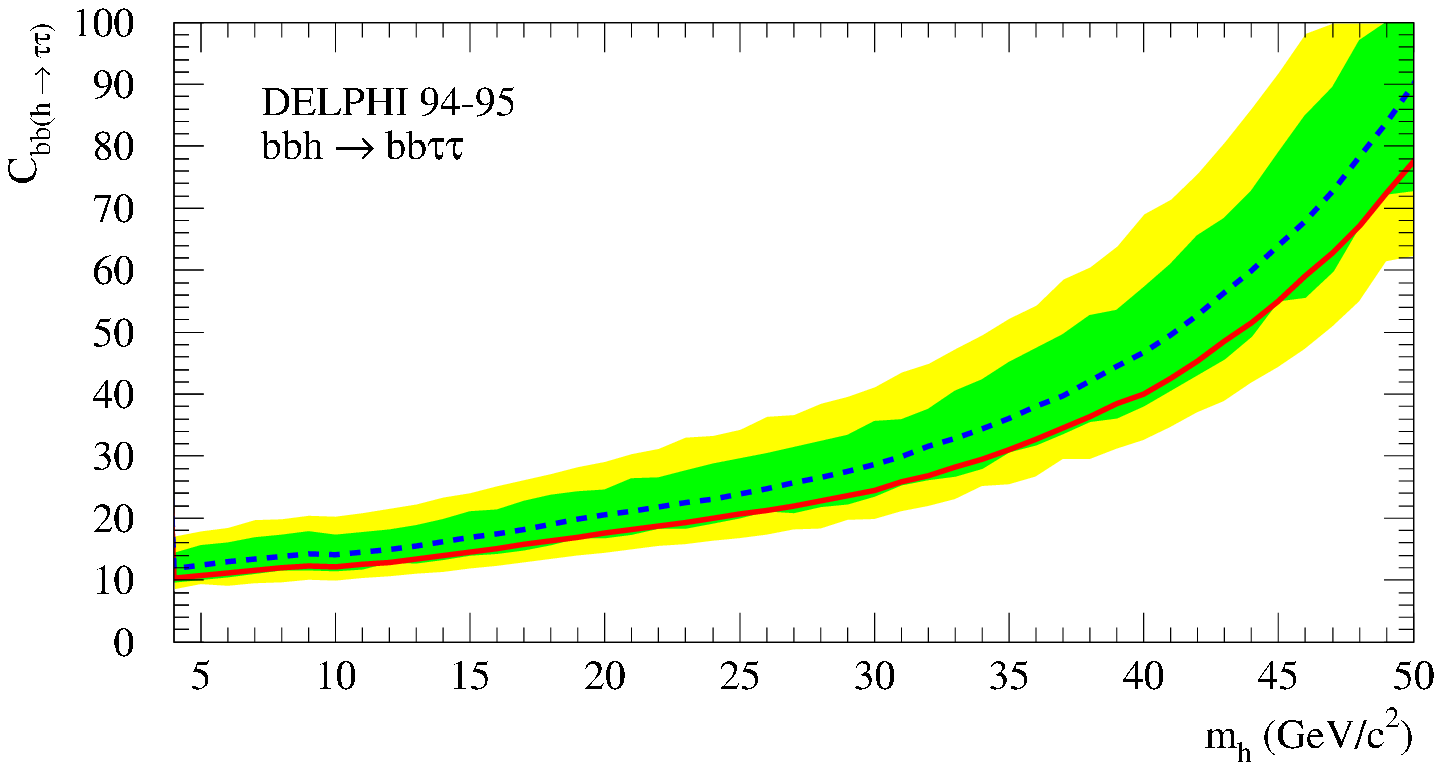}\hfill
\includegraphics[scale=0.53]{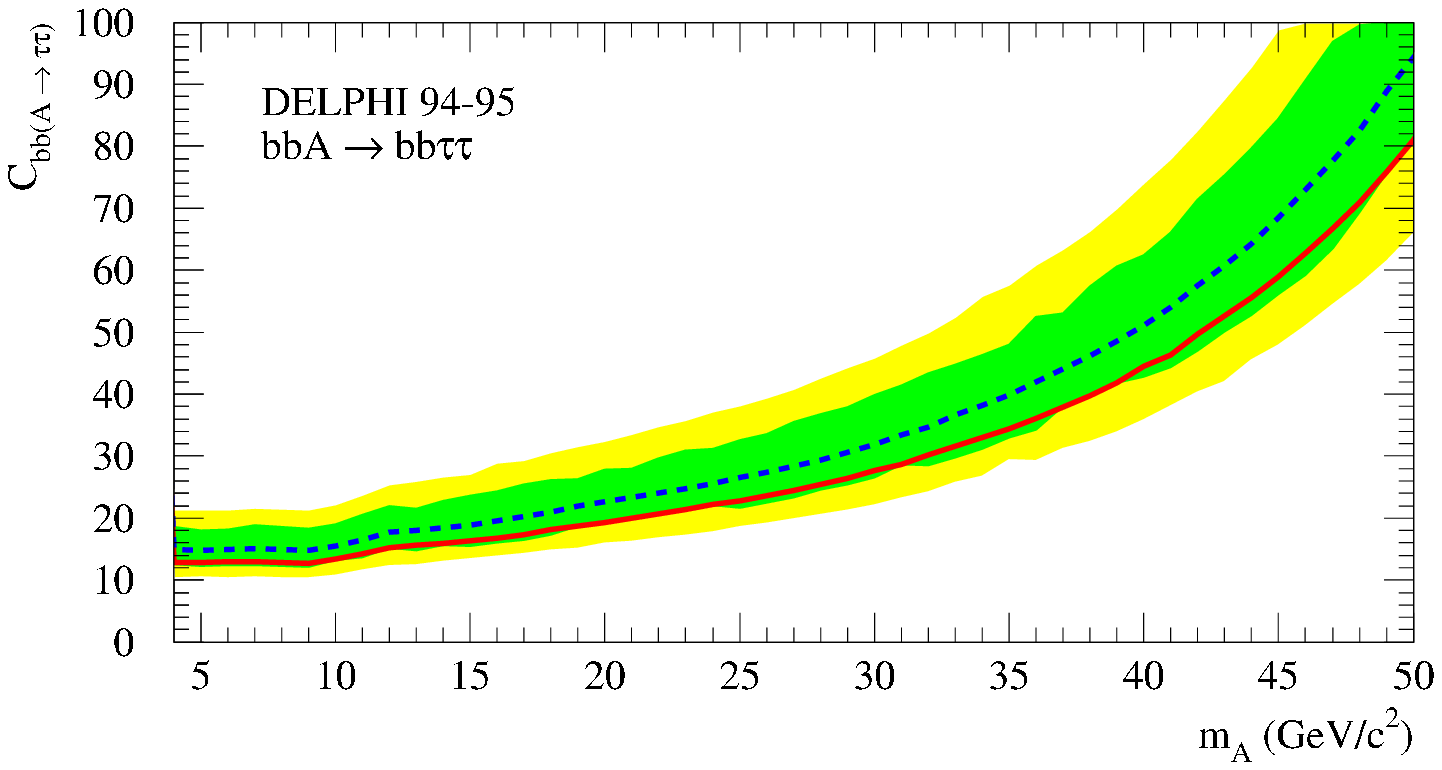}
\end{center}
\vspace*{-0.7cm}
\caption{Observed (solid line) and expected (dotted line) mass limits from searches 
           for the Yukawa processes 
           $\rm e^+e^-\rightarrow b\bar b\rightarrow b\bar bh,~b\bar bA$.
The $C$ factors include vertex enhancement factors and decay branching
fractions. The 1$\sigma$ and 2$\sigma$ error bands on the expected limit for background
are indicated.
\label{fig:yukawa}}
\end{figure}

Figure~\ref{fig:yukawa} shows mass limits from searches for the Yukawa processes 
$\rm e^+e^-\rightarrow b\bar b \rightarrow b\bar bh,~b\bar bA$.

\section{Singly-Charged Higgs Bosons}

\begin{figure}[htb]
\begin{center}
\vspace*{-0.8cm}
\includegraphics[scale=0.3]{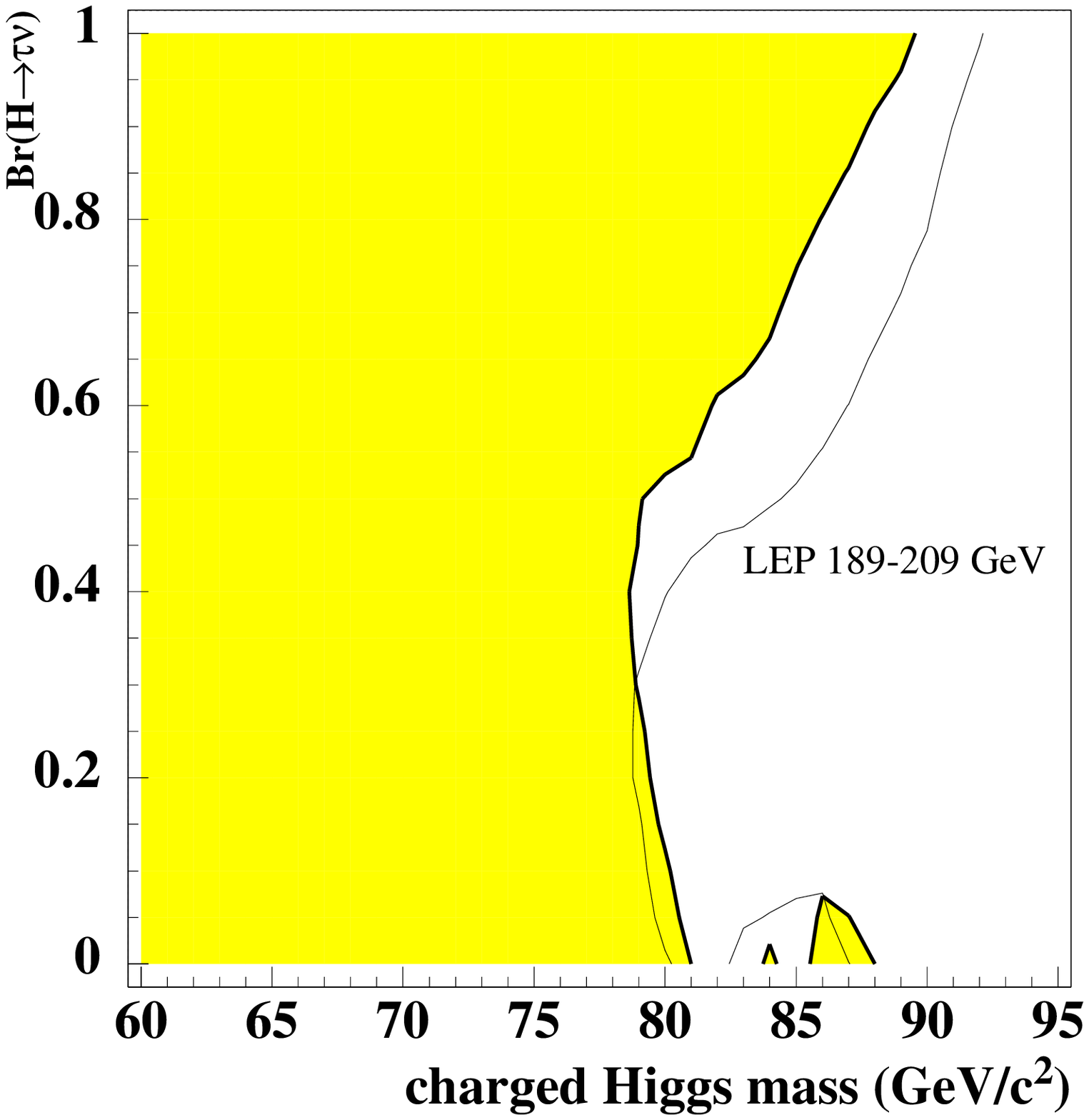}\hfill
\includegraphics[scale=0.27]{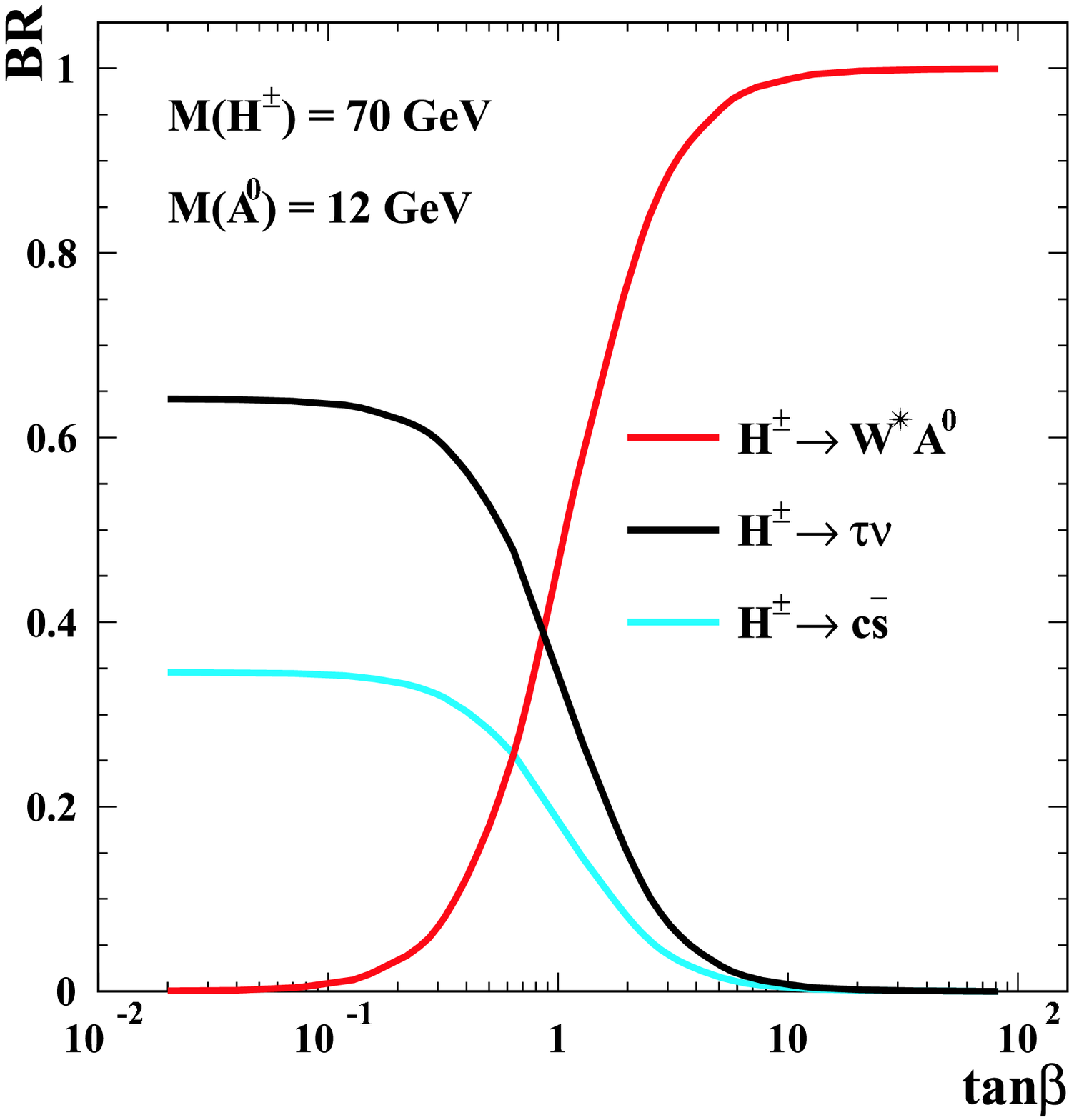}\hfill
\includegraphics[scale=0.27]{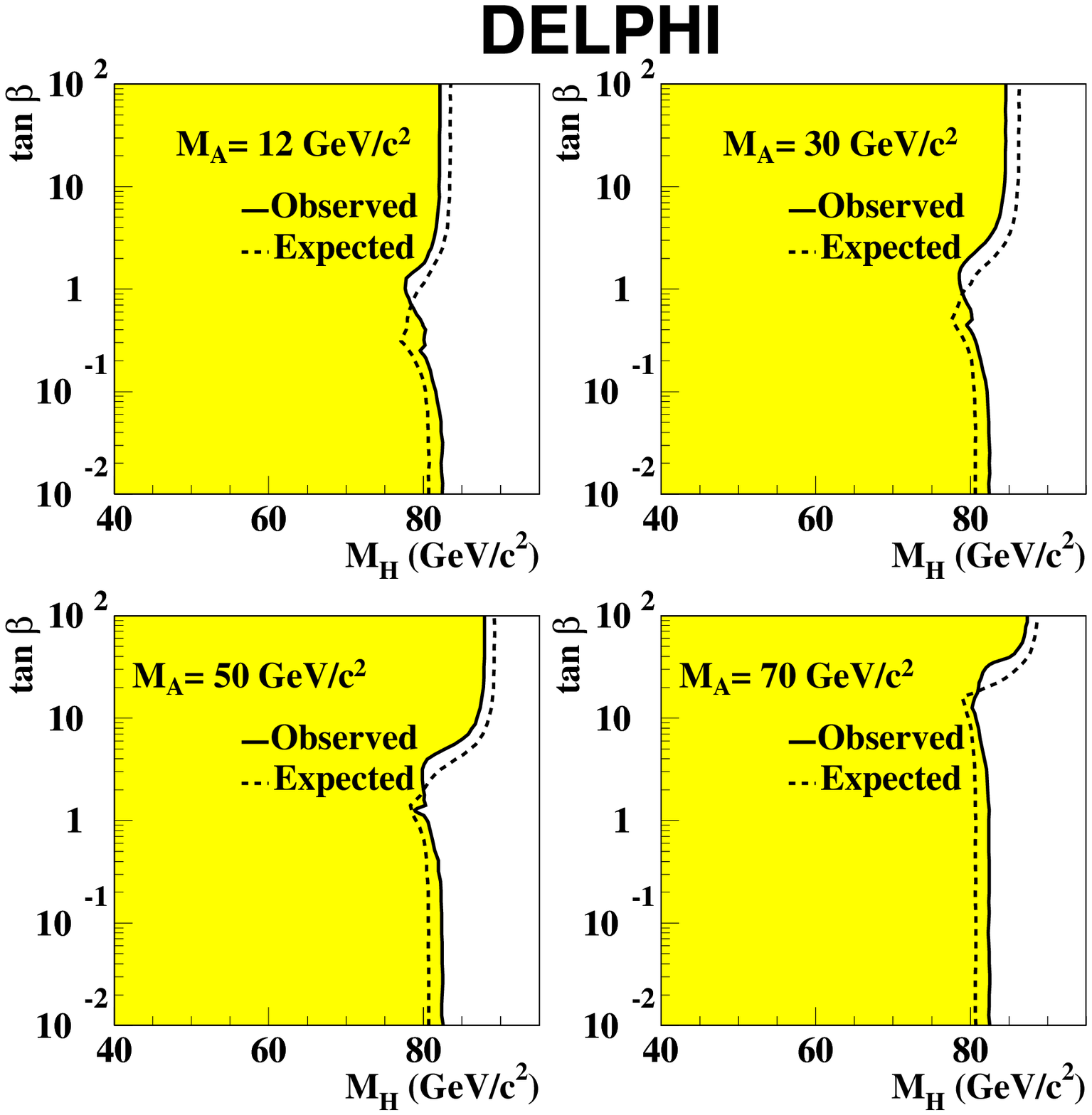}
\end{center}
\vspace*{-0.7cm}
\caption{Left: excluded mass region (shaded area) from searches for
           $\rm \ee\ra H^+H^- \ra c\bar s\bar c s,$ $\rm cs\tau\nu$ and
           $\rm \tau^+\nu \tau^-\bar\nu$.
           The thin line shows the expected limit.
           Center: $\rm H^\pm \ra W^\pm A$ decays could be dominant for light A 
           boson masses.
           Right: excluded mass region (shaded area) from searches for this process.
\label{fig:hpm}}
\vspace*{-0.5cm}
\end{figure}

Figure~\ref{fig:hpm} shows mass limits from searches for
$\rm \ee\ra H^+H^- \ra c\bar s\bar c s,~cs\tau\nu,~\tau^+\nu \tau^-\bar\nu$.
The decay $\rm H^\pm \ra W^\pm A$ could be dominant and limits from 
dedicated searches are set.

\section{Doubly-Charged Higgs Bosons}

\begin{figure}[htb]
\vspace*{-0.45cm}
\begin{center}
\includegraphics[scale=0.28]{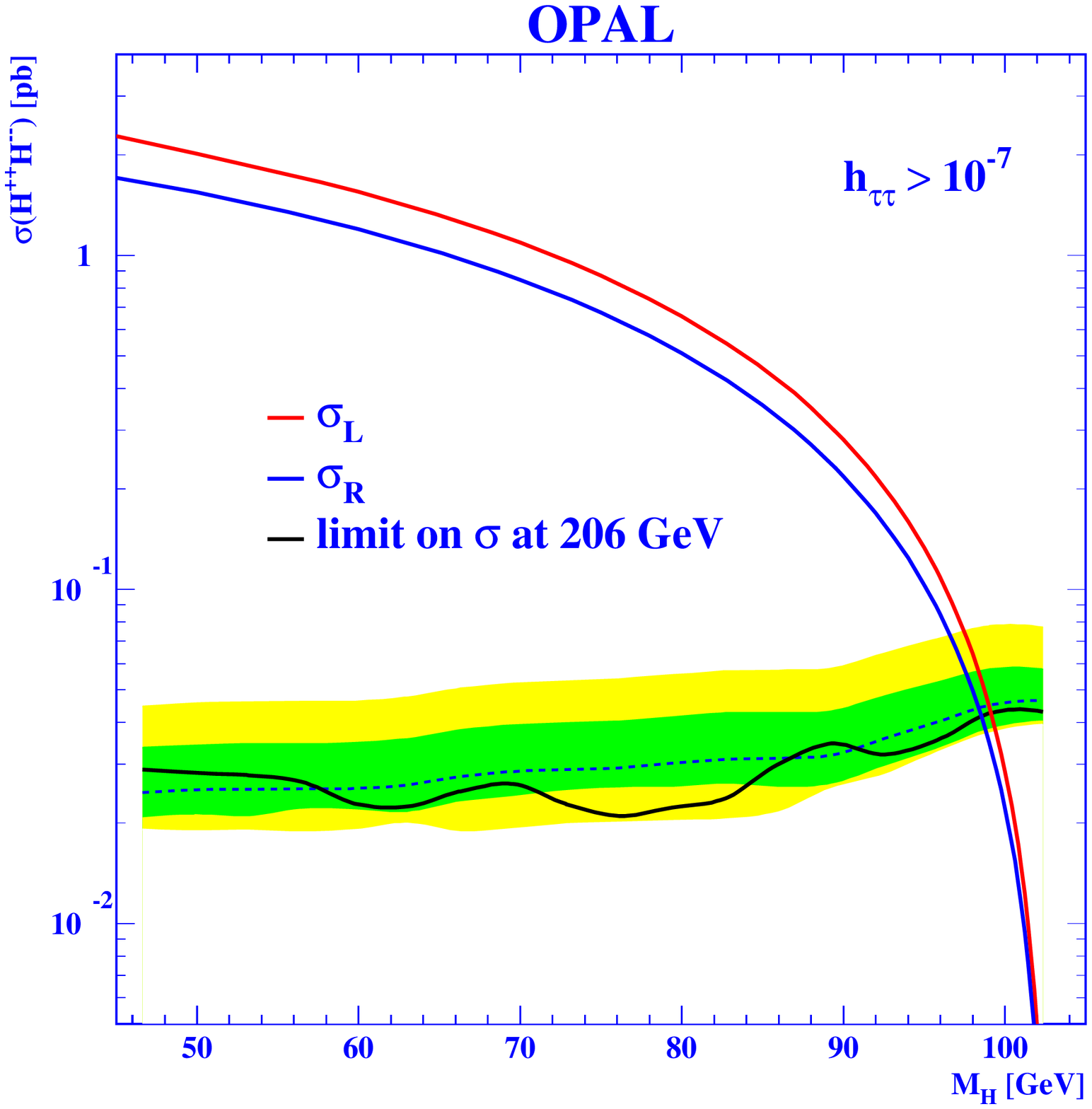}\hfill
\includegraphics[scale=0.36]{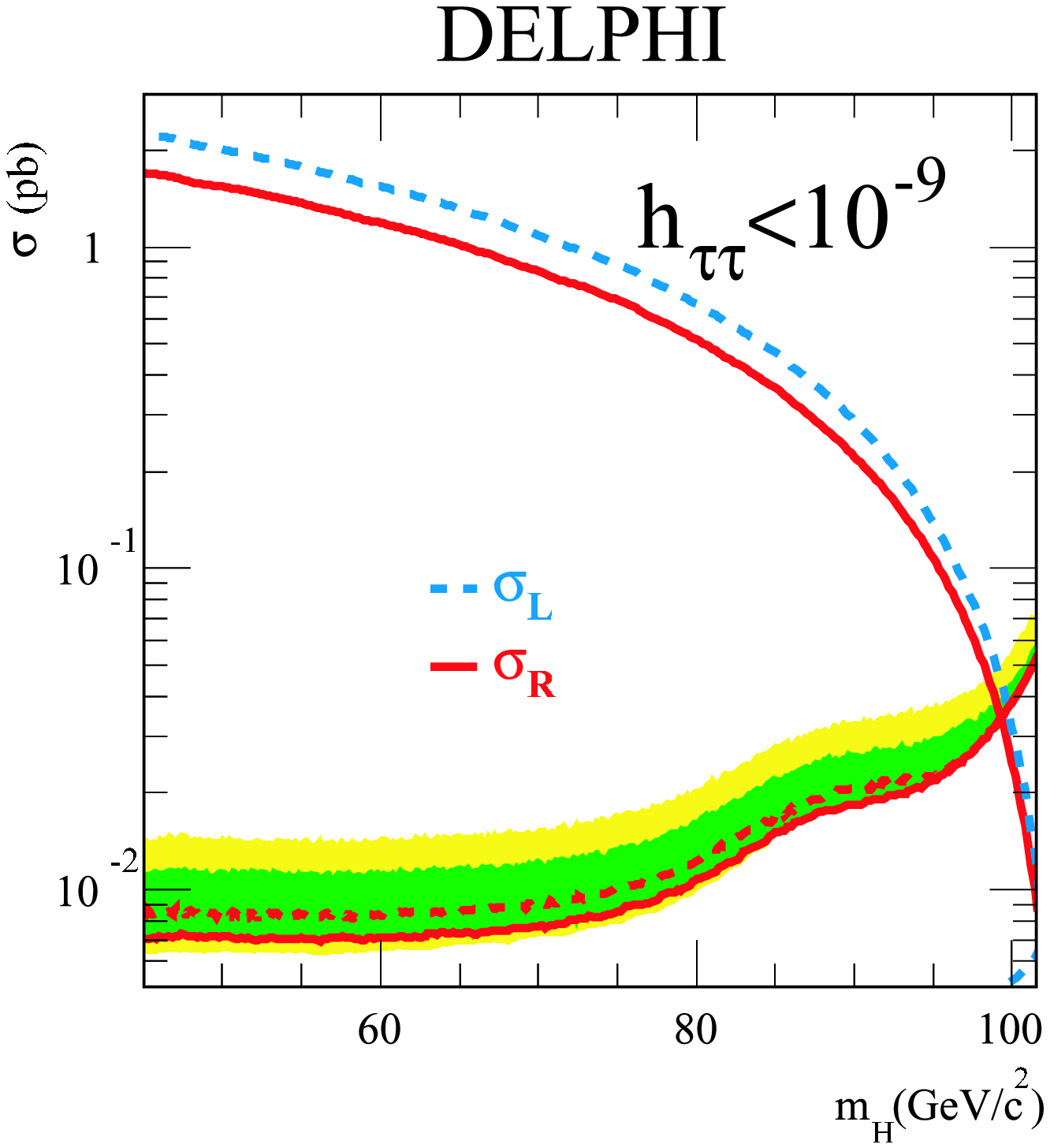}\hfill
\includegraphics[width=0.30\textwidth]{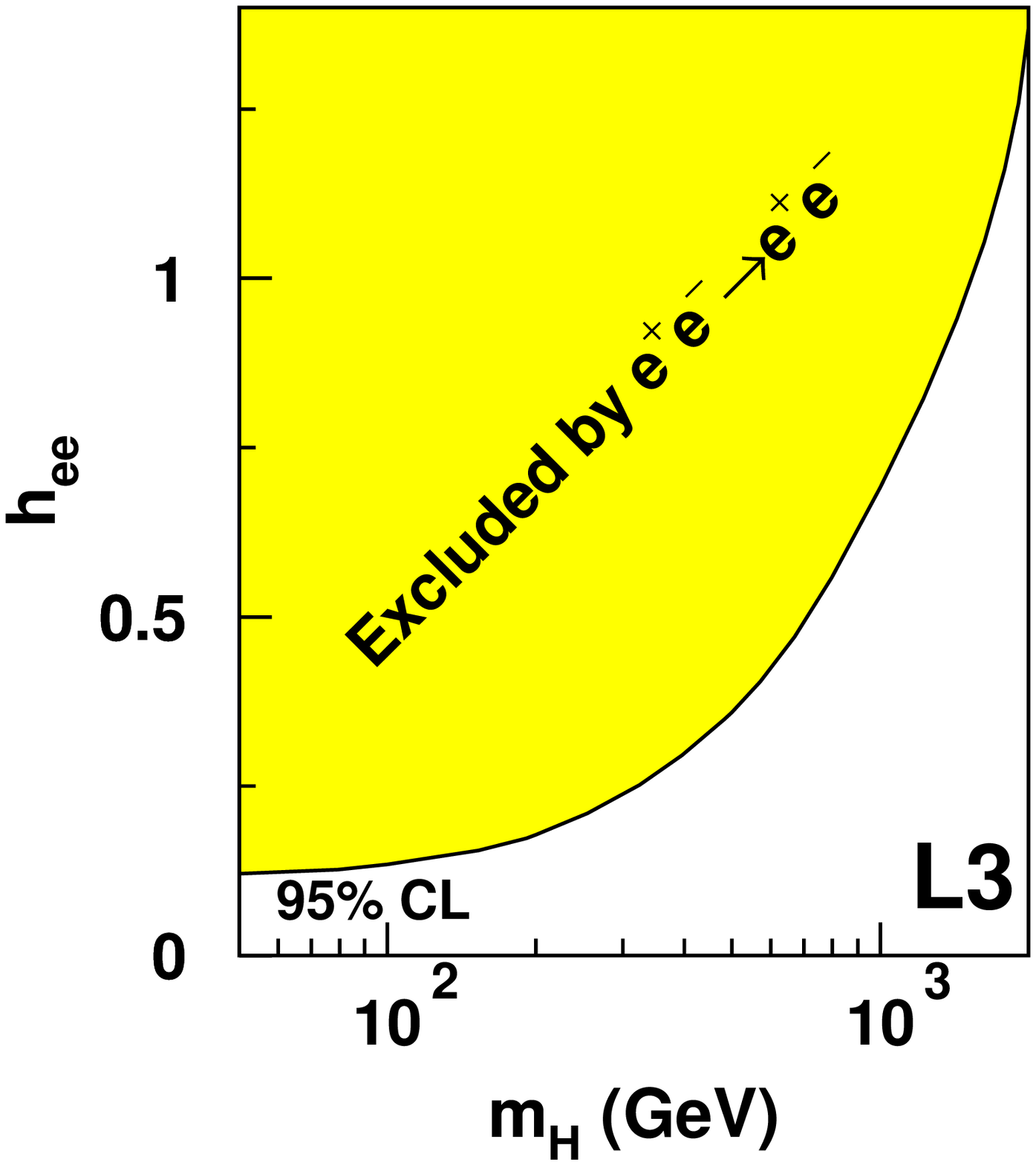}
\end{center}
\vspace*{-0.7cm}
\caption{Left and center: limits on the $\rm \ee\ra H^{++} H^{--}$ production cross section
           as a function of the doubly-charged Higgs boson mass. 
           The 1$\sigma$ and 2$\sigma$ error bands on the expected limit for background
           are indicated (shaded areas).
           Right: $\rm H^{--}$ limits from $\ee\ra\ee$ forward-backward asymmetry.
\label{fig:doubly}}
\vspace*{-0.5cm}
\end{figure}

The process $\rm \ee\ra H^{++} H^{--} \ra \tau^+\tau^+\tau^-\tau^-$ can lead
to decays at the primary interaction point 
($h_{\tau\tau}\geq 10^{-7}$) or a secondary vertex,
or to stable massive particle signatures.
Figure~\ref{fig:doubly} shows limits on the production cross section
and constraints by the forward-backward asymmetry of the process $\ee\ra\ee$.
New results from the Tevatron extend the LEP limits~\cite{d0_hpp,cdf_hpp}.

\clearpage

\section{Fermiophobic Higgs Boson Decays:
\boldmath$\rm h\ra$ WW,~ZZ,~$\gamma\gamma$\unboldmath}
\vspace*{-0.2cm}

\begin{figure}[htb]
\begin{center}
\includegraphics[scale=0.25]{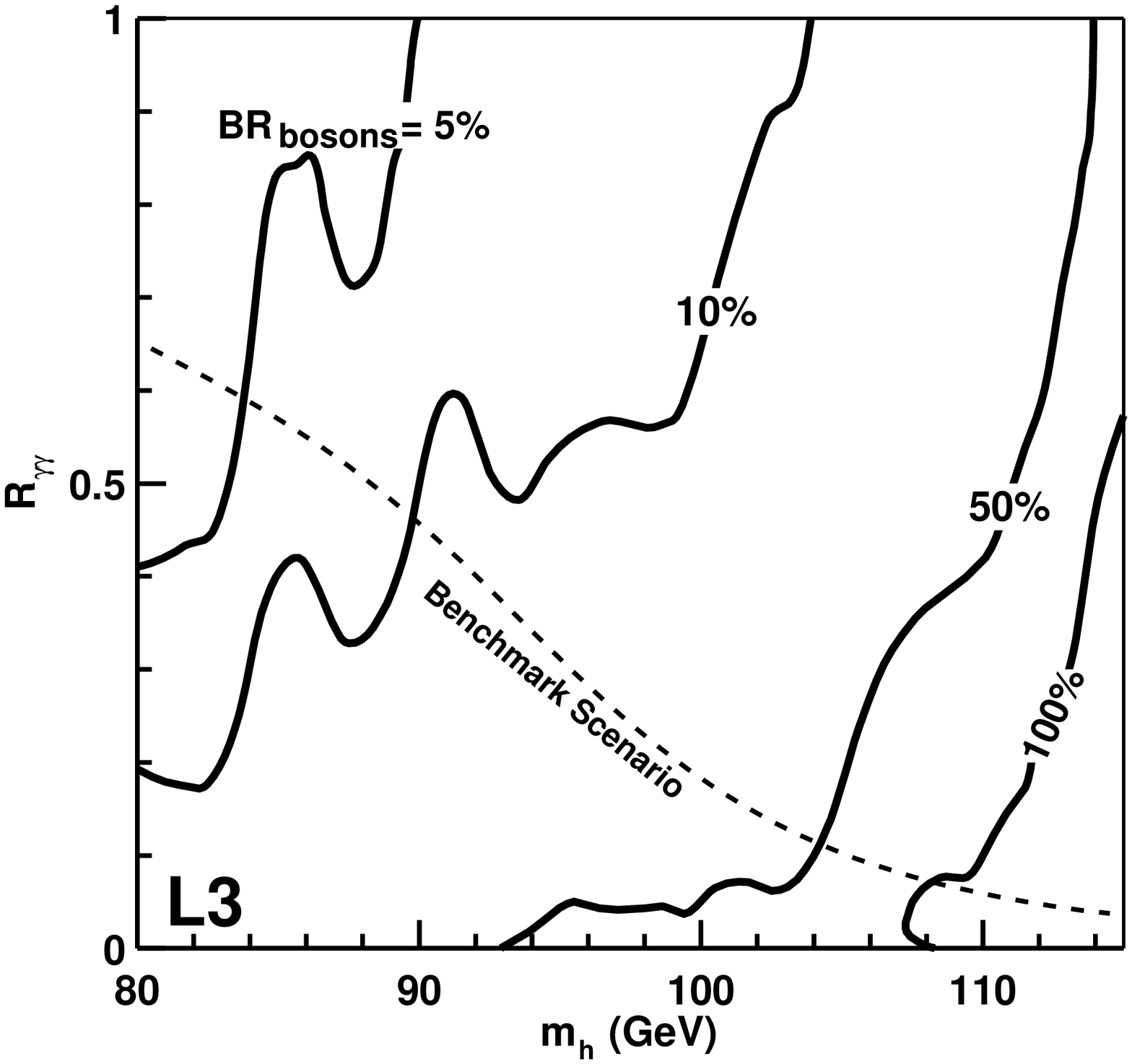}\hfill
\includegraphics[scale=0.26]{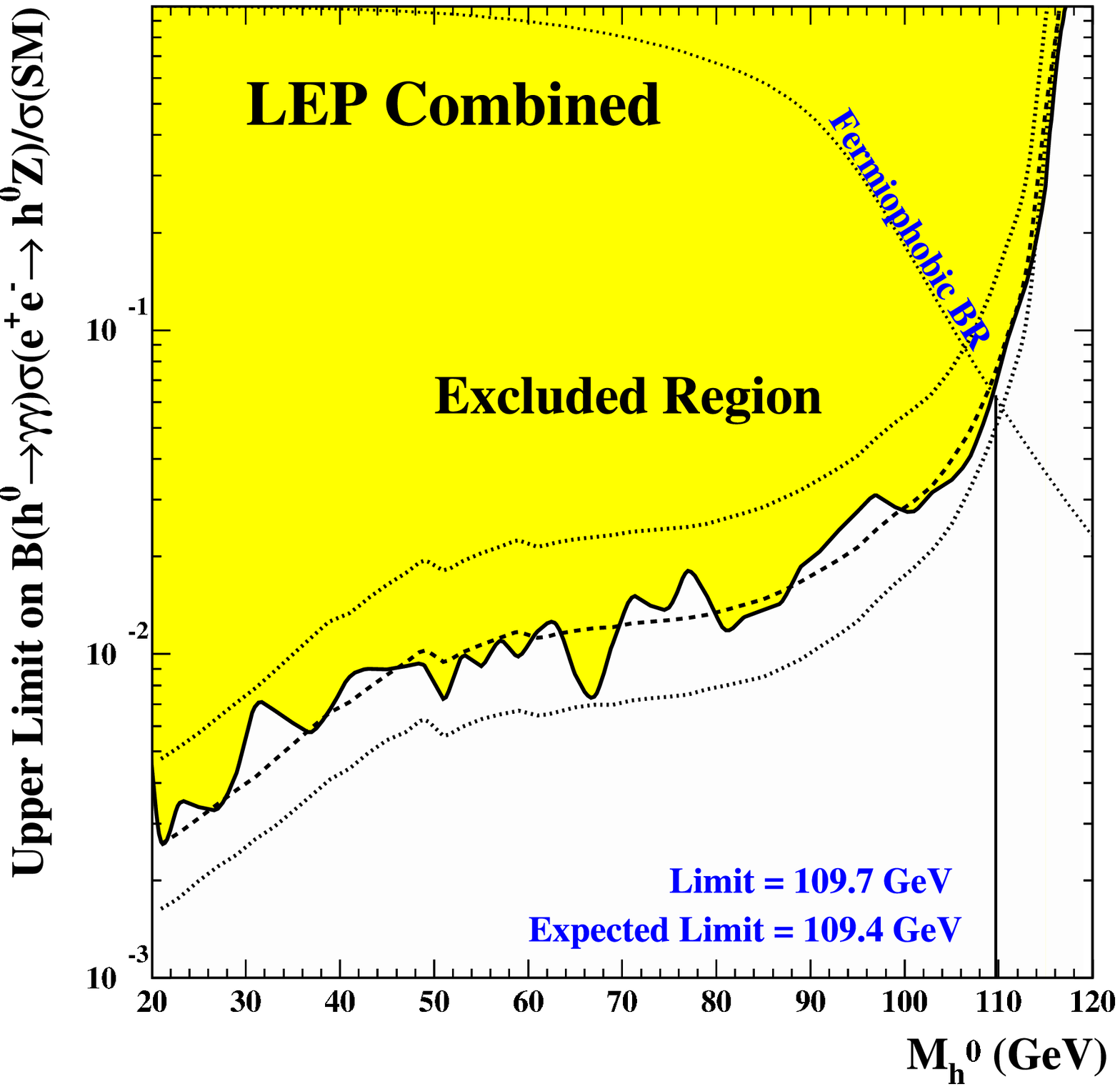}
\end{center}
\vspace*{-0.6cm}
\caption{Left: mass limits 
           from $\rm h\ra$ WW,~ZZ,~$\gamma\gamma$ searches.
           Right: mass limits from $\rm h\ra \gamma\gamma$ combined LEP
                  results.
\label{fig:fermio}}
\end{figure}

If Higgs boson decays into fermions are suppressed, 
$\rm h\ra$ WW,~ZZ,~$\gamma\gamma$ decays could be dominant.
Mass limits from dedicated searches are shown in Fig.~\ref{fig:fermio}.

\section{Uniform and Stealthy Higgs Boson Scenarios}

\begin{figure}[htb]
\vspace*{0.4cm}
\begin{minipage}{0.32\textwidth}
\includegraphics[width=1\textwidth]{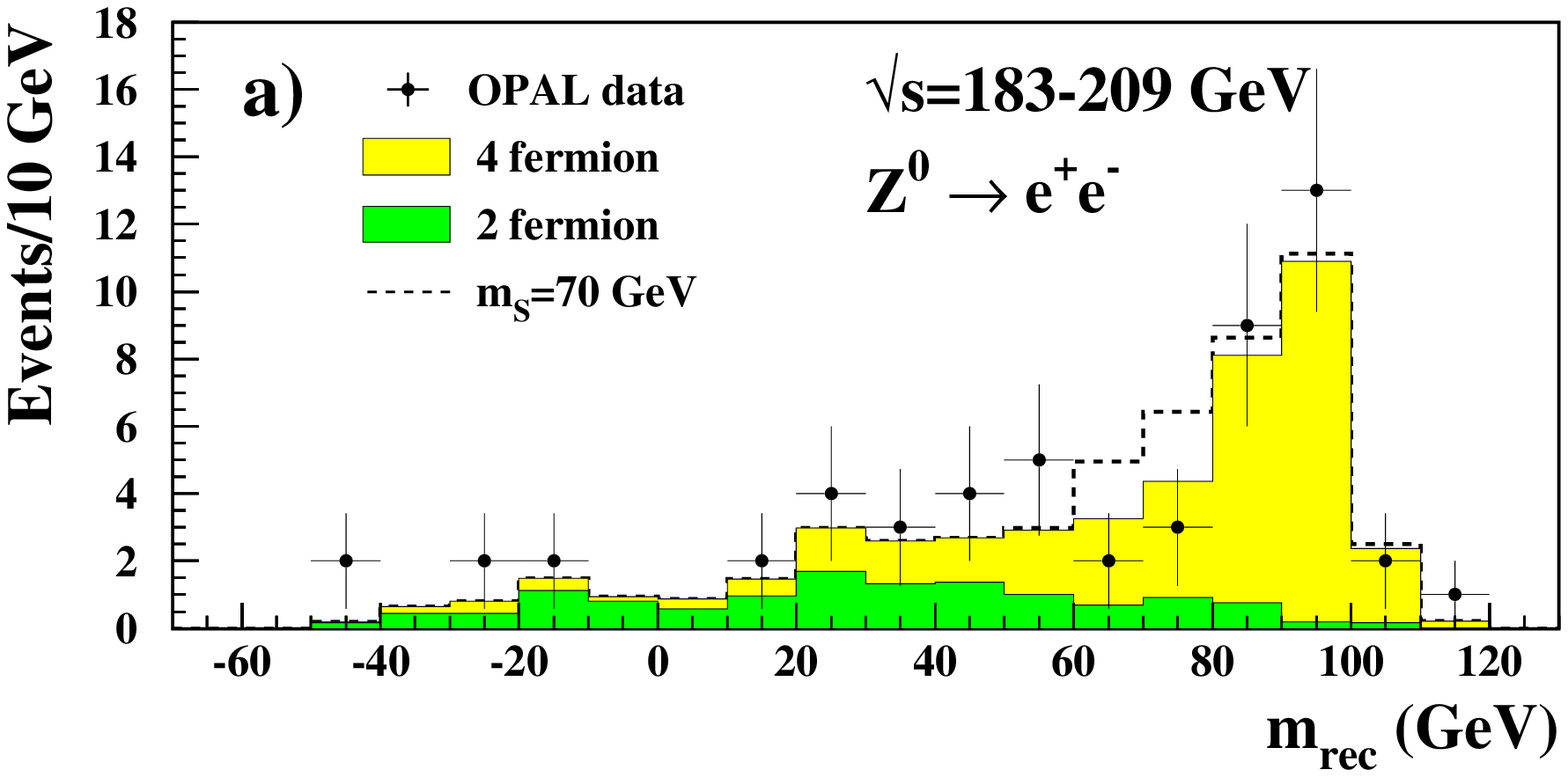}
\includegraphics[width=1\textwidth]{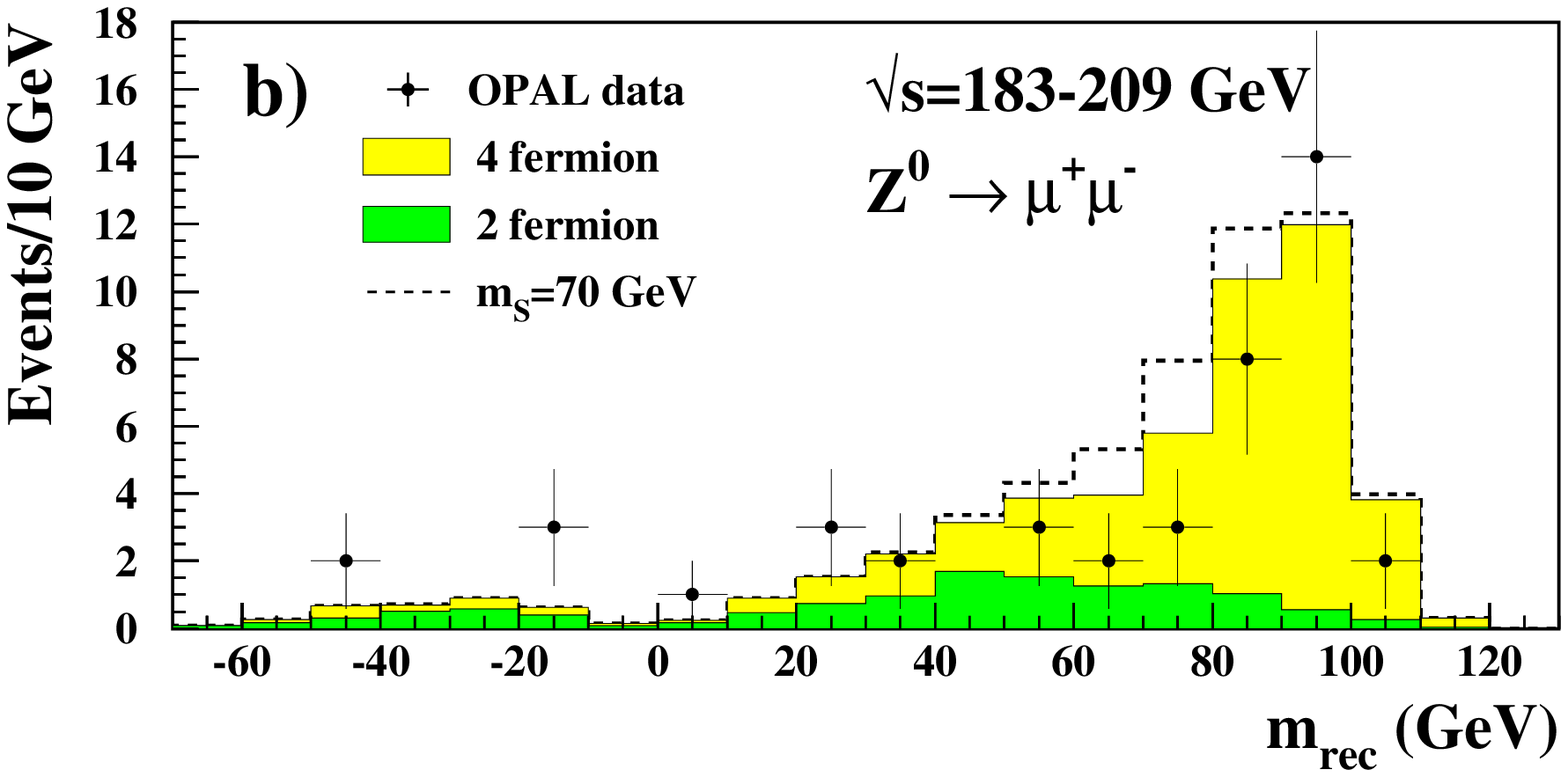}
\end{minipage}
\begin{minipage}{0.67\textwidth}
\includegraphics[width=0.50\textwidth]{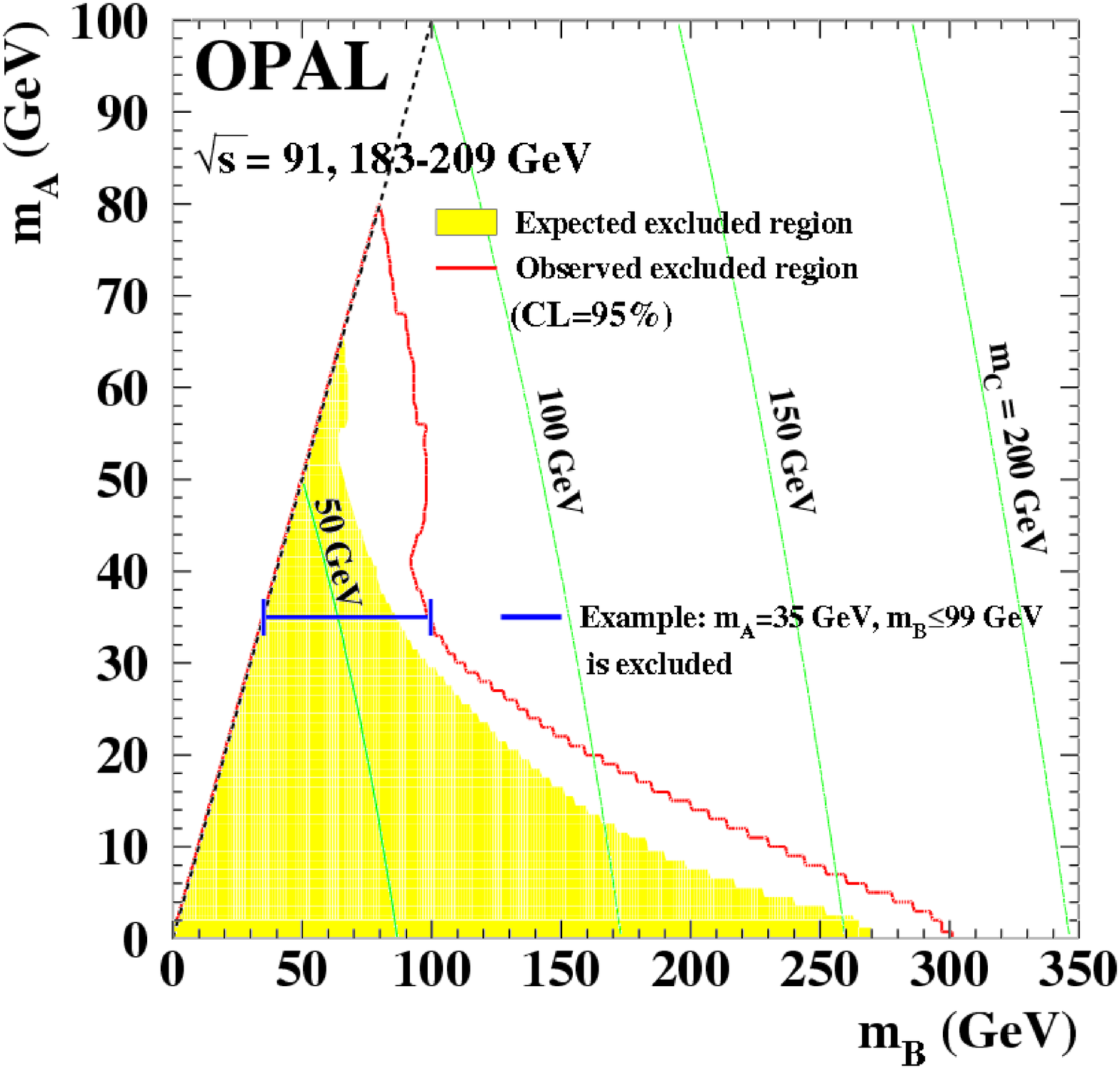}\hfill
\includegraphics[width=0.49\textwidth]{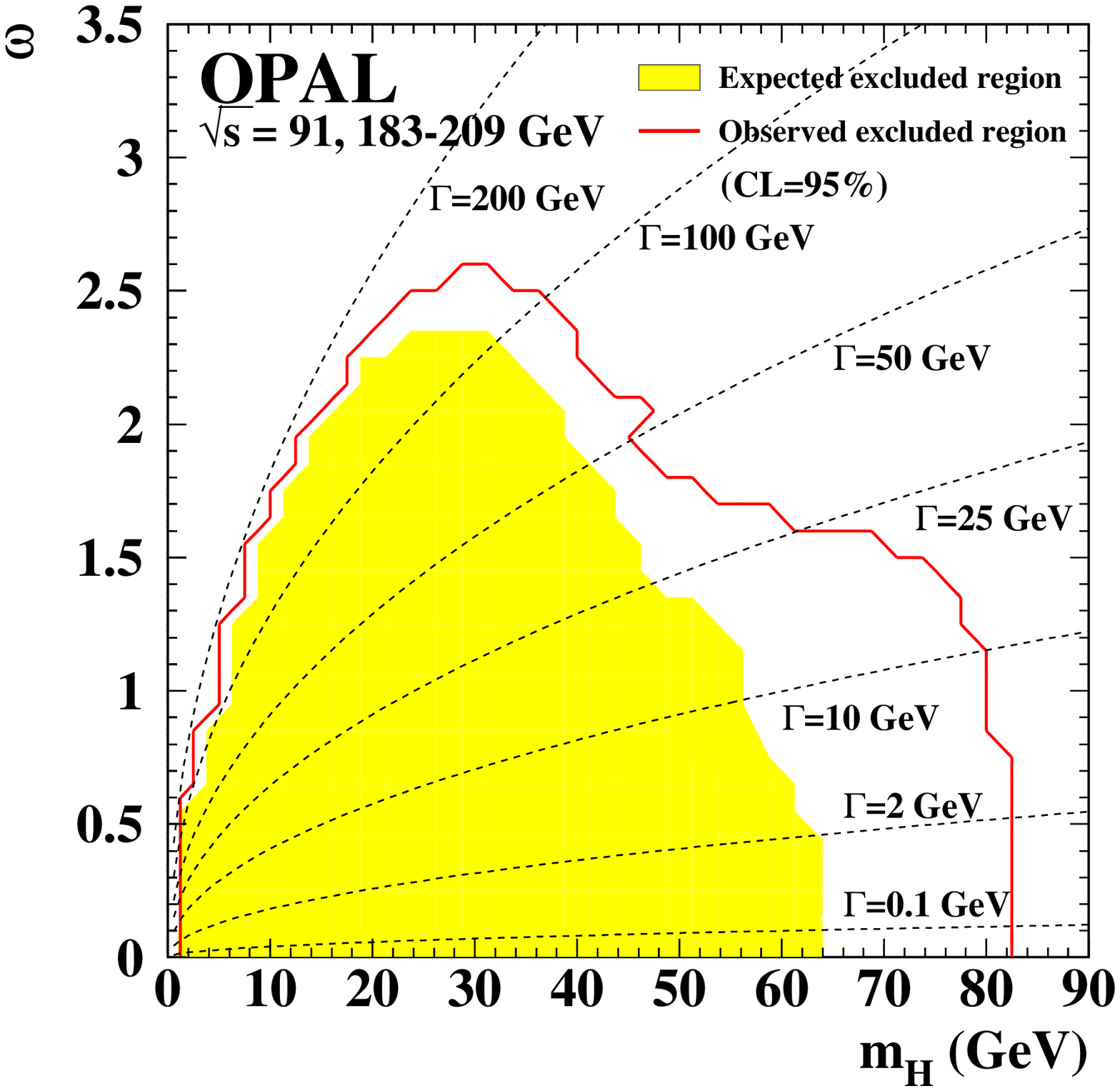}
\end{minipage}
\vspace*{-0.2cm}
\caption{Left: recoiling mass spectrum  of $\rm Z\to e^+e^-$ and $\rm Z\to\mu^+\mu^-$.
         Center: excluded mass range in the uniform Higgs model.
         Right: mass limits in the stealthy Higgs model.
\label{fig:reco_ee}}
\end{figure}

The recoiling mass of the Z boson in the reaction $\rm e^+e^- \to HZ$
allows to search for the Higgs boson independent of the Higgs boson decay mode.
No indication of a Higgs boson signal has been observed as shown in Fig.~\ref{fig:reco_ee}.
Mass limits are shown in the uniform Higgs boson model,
where many uniform Higgs boson states exist in the range between
$m_{\rm A}$ and $m_{\rm B}$.
Another result from the recoiling mass spectrum is shown, where a stealthy Higgs boson 
has a large decay width owing to extra Higgs boson singlets in the model. The decay width 
depends on the parameter $\omega$.

\clearpage
\section{LEP Summary}
\vspace*{-0.2cm}

Immense progress has been made over a period of about 15 years in 
searches for Higgs bosons and much knowledge has been gained in preparation 
for new searches. 
No 
signal has been observed and various stringent limits are set as summarized in Table~\ref{tab:summary}. 

\begin{table}[htb]
\renewcommand{\arraystretch}{1.2} 
\caption{Summary of Higgs boson mass limits at 95\% CL.
`LEP' indicates a combination of the results from ALEPH, DELPHI, L3 and OPAL.
If results from the experiments are not (yet) combined, examples
which represent the different research areas from individual experiments
are given.
\label{tab:summary} }
\begin{center}

\begin{tabular}{c|c|r}
 Search                     & Experiment & Limit \\\hline 
Standard Model              &   LEP  
   & $m^{\rm SM}_{\rm H} > 114.4$ GeV \\ 
Reduced rate and SM decay &       
  & $\xi^2>0.05:$ $ m_{\rm H} > 85$ GeV \\
& & $\xi^2>0.3:$ $ m_{\rm H} > 110$ GeV \\
Reduced rate and $\rm b\bar b$ decay  &   
  & $\xi^2>0.04:$ $ m_{\rm H} > 80$ GeV \\
& & $\xi^2>0.25:$ $ m_{\rm H} >110$ GeV \\ 
Reduced rate and $\tau^+\tau^-$ decay & 
  & $\xi^2>0.2:$ $ m_{\rm H} > 113$ GeV \\ 
\hspace*{-4mm} Reduced rate and hadronic decay\hspace*{-2mm} &
  & $\xi^2=1:$   $m_{\rm H} >112.9$ GeV\\ 
& & $\xi^2>0.3:$ $ m_{\rm H} > 97$ GeV \\ 
&ALEPH& $\xi^2>0.04:$ $m_{\rm H} \approx 90$ GeV \\ 
Anomalous couplings & L3 & $d,~\db,~\dgz,~\dkg$ exclusions \\ \hline
MSSM (no scalar top mixing) & LEP 
  & almost entirely excluded\\ 
General MSSM scan & DELPHI &  $m_{\rm h} > 87$ GeV, $m_{\rm A} >90$ GeV\\ 
Larger top-quark mass     & LEP & strongly reduced $\tan\beta$ limits \\ \hline
CP-violating models   & LEP    &  strongly reduced mass limits  \\ \hline
Visible/invisible Higgs decays & DELPHI & $m_{\rm H} >111.8$ GeV\\ 
Majoron model (max. mixing) &  & $m_{\rm H,S} >112.1$ GeV\\ \hline
Two-doublet Higgs model   & DELPHI
  & $\rm hA\ra b\bar b b\bar b:$
    $m_{\rm h}+m_{\rm A} >  150$ GeV\\
(for $\sigma_{\rm max}$) & 
  & $\tau^+\tau^-\tau^+\tau^-:$
    $m_{\rm h}+m_{\rm A} >  160$ GeV\\
& & $\rm (AA)A\ra 6b:$ $m_{\rm h}+m_{\rm A} >  150$ GeV\\
& & $\rm (AA)Z\ra 4b~Z:$ $m_{\rm h} >  90$ GeV\\
& & $\rm hA\ra q\bar q q\bar q:$ 
      $m_{\rm h}+m_{\rm A} >  110$ GeV\\
Two-doublet model scan & OPAL
  & $\tan\beta > 1:$ $ m_{\rm h} \approx m_{\rm A} > 85$ GeV \\\hline 
Yukawa process & DELPHI & $C > 40:$ $m_{\rm h,A} > 40$ GeV \\\hline 
Singly-charged Higgs bosons & LEP 
  & $m_{\rm H^\pm} > 78.6$ GeV \\
$\rm W^\pm A$ decay mode & DELPHI& $m_{\rm H^\pm} > 76.7$ GeV \\ \hline
Doubly-charged Higgs bosons & \hspace*{-2.5mm} DELPHI/OPAL \hspace*{-2.5mm} 
  & 
$m_{\rm H^{++}} > 99$ GeV \\
$\ee\ra\ee$ &L3 &$h_{\rm ee} > 0.5:$ $m_{\rm H^{++}} > 700$ GeV \\ \hline
Fermiophobic $\rm H\ra WW, ZZ, \gamma\gamma$ & L3 
  &  $m_{\rm H} > 108.3$ GeV \\
$\rm H\ra \gamma\gamma$ &LEP &  $ m_{\rm H} > 109.7$ GeV \\ \hline
Uniform and stealthy scenarios & OPAL & depending on model parameters
\end{tabular}
\end{center}
\vspace*{-0.6cm}
\end{table}

\clearpage


\section{International Linear Collider (ILC)}

The next generation $\rm e^+e^-$ collider will be a linear collider. 
In summer 2004 as a major step forward in the ILC programme a world-wide
agreement has been reached that the ILC will operate with the cold 
technology (superconducting cavities).
Since the TESLA Technical Design Report (TDR)~\cite{tdr},
the physics programme of an electron-positron linear collider (LC)
has been further developed and a wide consensus has been reached on the
physics case and the need for a high luminosity LC with center-of-mass
energy up to about 1 TeV as the next worldwide high-energy physics project.
Recent milestones were set with 
the Snowmass Study~\cite{snowmass} and 
the international workshops on LC in Korea~\cite{korea} 
and Paris~\cite{paris04}.

The study of the Higgs boson properties represents a significant part of
this physics programme. 
A linear collider of at least 500~GeV and a total luminosity of at least
1000~fb$^{-1}$ has much potential for studying Higgs bosons and 
understanding the electroweak symmetry breaking and mass generation. 

The SM physics is addressed and it is discussed how precisely a future LC 
can determine the Higgs boson production mechanism.
Indirect and direct branching ratio measurements are reviewed. 
Then the characterization of the Higgs boson potential, which 
will contribute to establishing the underlying mechanism of mass generation,
is addressed.
Higgs bosons could also be produced via Higgs-strahlung off top quarks. 
In the general two Higgs doublet model (2HDM) charged Higgs bosons are
prominent. 
Various methods to determine the ratio of the vacuum expectation values
of the two doublets, $\tan\beta$, are discussed.
In the framework of the Minimal Supersymmetric extension of the SM (MSSM) 
or beyond, invisibly decaying Higgs bosons could be produced and their 
properties measured. It has recently been emphasized that the LC collider
precision is essential in the scenario of the decoupling limit 
(MSSM$\rightarrow$SM)~\cite{haber}.
Furthermore, the measurement of the Higgs boson parity is discussed as
well as important possibilities to distinguish Higgs boson models.
For several LC studies the relation to the LHC potential is addressed.
Future LC Higgs studies will concentrate on
more detailed detector simulations reflecting the progress in detector 
technologies,
the second phase of a LC with higher center-of-mass energies,
and new theoretical developments, such as extra dimensions.

\vspace*{-0.1cm}
\section{Standard Model Physics}

\subsection{Higgs boson production mechanism}
\vspace*{-0.1cm}

\begin{figure}[bp]
\vspace*{-1.3cm}
\includegraphics[scale=0.26]{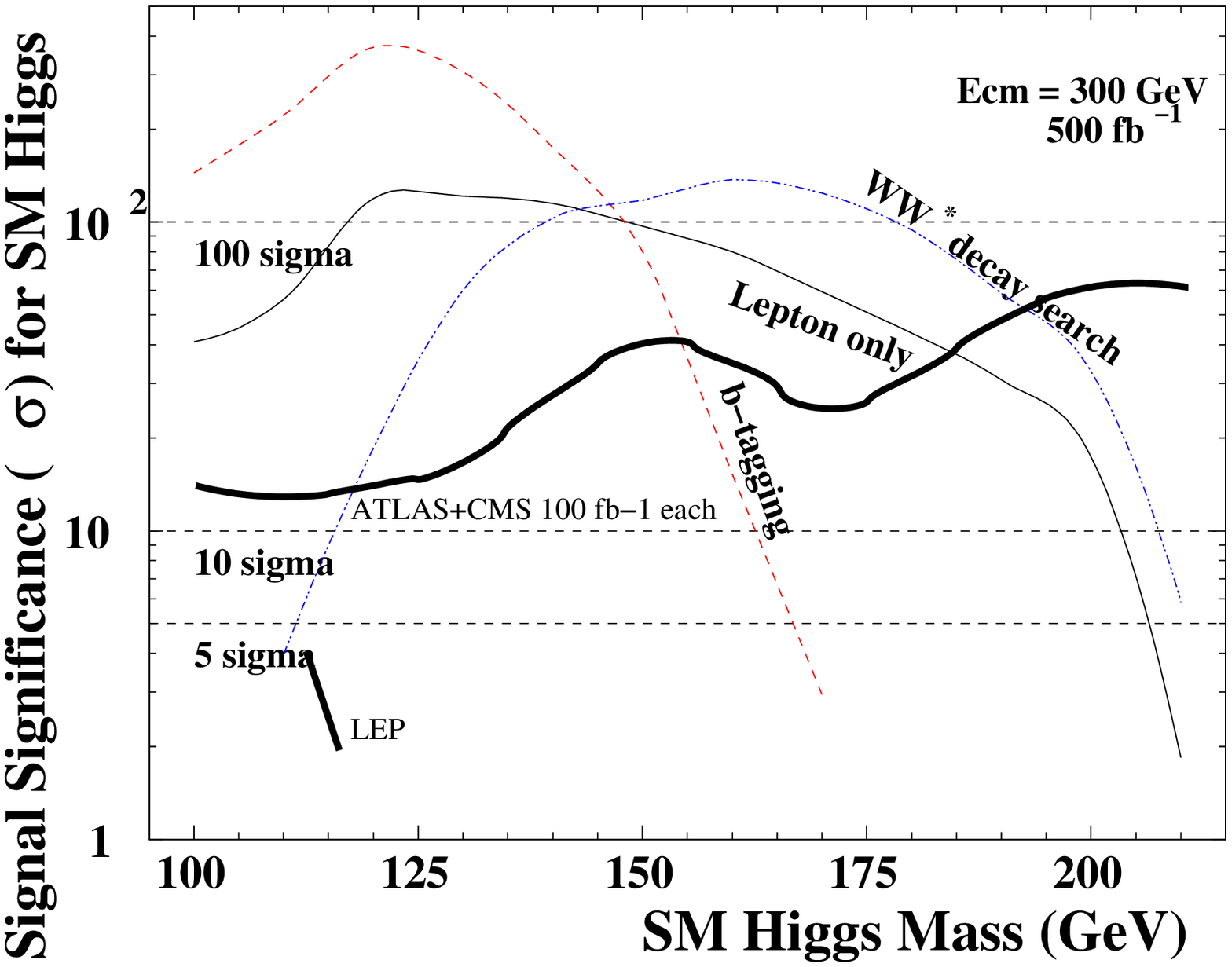}
\includegraphics[scale=0.26]{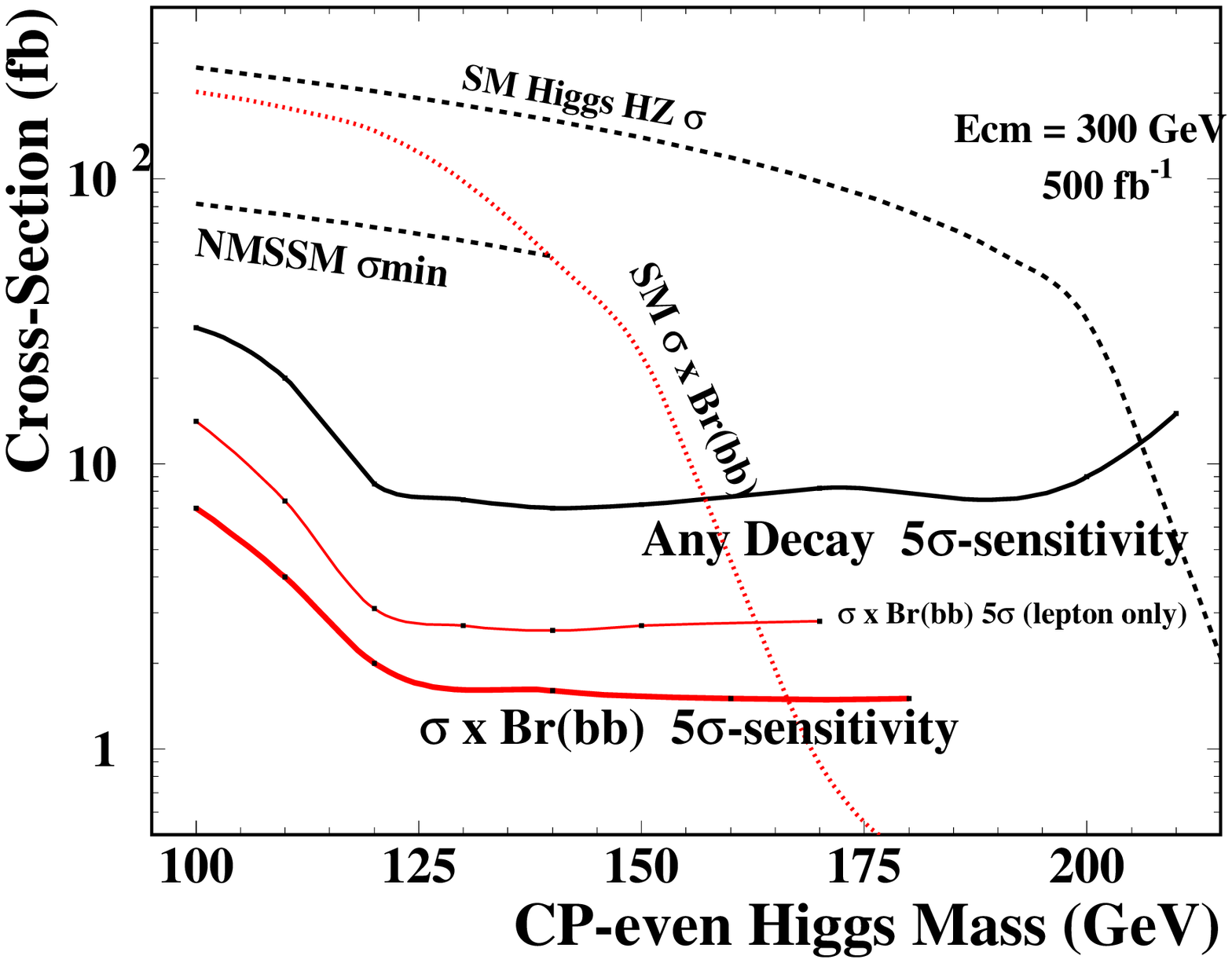}
\includegraphics[scale=0.7, bbllx=284pt,bblly=479pt,bburx=515pt,bbury=631pt,clip=]{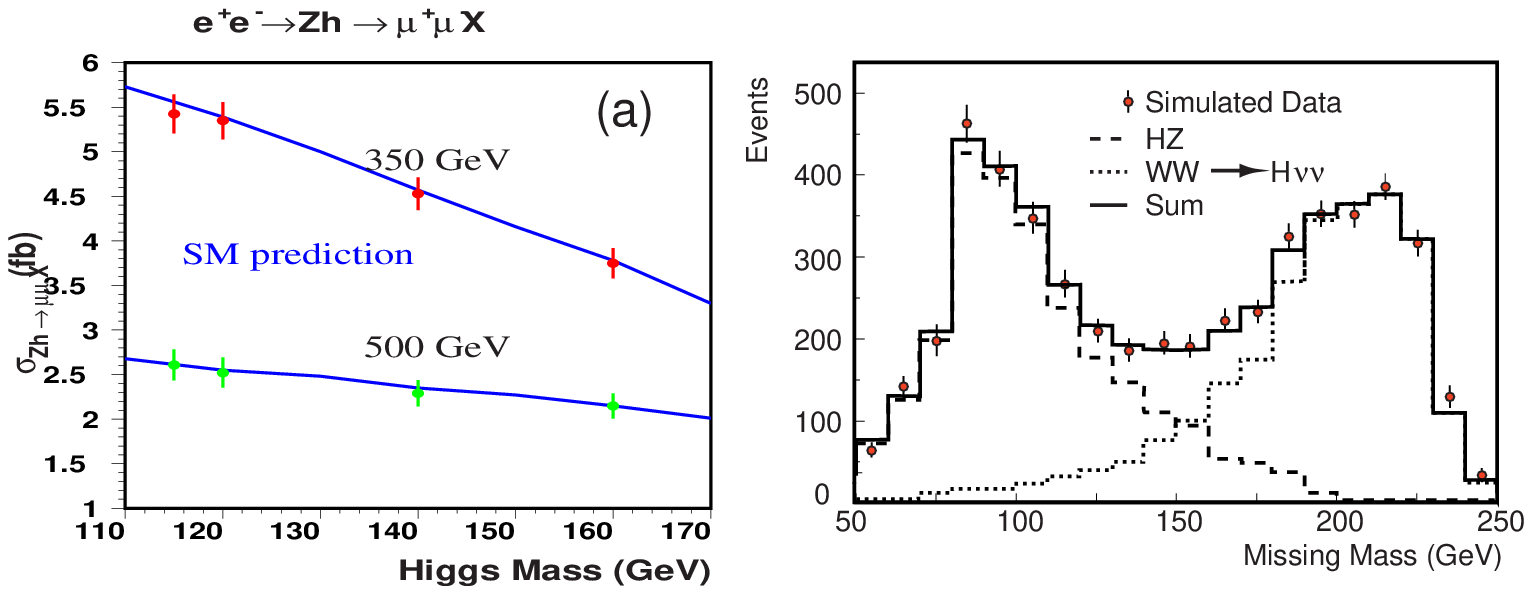}
\vspace*{-0.2cm}
\caption{
Left: SM Higgs boson signal significance.
Center: expected production cross sections (dashed lines) and simulated experimental 
      sensitivities (solid lines).
Right: expected missing mass distributions for Higgs-strahlung 
         and WW fusion. Their sum is shown in comparison with 
         simulated data.
}
\vspace*{-0.4cm}
\label{fig:smmass}
\end{figure}

The expected SM Higgs production rate
has a very large significance over the background
($\sigma\equiv N_{\rm sig}/\sqrt{N_{\rm bg}}$).
More than $100\sigma$ is obtainable in the $\rm H\rightarrow b\bar b$ 
decay channel.
For heavier Higgs bosons, the WW decay mode takes over.
In relation to the LHC, a LC has a much larger
sensitivity in the lower mass range, 
while the LHC can probe heavier Higgs bosons.
At about 115~GeV mass, the LEP sensitivity reduces from 
about $4\sigma$ to $2\sigma$.
Figure~\ref{fig:smmass} (left plot) shows the SM Higgs boson 
detection significance.

The sensitivity to the general production cross section has been
studied. The expected SM production cross section is much larger than the 
$5\sigma$ sensitivity to the cross section over a wide mass range. 
For lower masses the Higgs boson
decay mode into a pair of b-quarks gives the largest sensitivity.
Figure~\ref{fig:smmass} (center plot) shows the cross section sensitivity for
any Higgs boson decay mode, 
where the Higgs boson mass 
is reconstructed as the recoiling mass to the Z decay products. 
Even for models beyond the SM, the sensitivity to the
cross section is much better than the expected minimal 
production cross section.

The Higgs boson will be produced via Higgs-strahlung and 
WW fusion by the two processes 
$$\rm e^+e^-\rightarrow HZ \rightarrow H\nu\bar\nu \rightarrow b\bar b\nu\bar\nu~~~and
$$
$$\rm e^+e^-\rightarrow WW\nu\bar\nu \rightarrow \nu\bar\nu H\rightarrow \nu\bar\nu b\bar b.
$$
These production mechanisms can be distinguished by fitting the missing
mass distribution as shown in Fig.~\ref{fig:smmass} (right plot).
Higgs-strahlung gives a shape like the dashed line, while Higgs boson fusion is represented 
by the dotted line. 
Their sum is given by the solid line and the simulated data is shown with
error bars.

\subsection{Indirect and direct branching ratio measurements}

A LC will perform very high precision measurements of the Higgs boson
branching fractions. The underlying production process is the 
Higgs-strahlung, 
$$\rm e^+e^-\rightarrow HZ\rightarrow H\ell^+\ell^-,$$ 
where the associated Z boson decays into a lepton pair.
Two methods are discussed to determine the Higgs boson branching ratio. 

In the indirect method, the inclusive production cross section as the 
product of the Higgs boson production cross section 
and the branching fraction of the
Z into leptons is determined: 
$$\sigma_{\rm inc} = \sigma_{\rm HZ}
            BR(\rm Z\rightarrow \ell^+\ell^-).$$
This measurement is independent of the
Higgs boson decay mode. The mass recoiling to the lepton pair corresponds 
to the Higgs boson mass. An individual Higgs boson decay cross section
is measured, which is the product of the Higgs boson production cross
section and the Higgs and Z boson decay branching fractions:
\vspace*{-1mm}
$$\sigma({\rm X}) = \sigma_{\rm HZ}
            BR({\rm Z\rightarrow Y}) 
            BR({\rm H\rightarrow X}).$$
By taking the ratio
of both cross sections the Higgs branching ratio can be determined, since
the LEP experiments measured the Z decay branching fractions with
high precision.

\begin{figure}[htb]
\vspace*{-2cm}
\includegraphics[scale=0.26]{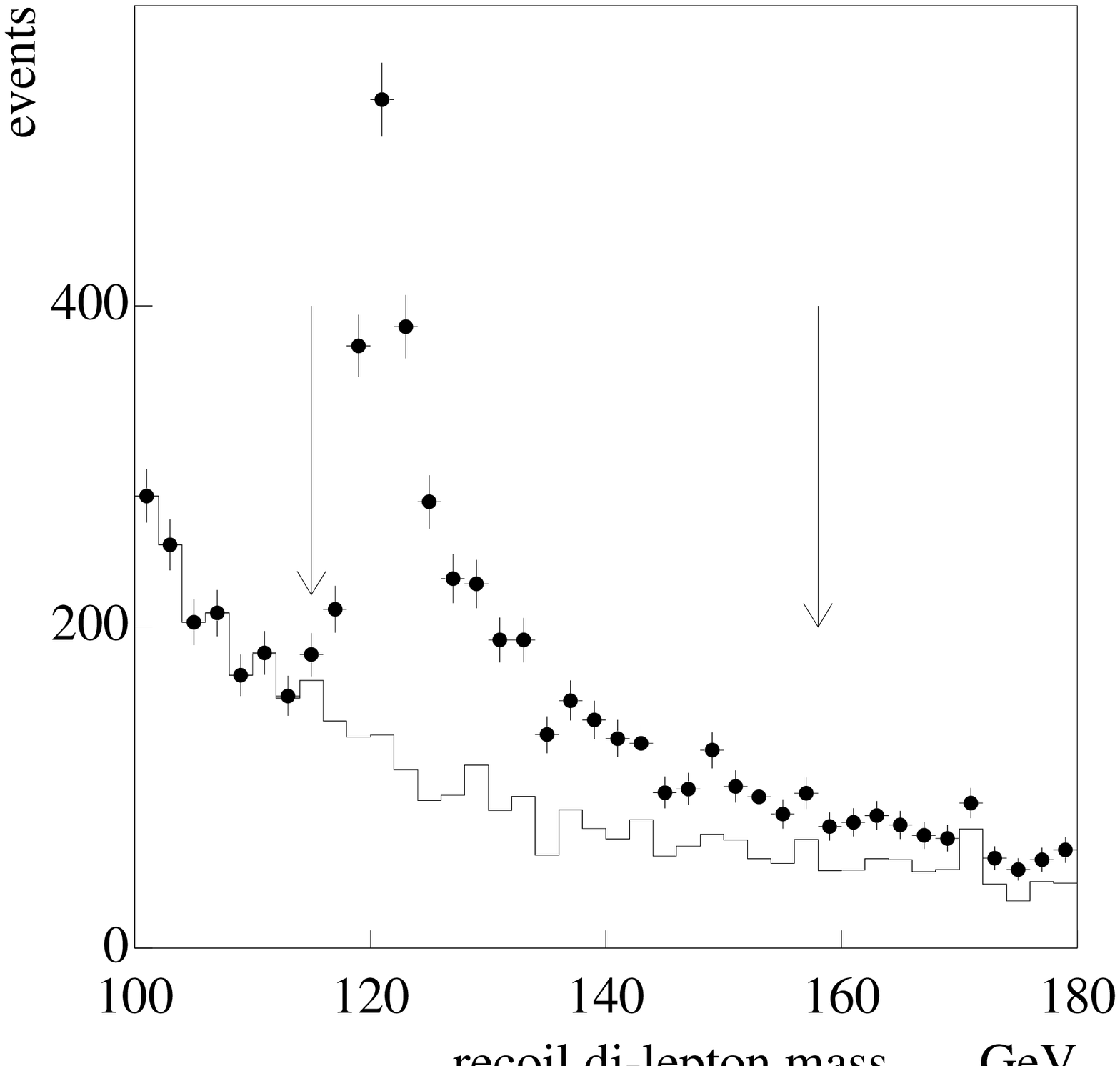}
\includegraphics[scale=0.5]{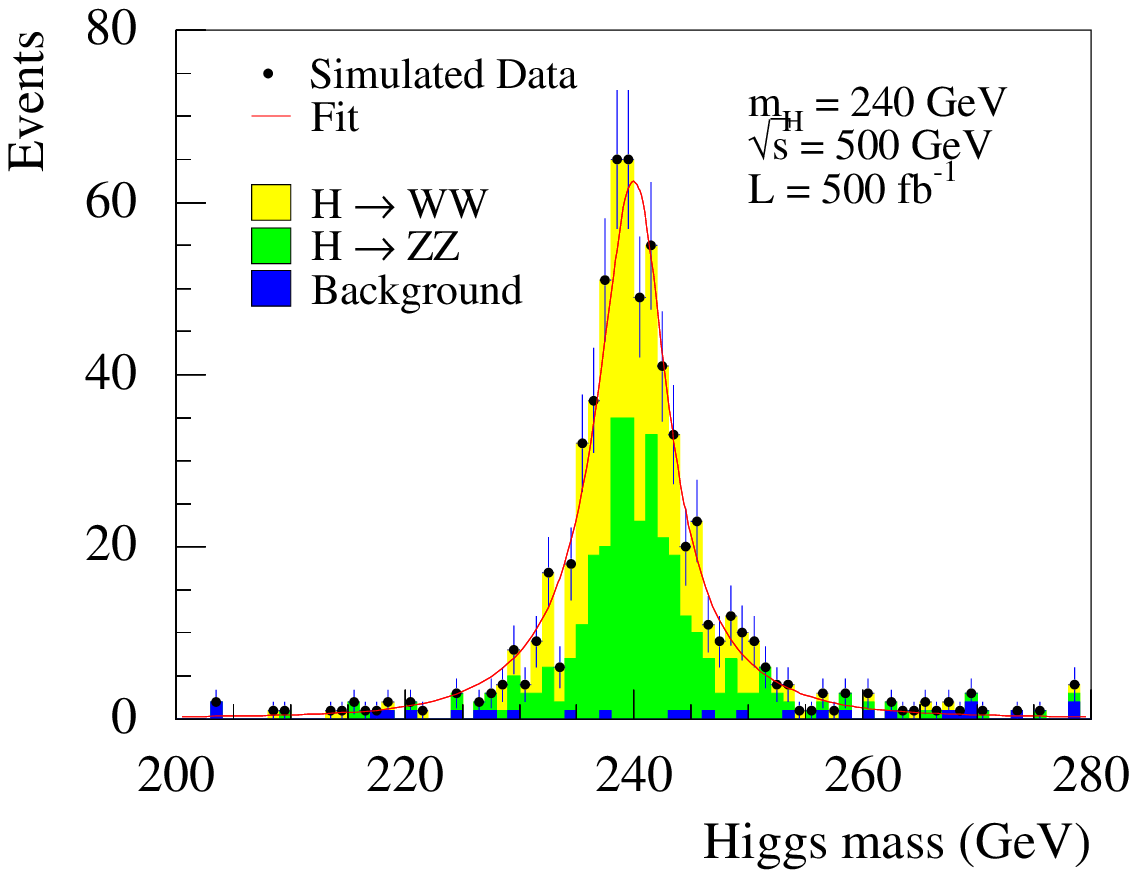}
\includegraphics[scale=0.28]{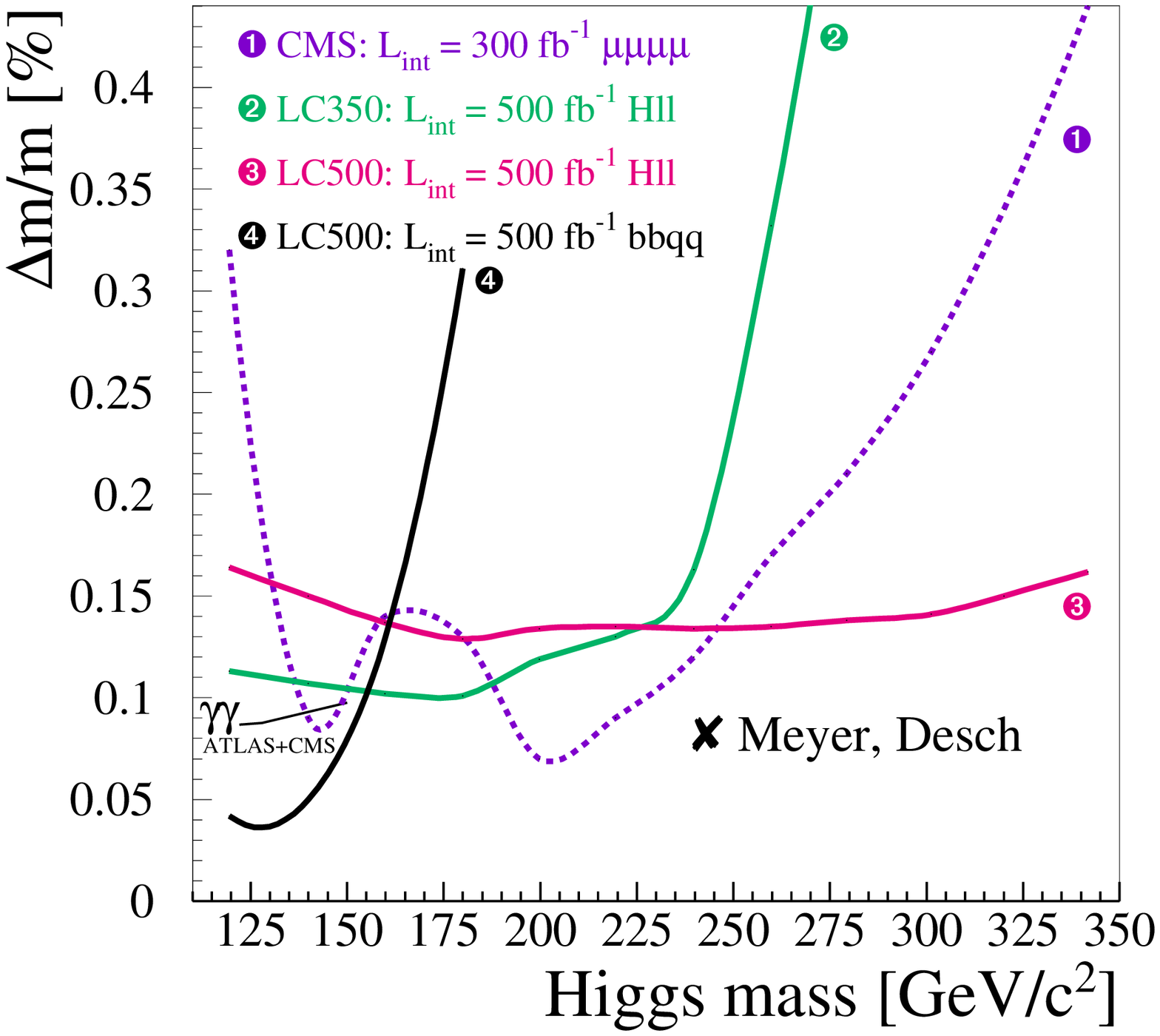}
\vspace*{1.7cm}
\caption{
Left: simulated events for a direct determination of the 
      Higgs boson decay branching ratios.
Center: mass and width determination.
Right: comparison of different methods to determine the Higgs boson mass.
}
\label{fig:br}
\vspace*{-0.1cm}
\end{figure}

The Higgs boson branching ratios can also be determined directly 
from the HZ($\rm Z\rightarrow \ell^+\ell^-$) 
event sample. The Higgs boson mass is reconstructed 
as the recoiling mass of the lepton pair: 
$$m_{\rm H}=m_{\ell^+\ell^-}^{\rm recoil}.$$
The simulation was performed for $\sqrt s=360$~GeV and 
${\cal L}=500$~fb$^{-1}$.
Figure~\ref{fig:br} (left plot) shows this event sample which is enriched with Higgs
bosons by the indicated cuts.
The expected signal distribution 
is given with
error bars and the expected background as a histogram.
In this event sample individual Higgs boson decay modes are selected. 
The resulting
precision on the Higgs boson decay branching ratios 
as well as a preliminary combination of both methods are 
listed in Table~\ref{table:br}. For the c-quark decay mode, recently
a larger uncertainty of 12.1\% was reported~\cite{kuhl}. 
The variation of the branching ratio precision as a function of 
the Higgs boson mass is shown as well in Table~\ref{table:br}~\cite{barklow}.

\begin{table}
\caption{Left: expected precision on branching ratios (in \%) for a 
         120 GeV Higgs boson from the direct method (d) and a 
         preliminary combination (c) with the indirect method.
         Right: variation of the $BR({\rm H\to b\bar b})$ precision (in \%) 
         as a function of the Higgs boson mass.}
\label{table:br}
\renewcommand{\arraystretch}{1.2} 
\begin{center}
\begin{tabular}{cccc}
Decay    &SM $BR$  &$\Delta BR_{\rm d} /BR_{\rm d}$&$\Delta BR_{\rm c} /BR_{\rm c}$ \\\hline
bb       & 68       &1.9 & 1.5  \\
$\tau\tau$ & 6.9    &7.1 & 4.1  \\
cc       & 3.1      &8.1 & 5.8  \\
gluons   & 7.0      &4.8 & 3.6  \\
$\gamma\gamma$&0.22 &35  & 21   \\ 
WW$^\star$ & 13     &3.6 & 2.7
\vspace*{-0.8cm}
\end{tabular} \hfill
\begin{tabular}{ccccc}
$m_{\rm H}$ (GeV)              & 120  &  140  &   160  & 160 \\\hline
$\Delta BR_{\rm b} /BR_{\rm b}$& 1.6  &  1.8  &   2.0  & 9.0 \\
\vspace*{-0.8cm}
\end{tabular}
\end{center}
\end{table}

In relation to the LHC, a LC will achieve a much higher precision
and will cover all decay modes. 
This would also allow precision testing of the fundamental
relation between the Yukawa coupling and the Higgs boson mass:
\vspace*{-1mm}
$$g_{\rm Hff}\propto m_{\rm f}.
\vspace*{-3mm}
$$

\subsection{Mass and width determination}

Further properties of the Higgs boson, like its mass and decay 
width, can be determined with high precision.
A study of a 240~GeV SM Higgs boson in the reactions
$$\rm e^+e^-\rightarrow HZ\rightarrow WWZ~~~and$$
$$\rm e^+e^-\rightarrow HZ\rightarrow ZZZ$$
has been performed. A very clear signal with only
a few background events is expected. 
Figure~\ref{fig:br} (center plot) shows the expected signal and background mass
distribution.
The fit (center plot) of the reconstructed
mass peak leads to a mass resolution of 0.08\% and an 11\% error
on the total decay width.
Figure~\ref{fig:br} (right plot) shows results from a CMS study (curve 1)
where the Higgs boson decays into a pair of Z bosons, which subsequently
decay into muon pairs. The sensitivity reduction near 160 GeV is due
to dominant decays into W pairs.
The curves 2 and 3 show extrapolated results according to the cross 
section and branching ratio expectations for a LC operation at 350 and 
500~GeV.
In the low mass region indirect methods can be applied (curve 4)
and for very low masses, the LHC has high sensitivity in the photon decay mode.

High precision $\Delta E/E_{\rm beam} < 10^{-4}$
is needed to achieve a mass precision of 
$m_{\rm H}=120.00\pm0.04$~GeV~\cite{raspereza}.
It has been pointed out~\cite{heinemeyer} that the expected experimental 
precision $\Delta m^{\rm exp}_{\rm h} \approx 0.05$~GeV is difficult to 
match in the theoretical prediction 
$\Delta m^{\rm th}_{\rm current} \approx 3$~GeV and 
$\Delta m^{\rm th}_{\rm future} \approx 0.5$~GeV.

\begin{figure}[htb]
\vspace*{-0.1cm}
\includegraphics[scale=0.6]{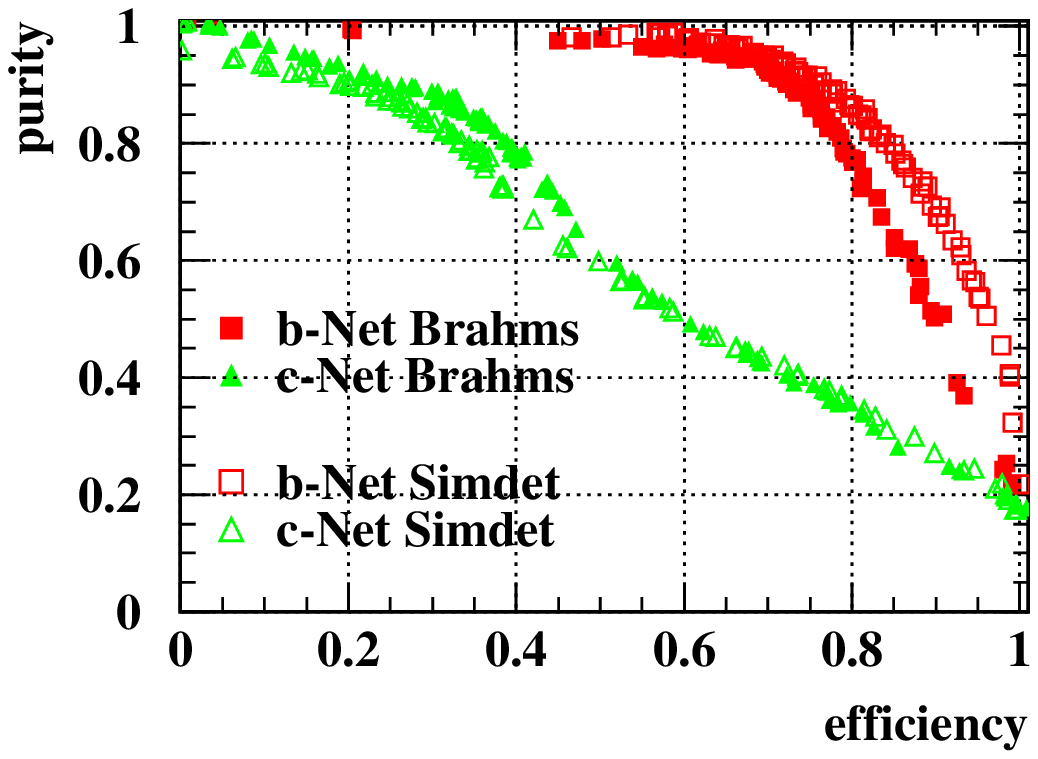}\hfill
\includegraphics[scale=0.2]{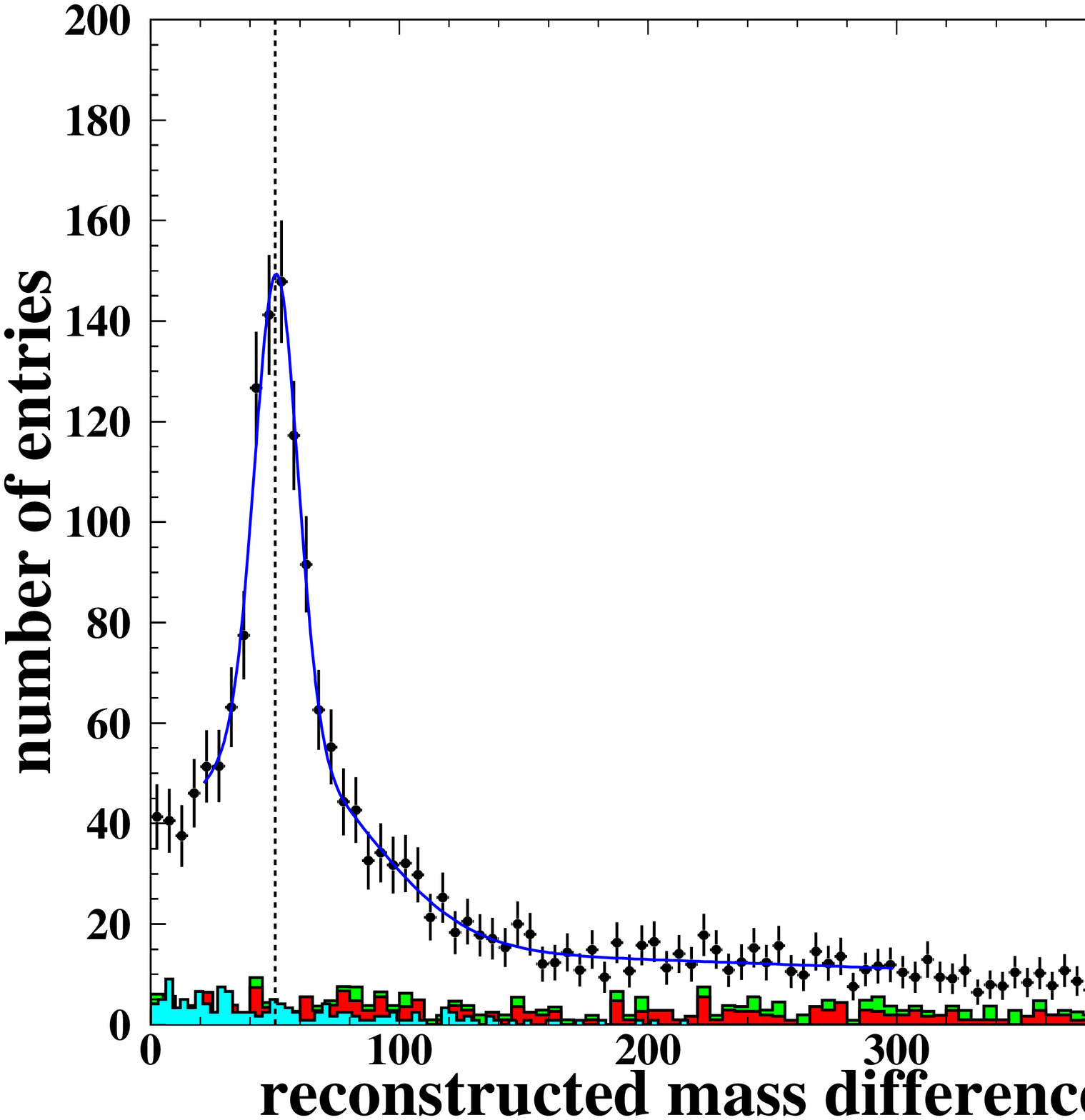}
\vspace*{-0.3cm}
\caption{
Left: c- and b-tagging efficiency versus purity for the LCFI vertex detector.
Right: reconstructed mass difference for the reaction
       $\rm HA\to b\bar b b\bar b$ at $\sqrt{s}=800$~GeV with Higgs boson masses
       of 250 and 300~GeV.
}
\label{fig:btag}
\end{figure}

The tagging of b- and c-quarks are crucial for the Higgs boson studies
as the Higgs boson decays predominantly into heavy quarks. 
Figure~\ref{fig:btag} (left plot)
shows the efficiency versus purity for a CCD 
vertex detector simulation~\cite{damerell}.
With this vertex detector the process $\rm HA\to b\bar b b\bar b$ at $\sqrt{s}=800$~GeV
has been studied and Fig.~\ref{fig:btag} (right plot) shows that a precision of
$\Delta m_{\rm H}\approx \Delta m_{\rm A}\approx 0.45$~GeV can be achieved~\cite{desch}.

\subsection{Characterization of the Higgs boson potential}

\begin{figure}[htb]
\vspace*{-0.1cm}
\begin{minipage}{0.32\textwidth}
\begin{Feynman}{60,33}{1,27}{0.8}
\put(25,40){\fermionul}        \put(5,22){${\rm e^-}$}
\put(25,40){\fermionur}        \put(5,55){${\rm e^+}$}
\put(15,30){\photonrightthalf} \put(32,30){$\gamma$}
\put(25,40){\fermionrighthalf}  \put(30,41){${\rm Z^*}$}
\put(40,40){\gaugebosonurhalf} \put(57,55){${\rm H}$}
\put(40,40){\gaugebosondrhalf} \put(57,22){${\rm Z}$}
\put(50,50){\gaugebosondrhalff} \put(57,40){${\rm H} $}
                                \put(40,45){${\rm H} $}
\end{Feynman}
\end{minipage}
\begin{minipage}{0.67\textwidth}
\includegraphics[scale=0.26]{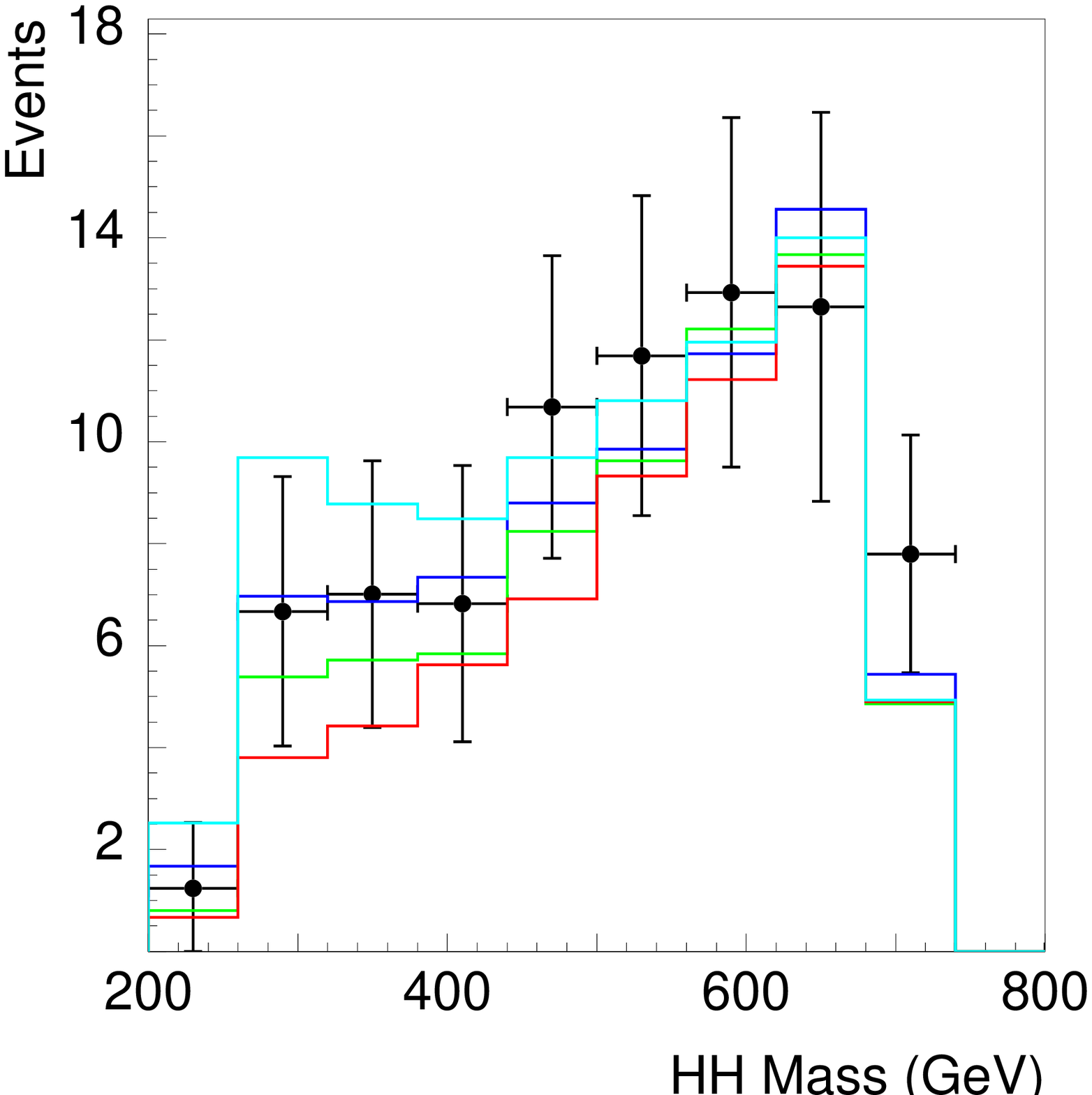}\hfill
\includegraphics[scale=0.26]{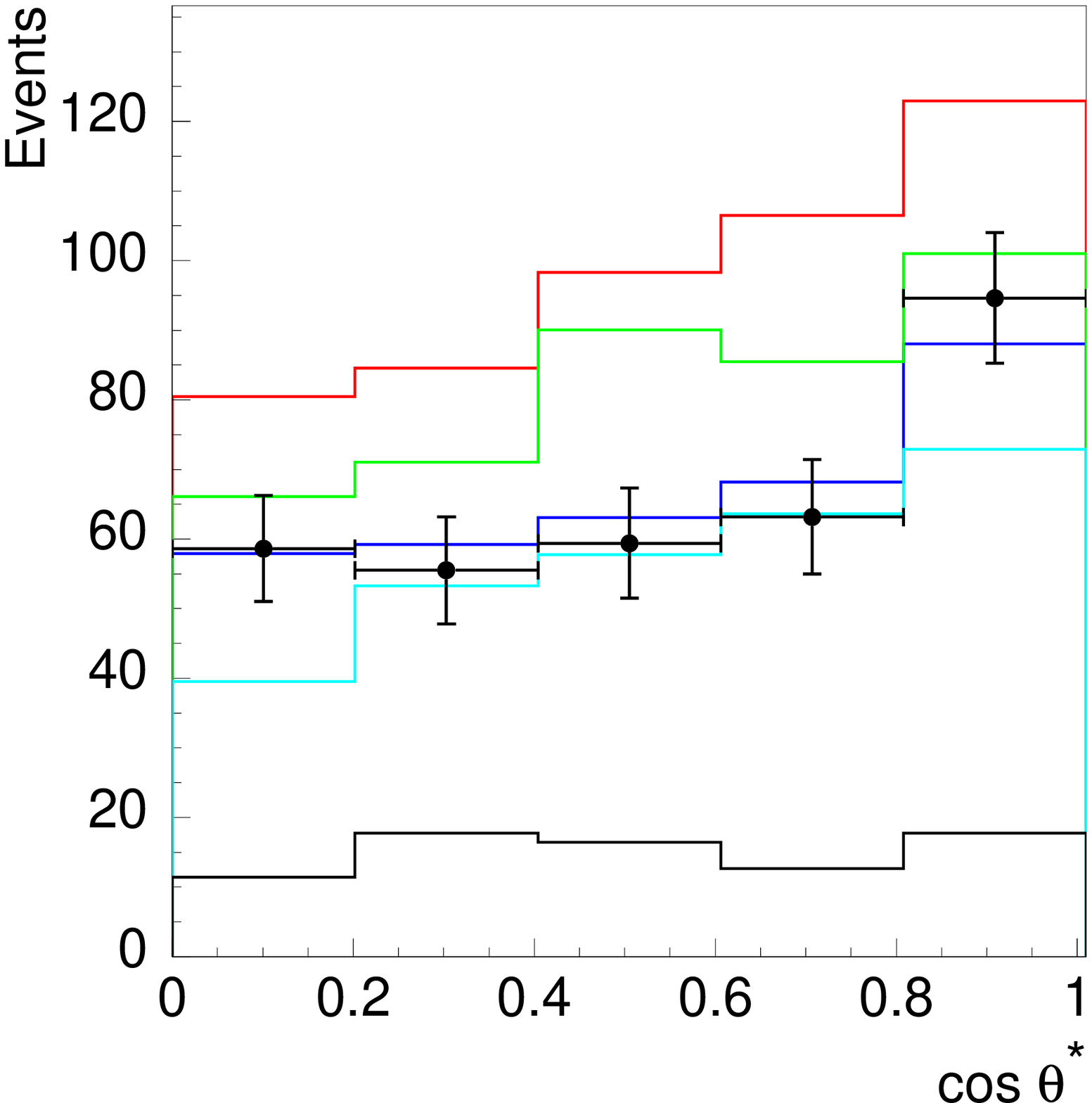}
\end{minipage}
\caption{
Left: Higgs boson self-coupling reaction.
Center: reconstructed invariant mass of the HH pairs.
Right: reconstructed angle between the H and HH directions.
}
\label{fig:hhh}
\end{figure}

A LC will be able to measure fundamental properties of the Higgs boson
potential. The Higgs boson can decay into a pair of Higgs bosons:
$$\rm e^+e^-\rightarrow HZ\rightarrow HHZ.$$
Figure~\ref{fig:hhh} illustrates this Higgs boson production reaction.

The self-coupling interaction can probe the shape of the Higgs boson
potential through the relation $$g_{\rm HHH} = 3m_{\rm H}^2/2v,$$
where $v=246$~GeV.
A sensitivity of $\Delta g/g=29$\% was obtained for
$m_{\rm H} =120$~GeV, $\sqrt s=800$~GeV and 
${\cal L}=1000$~fb$^{-1}$.
Figure~\ref{fig:hhh} shows the precision on 
the Higgs boson self-coupling where the lines indicate 
$g_{\rm HHH}/g_{\rm HHH}^{\rm SM} =1.25,~1.00,~0.75,~0.50$.
Higher sensitivity of $\Delta g/g=7$\% could be reached for 
a LC with $\sqrt s=3$~TeV and ${\cal L}=5000$~fb$^{-1}$ as 
shown in Fig.~\ref{fig:hhh}.

\subsection{Higgs-strahlung from top quarks}

Owing to the strong coupling of Higgs bosons to top
quarks, the radiation of Higgs boson off top quarks is
a possible production mechanism.
The decay modes involving a pair of b quarks and W bosons 
were studied in the reactions 
$$\rm e^+e^-\rightarrow t\bar tH\rightarrow t\bar t b\bar b~~~and$$
$$\rm e^+e^-\rightarrow t\bar tH\rightarrow t\bar t WW.$$
A challenge is the precision determination of the background
to a level of 5\% uncertainty, leading to
$$\Delta g_{\rm ttH}/g_{\rm ttH}=7.5\%~{\rm to}~20\%.$$
Figure~\ref{fig:hphm} (left plot) shows the resulting precision for the $\rm b\bar b$
and WW decay modes, as well as their statistical combination
as a function of the SM Higgs boson mass.
A similar sensitivity was recently confirmed~\cite{yamashita}.
In addition, a previous study at 120~GeV with slightly higher sensitivity 
is indicated. The reaction $\rm H\rightarrow ZZ$ is currently under study~\cite{gay}.

\section{Beyond the Standard Model}

\subsection{Charged Higgs bosons}

\begin{figure}[htb]
\vspace*{-0.3cm}
\includegraphics[scale=0.22]{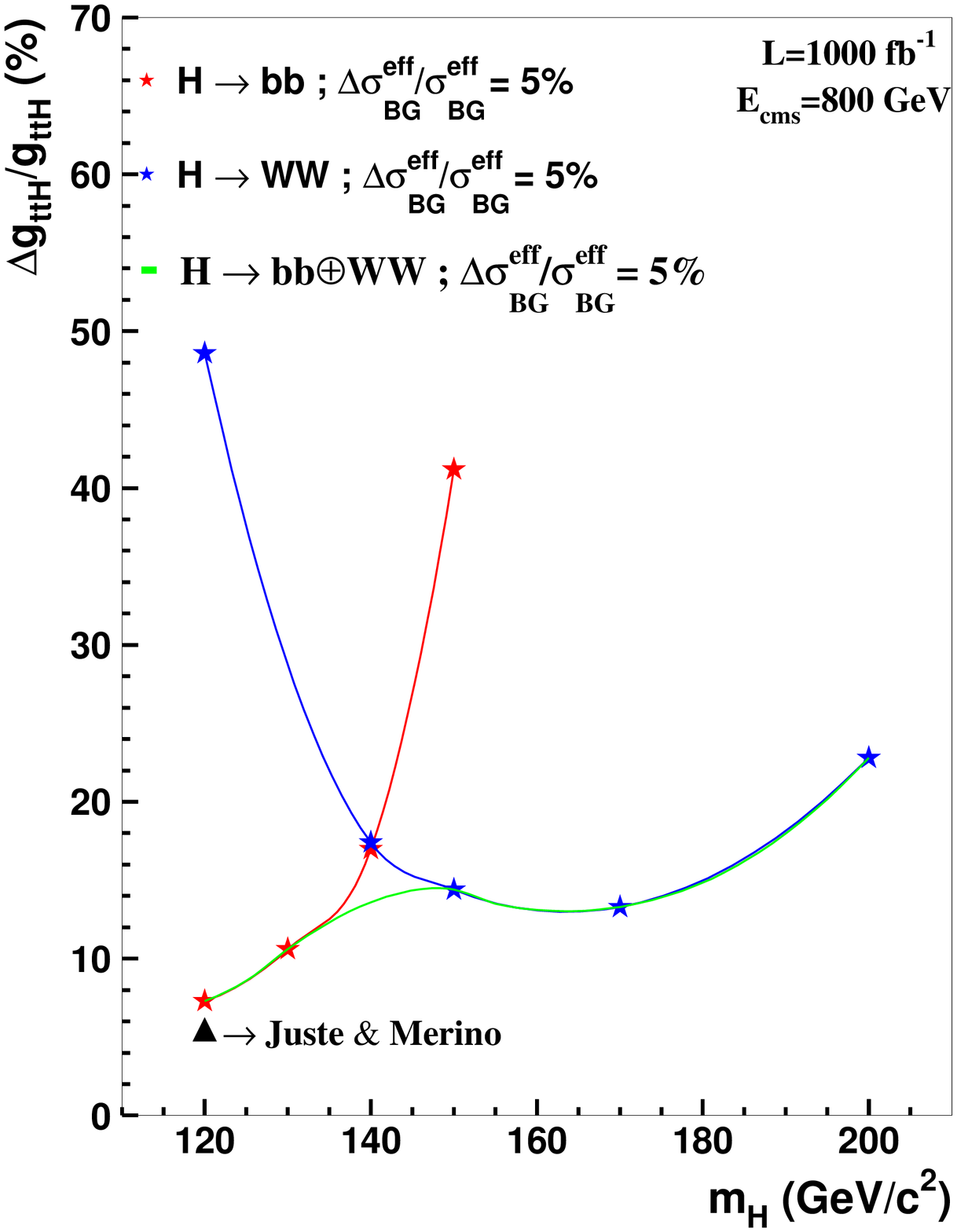}\hfill
\includegraphics[scale=0.55]{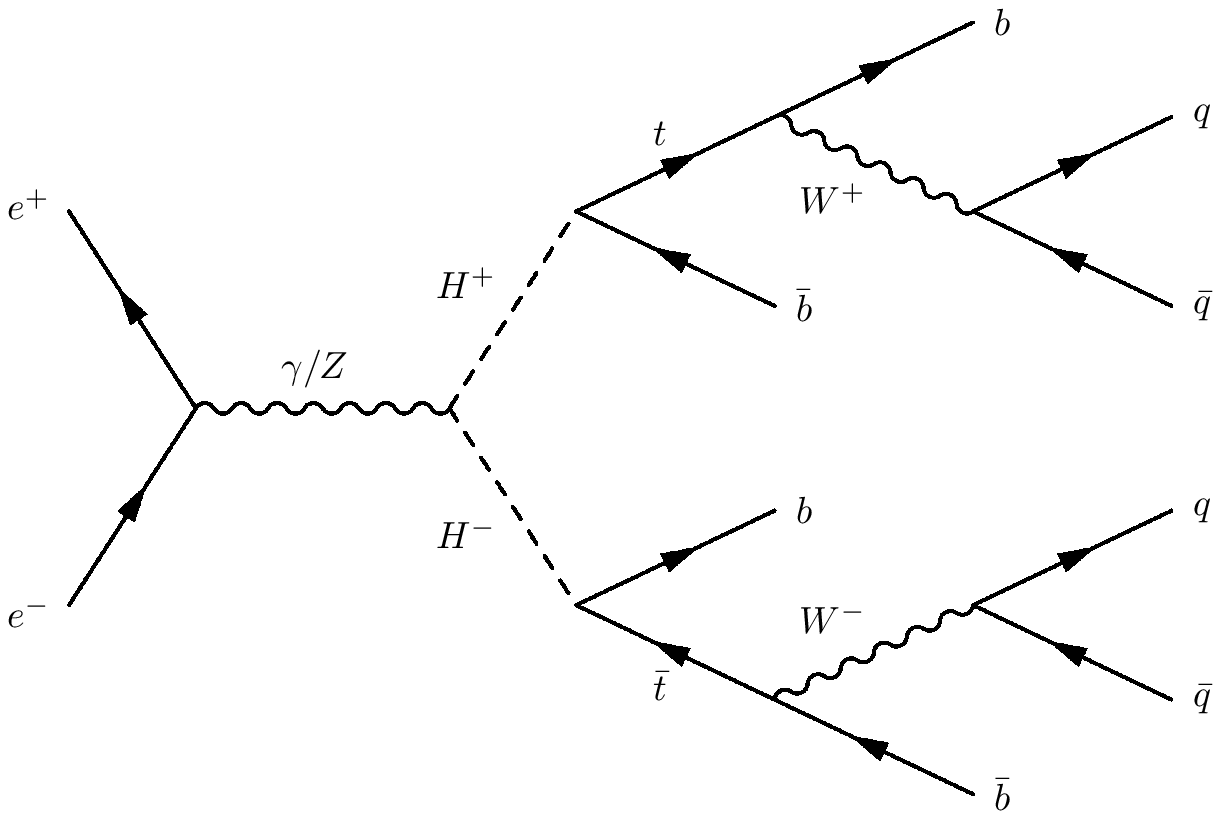}\hfill
\includegraphics[scale=0.26]{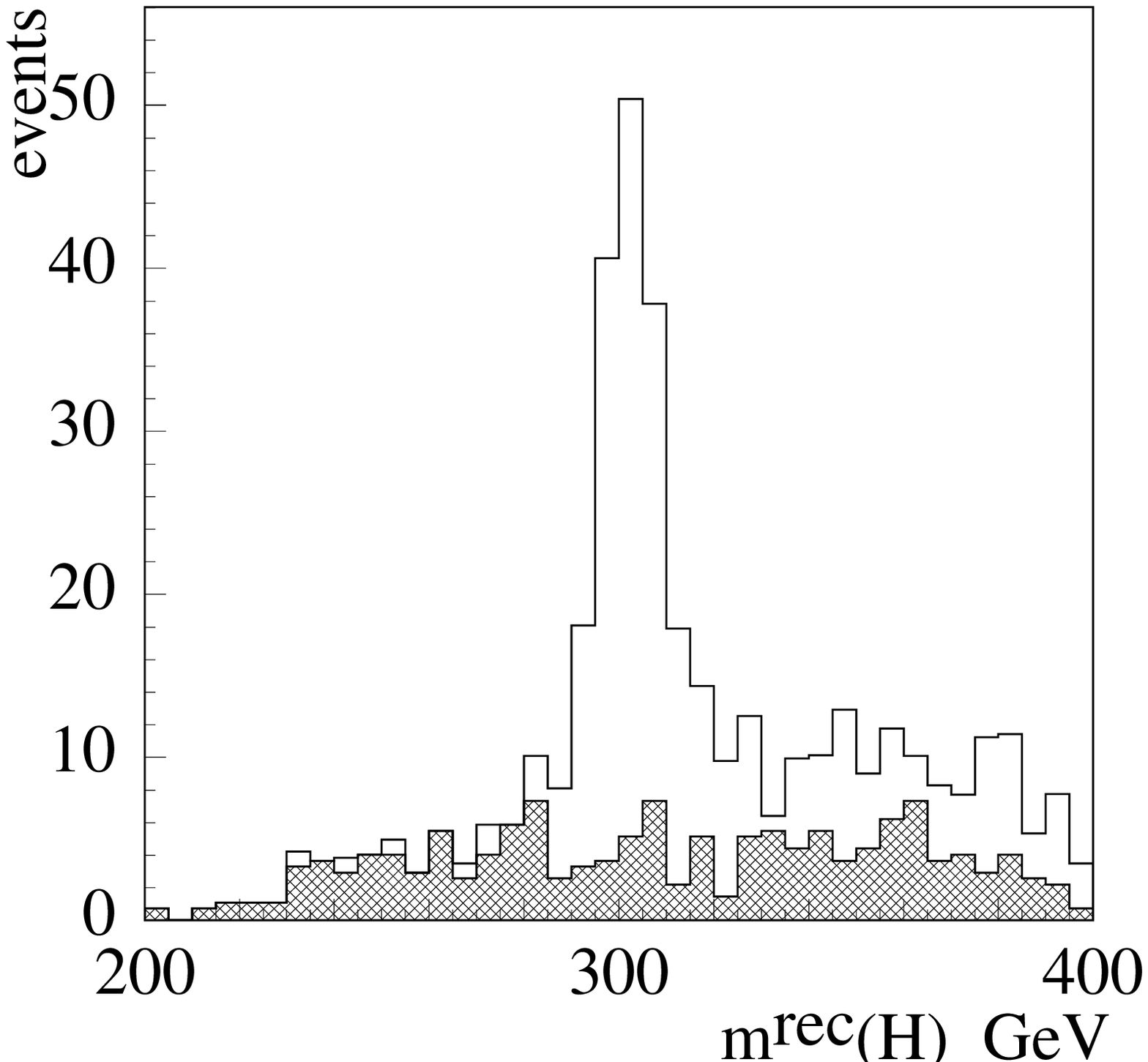}
\vspace*{-0.3cm}
\caption{
Left: expected precision on the Yukawa coupling $g_{\rm ttH}$.
Center: charged Higgs boson production and decay.
Right: expected charged Higgs boson signal and background.
}
\label{fig:hphm}
\vspace*{-0.2cm}
\end{figure}

The discovery of charged Higgs bosons would immediately prove that
physics beyond the SM exists. 
The reaction 
$$\rm e^+e^-\rightarrow Z \rightarrow H^+H^-\rightarrow t\bar b\bar t b$$
can be observed at a LC and recent high-luminosity 
simulations show that the production cross section times
branching ratio can be measured very precisely:
$$\Delta(\sigma BR({\rm H^+\rightarrow t\bar b}))/
        \sigma BR({\rm H^+\rightarrow t\bar b})=8.8\%$$
for $m_{\rm H^\pm}=300$~GeV, $\sqrt s = 800 $~GeV and 
${\cal L}=1000$~fb$^{-1}$.
A detailed reconstruction of the entire decay chain, as
illustrated in Fig.~\ref{fig:hphm} (center plot), is possible.
Figure~\ref{fig:hphm} (right plot) shows also a clear expected 
charged Higgs boson signal and small background.

\subsection{Determination of \boldmath$\tan\beta$\unboldmath}

\begin{figure}[htb]
\includegraphics[scale=0.21]{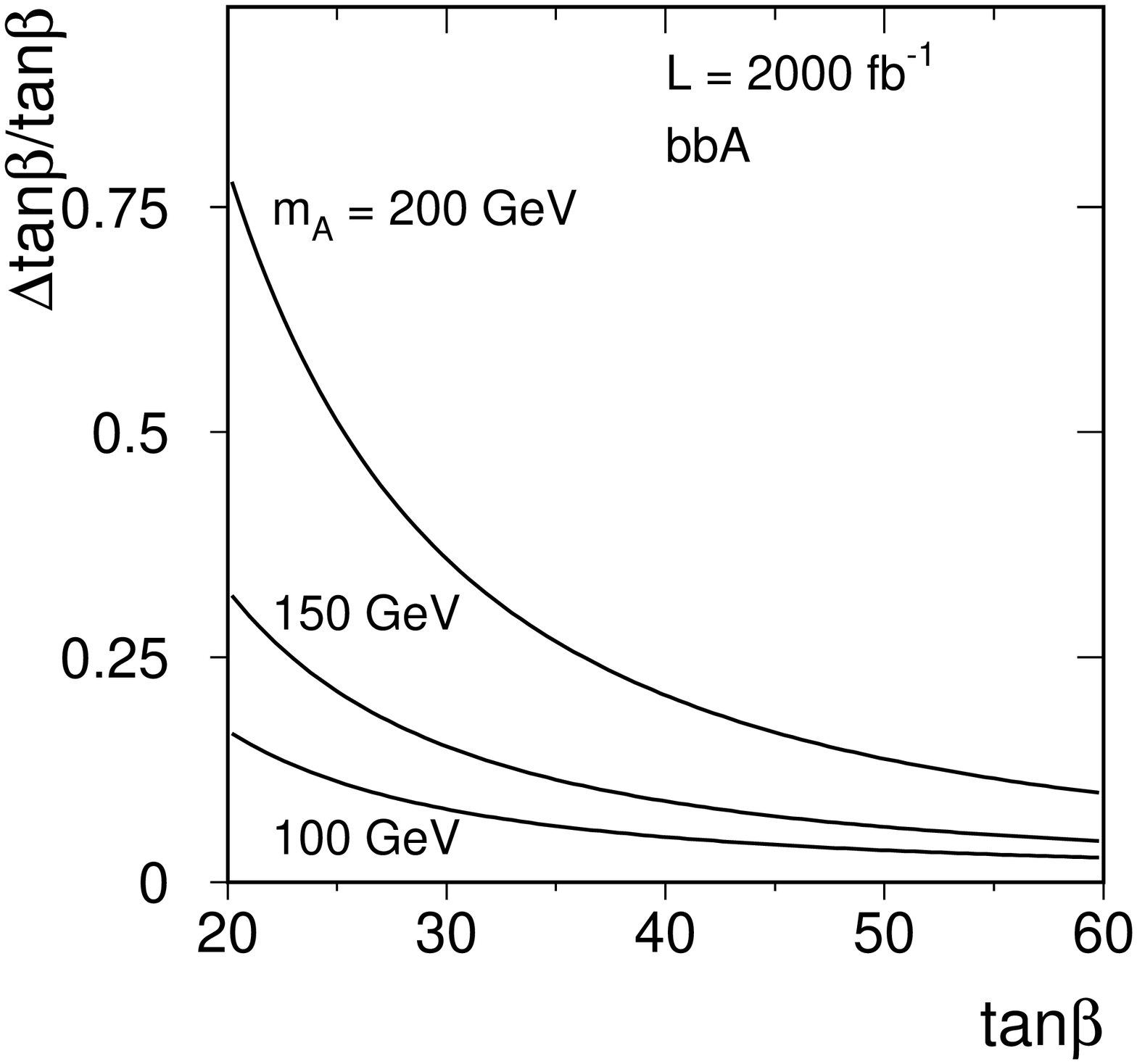} \hfill
\includegraphics[scale=0.26,angle=90]{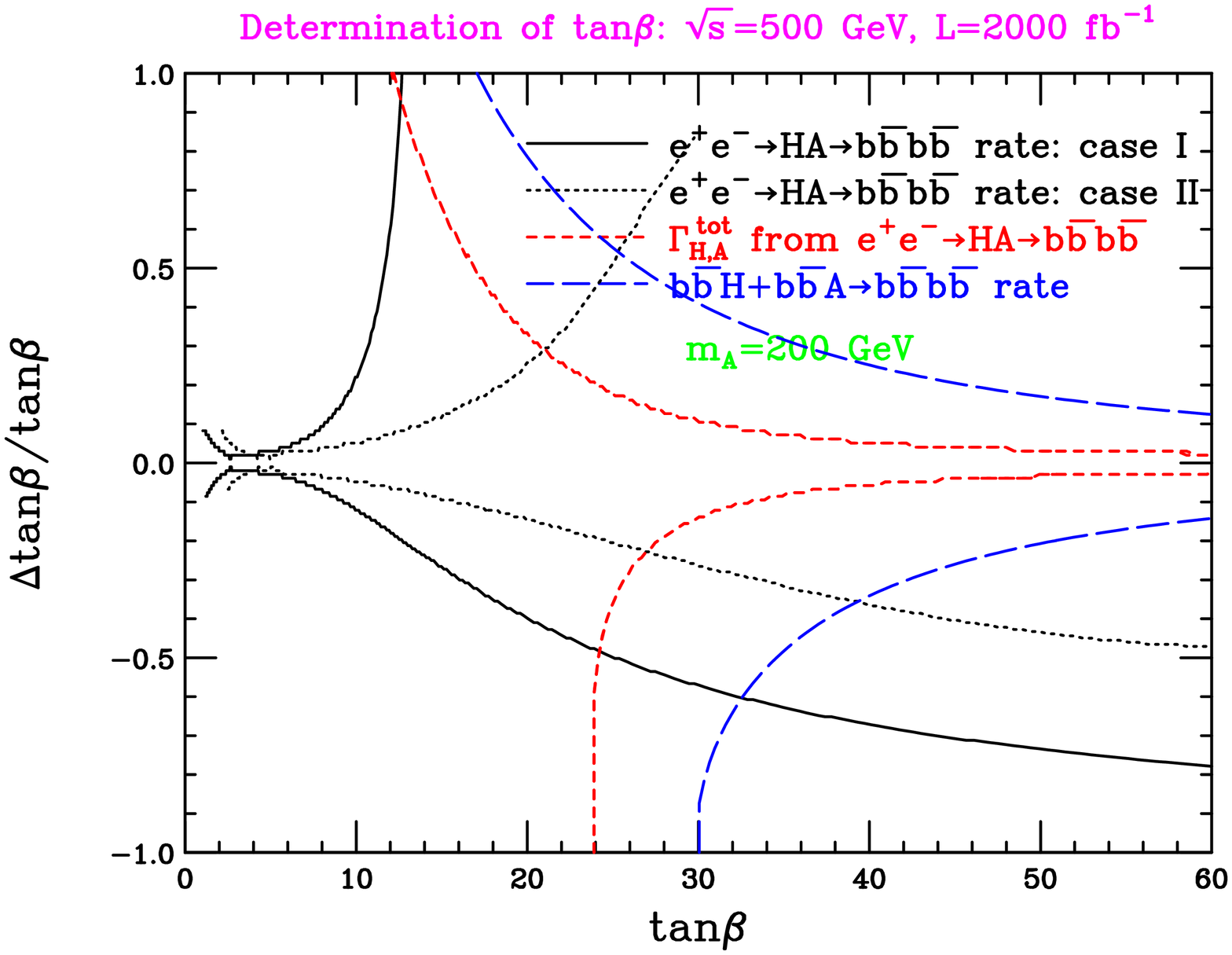}\hfill
\includegraphics[scale=0.31]{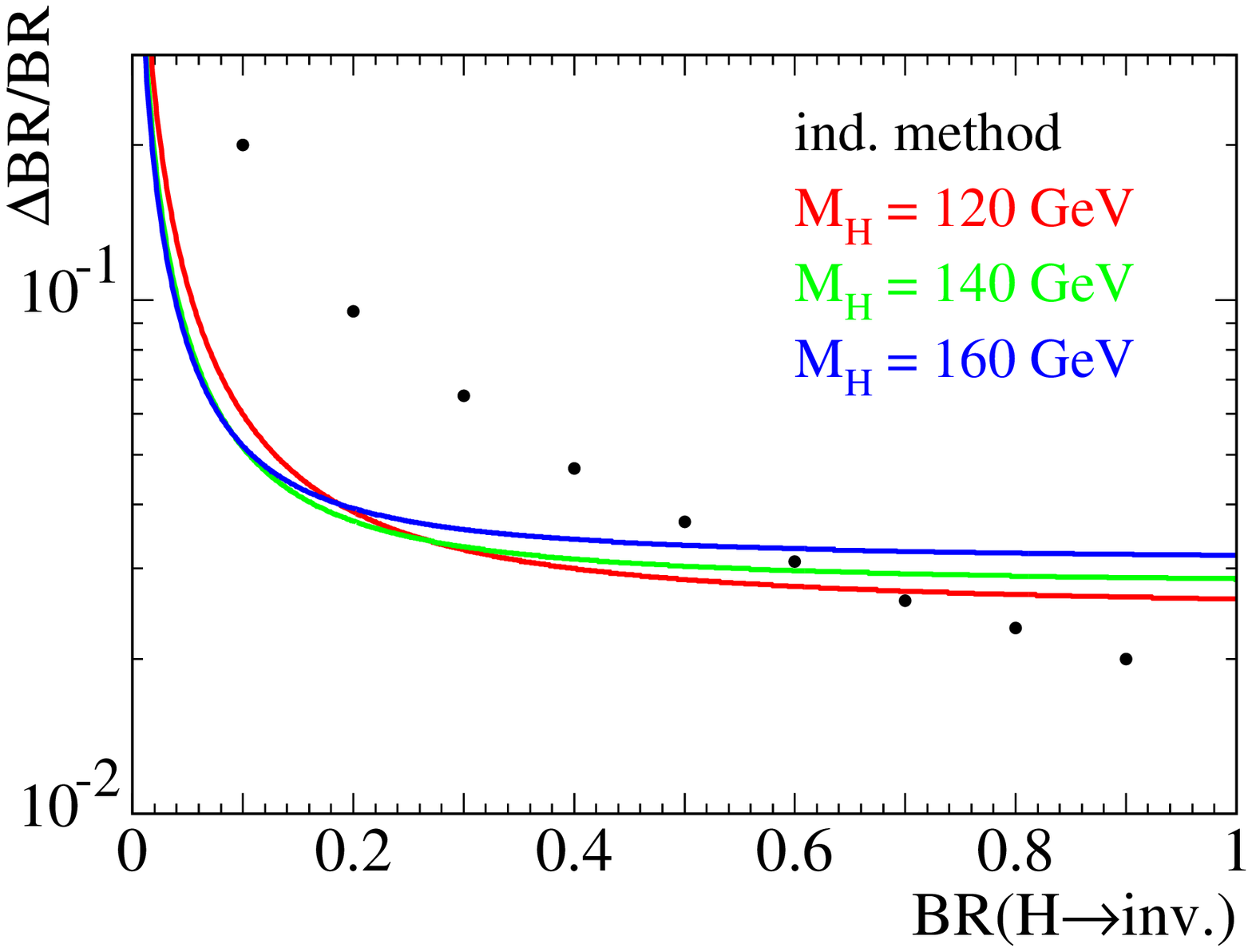}
\vspace*{-0.4cm}
\caption{
Left: 2HDM $\tan\beta$ determination.
Center: MSSM $\tan\beta$ determination.
Right: expected precision on the invisible Higgs boson 
       branching fraction.
       The dots represent the indirect method.
}
\label{fig:bba}
\end{figure}

The ratio of the vacuum expectation values
$\tan\beta$ can be measured with several methods.
The pseudoscalar Higgs boson, A, could be produced
via radiation off a pair of b-quarks:
$$\rm e^+e^-\rightarrow b\bar b\rightarrow b\bar b A \rightarrow b\bar b b\bar b.$$
The $\rm b\bar bA$ coupling is proportional to $\tan\beta$ and thus 
the expected
production rate is proportional to $\tan^2\beta$.
A precision better than 10\% can be achieved for a
Higgs boson mass of 100~GeV and large $\tan\beta$ values. 
The sensitivity decreases with increasing Higgs boson masses and decreasing 
$\tan\beta$ values as shown in Fig.~\ref{fig:bba} (left plot). 
This study assumes a luminosity of 2000~fb$^{-1}$, corresponding 
to several years of data taking.
There are further methods to determine $\tan\beta$:
\begin{itemize}
\item The $\rm b\bar b\bar b b$ rate 
      from the pair-production of the heavier scalar in association 
      with the pseudoscalar Higgs boson,
      $$\rm e^+e^- \rightarrow HA \rightarrow b\bar b b\bar b,$$
      can be exploited.
      While the HA production rate is almost independent of $\tan\beta$ 
      the sensitivity is achieved owing to the variation of the 
      decay branching ratios with $\tan\beta$. 
\item The value of $\tan\beta$ can also be determined from the 
      H and A decay widths, which can be obtained from the previously
      described reaction.
\item The pair-production 
      rate and total decay width of charged Higgs bosons
      can contribute to the determination of $\tan\beta$.
      Charged Higgs boson production can also be used at the LHC 
      to measure $\tan\beta$.
\end{itemize}

The results from the methods involving the neutral
Higgs bosons are summarized in Fig.~\ref{fig:bba} (center plot)
for the MSSM. As the H and A decay rates depend on the MSSM parameters, 
two cases are considered. In case~(I) heavy Supersymmetric particles 
are expected, while in case~(II) the Higgs bosons could decay into light 
Supersymmetric particles.

\subsection{Invisible Higgs boson decays}

Among other possibilities, invisible Higgs boson decays could 
occur from the reaction involving neutralinos:
$$\rm e^+e^-\rightarrow ZH\rightarrow Z\tilde\chi^0\tilde\chi^0.$$ 
In this case the Higgs boson mass can be reconstructed from the recoiling
mass of the visible decay products:
$$m_{\rm H}=m_{\rm Z}^{\rm recoil}.$$
At LEP all Z decay modes involving charged fermions contributed to the
search,
while in a recent LC study so far only the hadronic 
decay mode $\rm Z\rightarrow q\bar q$ has been investigated.
This study for $\sqrt s = 350 $~GeV and ${\cal L}=500$~fb$^{-1}$
gives higher sensitivity compared to indirect methods
($1-$ sum of visible H decay modes).
For Higgs boson branching ratios $BR_{\rm i}$ into invisible decay 
products larger than 20\% and a SM Higgs boson production rate,
$$\Delta BR_{\rm i} /BR_{\rm i} <4\%$$ can be achieved.
Figure~\ref{fig:bba} (right plot) 
shows the resulting sensitivities as a function
of the branching ratio into invisible decays.

\subsection{Higgs boson parity}

After the discovery of one or several Higgs bosons,
it would be very important to determine the parity of the Higgs bosons and
distinguish a CP-even H boson from a CP-odd A boson.
This could be achieved by investigating the Higgs boson decay 
properties into $\tau$-leptons. 
The subsequent decay of the $\tau$'s into $\rho$'s and pions 
through the reaction
$$\rm H/A\rightarrow \tau^+\tau^-
\rightarrow\rho^+\bar\nu_\tau\rho^-\nu_\tau
\rightarrow \pi^+\pi^0\bar\nu_\tau\pi^-\pi^0\nu_\tau$$
 is studied.

\begin{figure}[htb]
\includegraphics[scale=0.6]{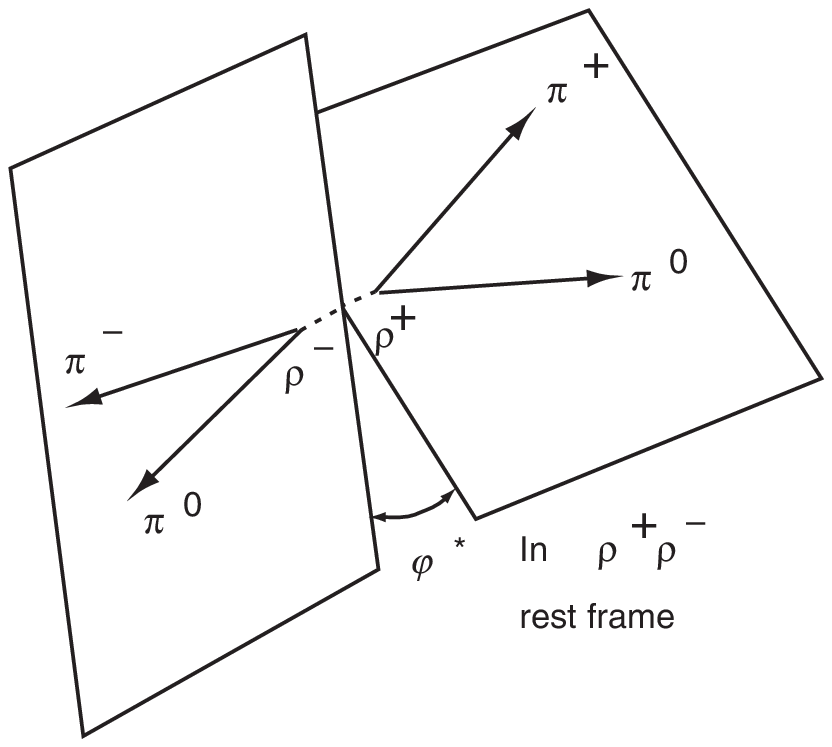}
\includegraphics[scale=0.35]{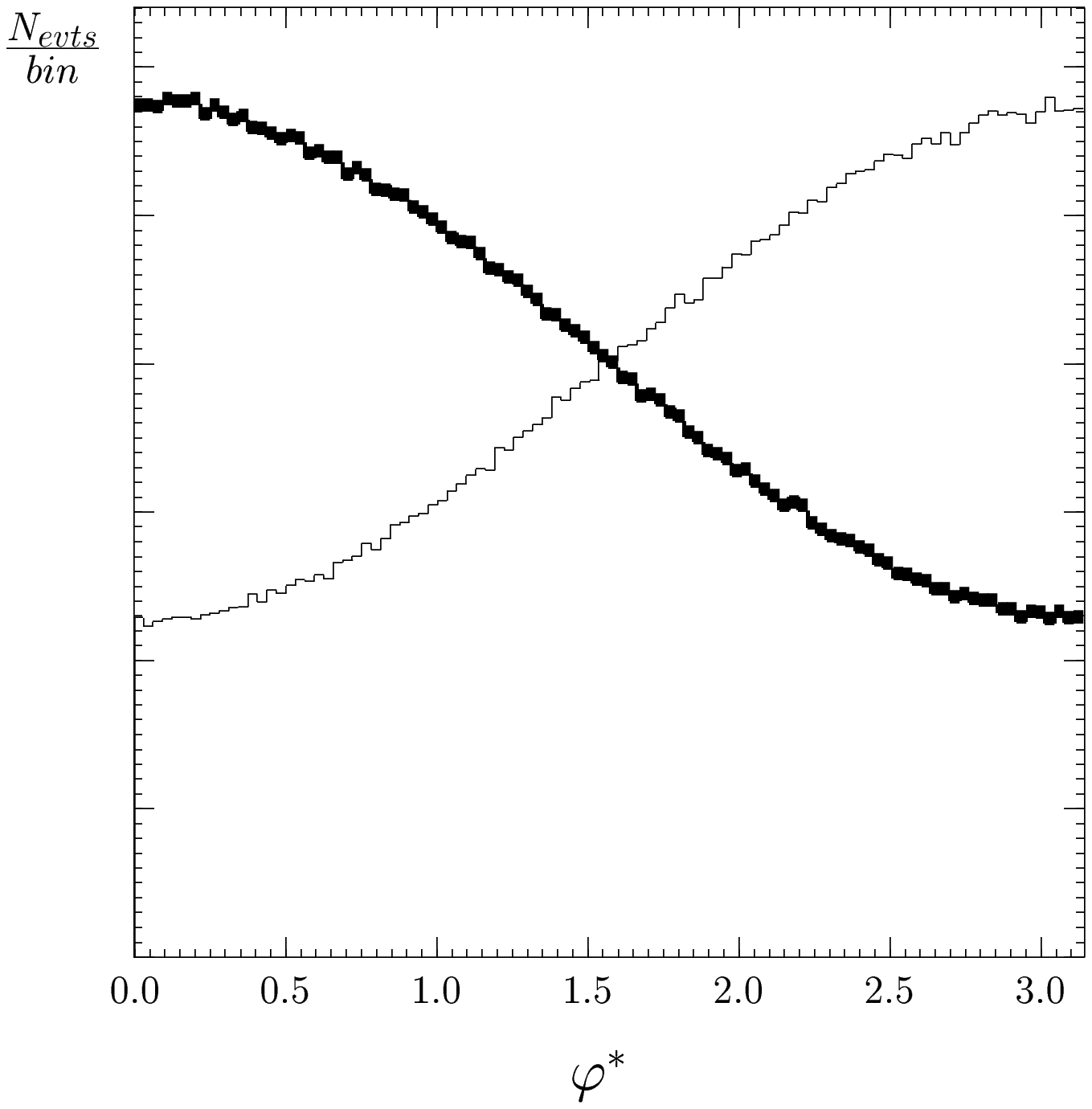}
\includegraphics[scale=0.35]{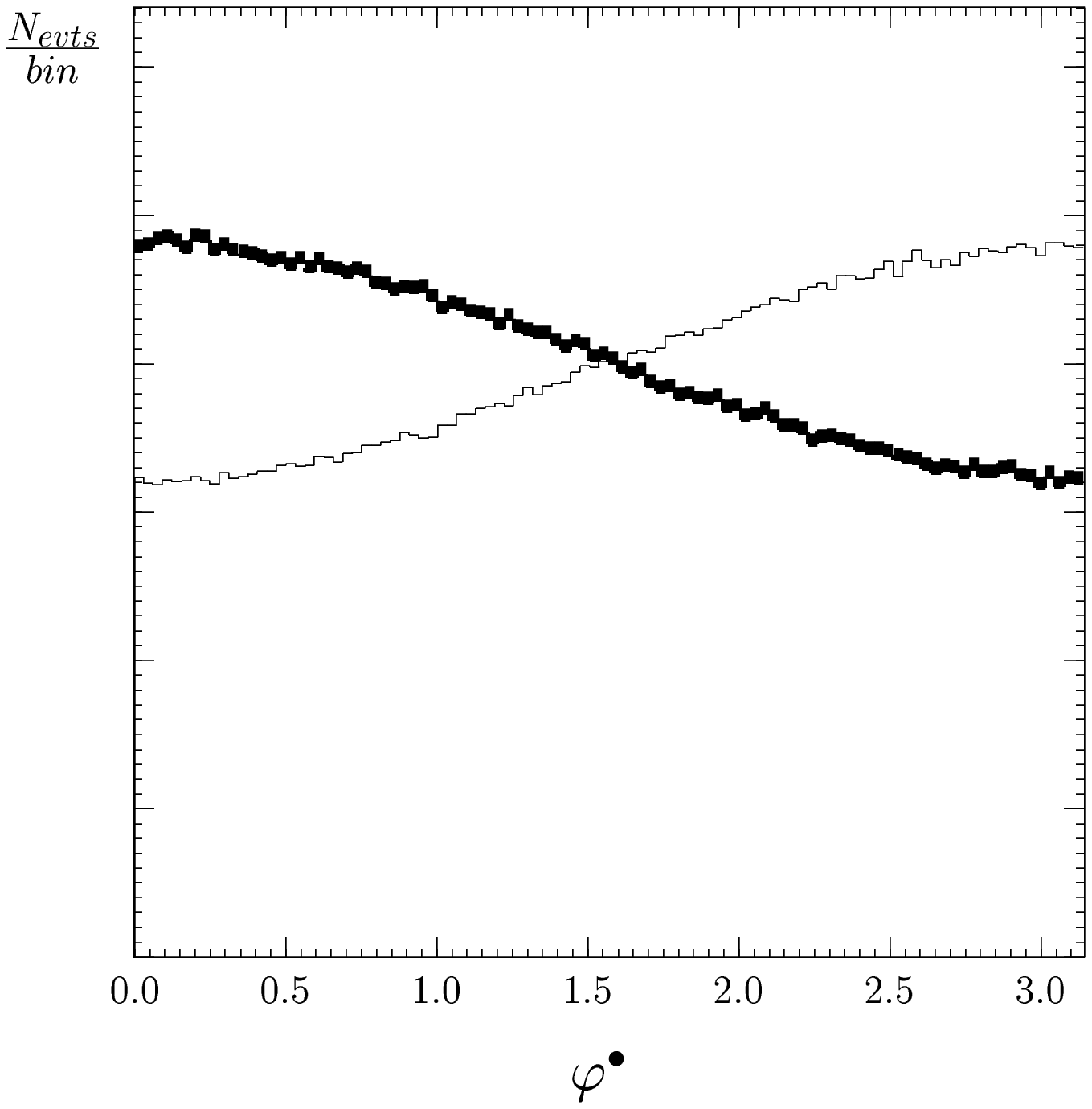}
\vspace*{-0.3cm}
\caption{
Left: acoplanarity angle definition.
Center: Higgs boson parity determination before detector simulation.
Right: Higgs boson parity determination after a preliminary detector simulation.
}
\vspace*{-0.2cm}
\label{fig:parity}
\end{figure}

The $\rho^+\rho^-$ acoplanarity angle is defined by the planes
of the pions in the rest frame of the $\rho$'s as illustrated
in Fig.~\ref{fig:parity} (left plot). 
The acoplanarity angle distribution clearly distinguishes the different parity
states. Figure~\ref{fig:parity} (center plot) shows the acoplanarity angle 
before detector effects are included, and Fig.~\ref{fig:parity} (right plot)
after a preliminary detector simulation.
The thick line is the expectation for scalar Higgs bosons and the 
thin line for pseudoscalar Higgs bosons.

\subsection{Distinction of Higgs boson models}

\begin{figure}[htb]
\includegraphics[scale=0.3]{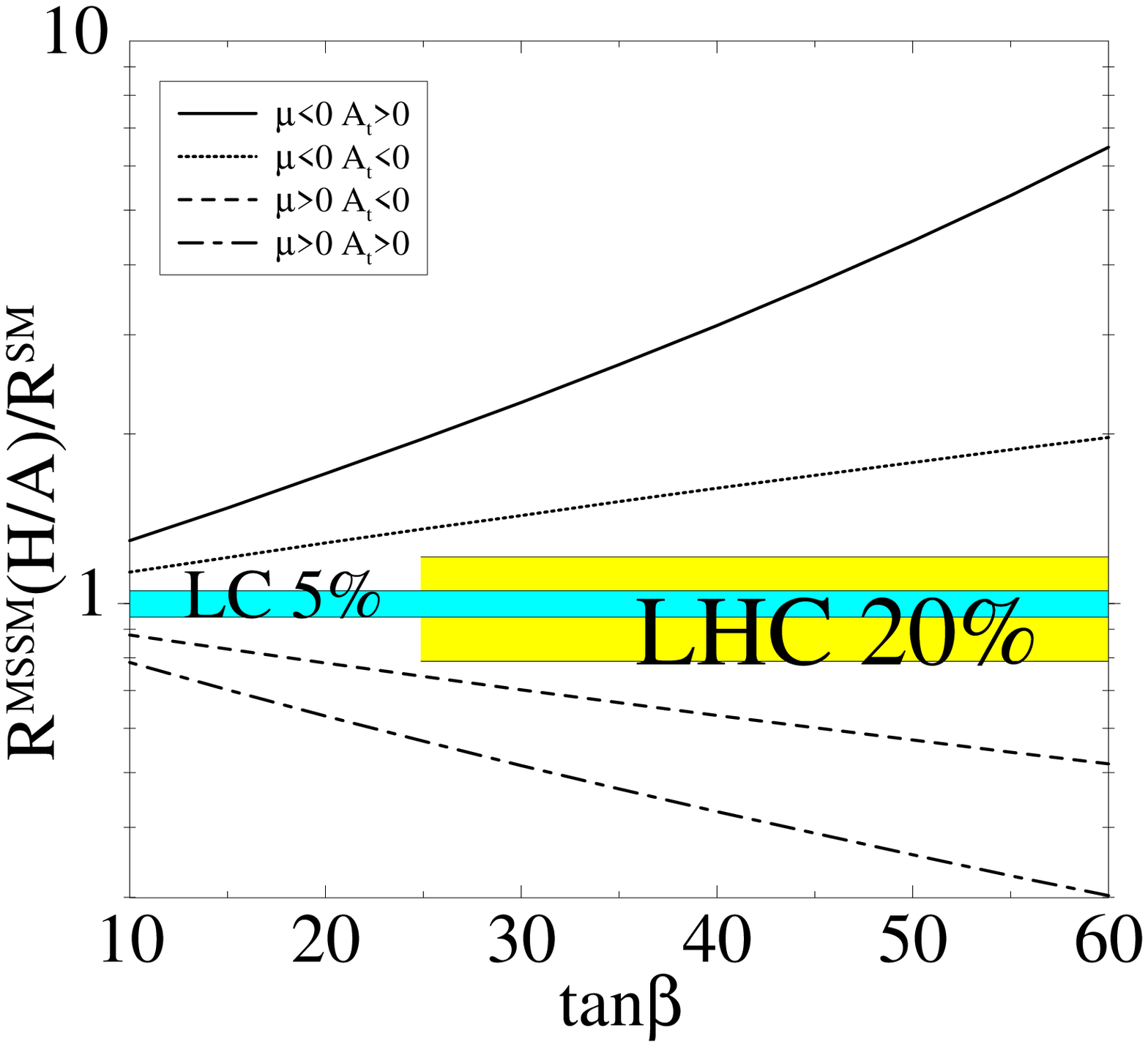}  \hfill
\includegraphics[scale=0.3]{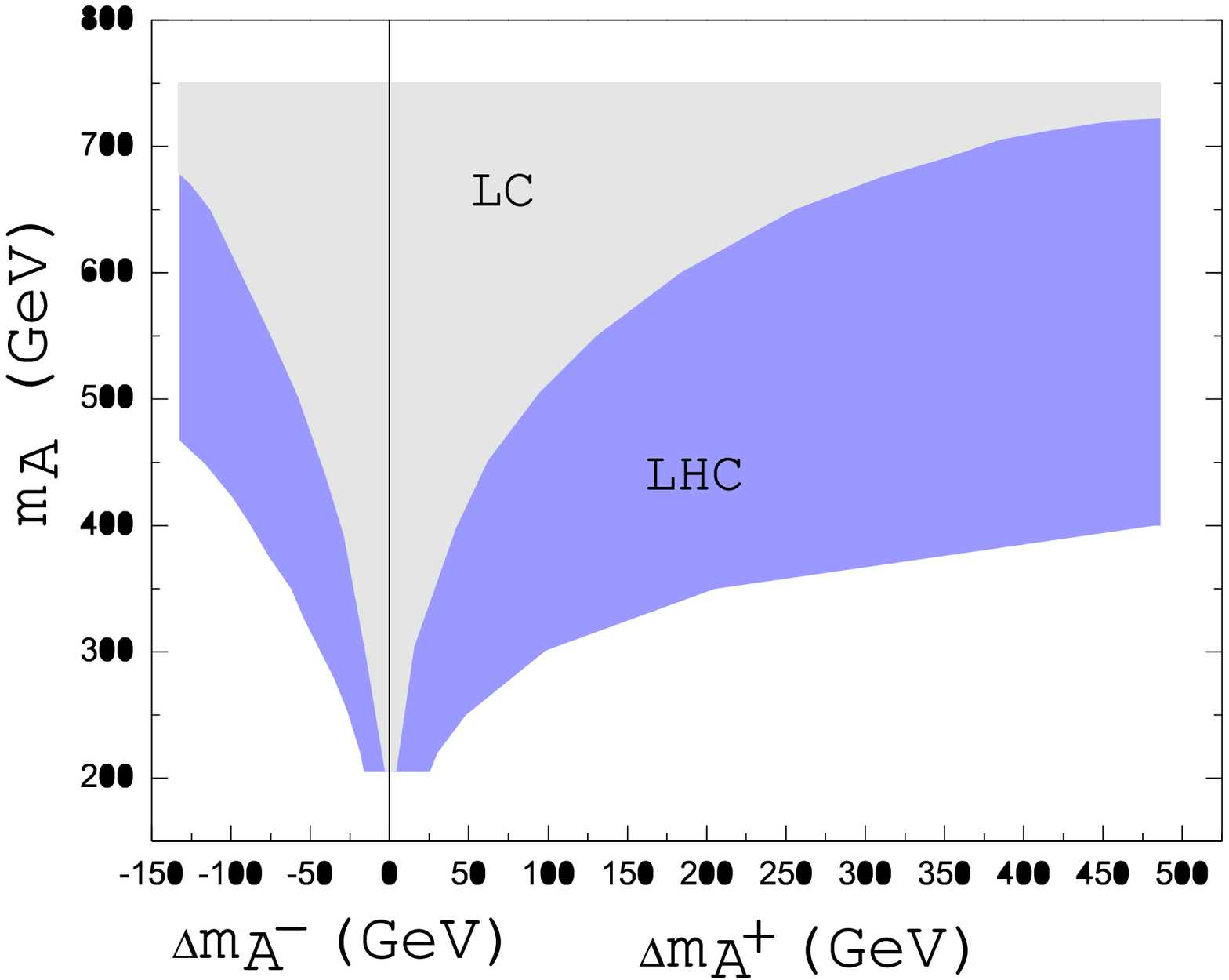} 
\vspace*{-0.3cm}
\caption{
Left: distinction of Higgs boson models from precision
      Higgs boson decay branching ratio determinations.
Right: comparison of LHC and LC A mass predictions from precision
       scalar Higgs boson decay branching ratio determinations.
}
\label{fig:models}
\end{figure}

The distinction of Higgs boson models is very important and could be based
on precision branching ratio measurements. 
The ratio of the Higgs boson decay rates into b-quarks and $\tau$-leptons
is defined by
$$R\equiv BR({\rm H/A\rightarrow b\bar b)}/
          BR({\rm H/A\rightarrow \tau^+\tau^-}).$$
The normalized value of $R$ to the SM expectation can be used 
to distinguish a general 2HDM from the MSSM. 
Large deviations from $R=1$ are
expected in the MSSM for several MSSM parameter combinations
as shown in Fig.~\ref{fig:models} (left plot).
In relation to the LHC, where models can only be distinguished for 
$\tan\beta>25$, a LC covers the entire $\tan\beta$ range.

Important predictions on the pseudoscalar Higgs boson mass
can be made from the precision measurements of the scalar Higgs boson
branching ratios:
$$BR({\rm H\rightarrow b\bar b}) / 
  BR({\rm H\rightarrow WW}).$$
A LC obtains a precision normalized to the SM
expectation of less than $3.5$\%, while 
the expected precision at the LHC is less than $20$\%.
For a 400~GeV pseudoscalar Higgs boson A, its mass can be predicted 
at a LC with an error of about 50~GeV, while at the LHC only 
a lower mass limit can be set.
Figure~\ref{fig:models} (right plot) shows the sensitivity on the prediction of
the A boson mass at the LHC and a LC for a wide range of A masses. 

Beyond the MSSM, the Higgs boson particle spectrum is enriched
for example in the framework of the Non-Minimal Supersymmetric extension
of the SM (NMSSM) where an
extra Higgs boson singlet $\lambda NH_1H_2$ is present.
Such a model could be distinguished from the MSSM by precision
measurements of the Higgs boson masses and comparison with the 
predictions. 
Moreover, additional light neutral Higgs bosons may be observed
and the mass sum rules are modified, leading, for example,
to a reduced mass of the charged Higgs boson.

\section{Conclusions}
\vspace*{-0.2cm}

At LEP various stringent limits from direct Higgs boson searches are set.
      Small but intriguing data accesses have been observed.
Much knowledge has been gained at LEP in preparation for new searches. 
After a first discovery and initial
      precision measurements in some decay modes
      at the Tevatron or the LHC,
      already in the first phase of a LC,
      many Higgs boson decay modes will be measured 
      with very high precision.
The precise LC data will allow the determination of the nature of the
      Higgs sector. 
      Models like the SM, the general 2HDM, the MSSM and the NMSSM 
      will be distinguished for a wide range of parameters.
The underlying mechanism of symmetry breaking and 
      mass generation will be tested.
Like for the top quark (LEP mass prediction, Tevatron observation),
      important consistencies of the model can be probed with
      combined LC and LHC physics. 
After about 12 years of preparatory studies the LC has a 
      solid case and
      the high-energy physics community is prepared to 
      answer fundamental questions
      over the coming decades.

\vspace*{-1mm}
\section*{Acknowledgments}
\vspace*{-1mm}

I would like to thank the organizers of QFTHEP'04 
for their kind hospi\-tality, and the organizers of WONP'05
for the invitation.

\end{document}